\documentclass[twocolumn]{aastex631}

\usepackage{CJK}
\usepackage{multirow}
\usepackage[mathlines]{lineno}
\usepackage{bm}
\usepackage{amsmath}

\usepackage[normalem]{ulem}

\tabletypesize{\scriptsize}

\newcommand{\lx}{\ensuremath{L_\mathrm{X}}}
\newcommand{\lbol}{\ensuremath{L_\mathrm{bol}}}

\newcommand{\xmm}{{XMM-Newton}}

\newcommand{\spitzer}{Spitzer}
\newcommand{\wise}{WISE}

\newcommand{\lsun}{\ensuremath{L_{\sun}}}
\newcommand{\msun}{\ensuremath{M_{\sun}}}

\newcommand{\lumcgs}{\ensuremath{\mathrm{erg}\,\mathrm{s}^{-1}}}
\newcommand{\fluxcgs}{\ensuremath{\mathrm{erg}\,\mathrm{cm}^{-2}\,\mathrm{s}^{-1}}}

\newcommand{\mstar}{\ensuremath{M_\star}}

\newcommand{\nh}{\ensuremath{N_\mathrm{H}}}
\newcommand{\nhcgs}{\ensuremath{\mathrm{cm}^{-2}}}

\newcommand{\lxobs}{\ensuremath{L_\mathrm{X,obs}}}
\newcommand{\lsix}{\ensuremath{L_\mathrm{6\mu m,AGN}}}
\newcommand{\specz}{\mbox{spec-$z$}}
\newcommand{\photoz}{\mbox{photo-$z$}}

\newcommand{\fir}{\ensuremath{f_\mathrm{24\mu m}}}
\newcommand{\mJy}{\ensuremath{\mathrm{mJy}}}
\newcommand{\uJy}{\ensuremath{\mathrm{\mu Jy}}}

\newcommand{\lir}{\ensuremath{L_\mathrm{IR}}}
\newcommand{\kev}{\ensuremath{\mathrm{keV}}}
\newcommand{\mum}{\ensuremath{\mathrm{\mu m}}}

\newcommand{\hard}{\ensuremath{\mathrm{CR}_\mathrm{H}}}
\newcommand{\soft}{\ensuremath{\mathrm{CR}_\mathrm{S}}}

\newcommand{\qfir}{\ensuremath{q_\mathrm{IR}}}
\newcommand{\qmir}{\ensuremath{q_{24}}}
\newcommand{\kbol}{\ensuremath{k_\mathrm{bol}}}
\newcommand{\lambedd}{\ensuremath{\lambda_\mathrm{Edd}}}
\newcommand{\fagn}{\ensuremath{f_\mathrm{AGN}}}

\newcommand{\sfrnorm}{\ensuremath{\mathrm{SFR_{norm}}}}
\newcommand{\sigmaNMAD}{\ensuremath{\sigma_\mathrm{NMAD}}}
\newcommand{\qznew}{\ensuremath{Q_z^\mathrm{good}}}
\newcommand{\qz}{\ensuremath{Q_z}}

\begin{document}

\begin{CJK*}{UTF8}{gbsn}

\title{Dust-Obscured Galaxies in the XMM-SERVS Fields: Selection, Multiwavelength Characterization, and Physical Nature}

\author[0000-0002-6990-9058]{Zhibo Yu (喻知博)}
\affiliation{Department of Astronomy and Astrophysics, The Pennsylvania State University, 525 Davey Lab, University Park, PA 16802, USA}
\email{Email: zvy5225@psu.edu}
\affiliation{Institute for Gravitation and the Cosmos, The Pennsylvania State University, University Park, PA 16802, USA}
\author[0000-0002-0167-2453]{W. N. Brandt}
\affiliation{Department of Astronomy and Astrophysics, The Pennsylvania State University, 525 Davey Lab, University Park, PA 16802, USA}
\affiliation{Institute for Gravitation and the Cosmos, The Pennsylvania State University, University Park, PA 16802, USA}
\affiliation{Department of Physics, 104 Davey Laboratory, The Pennsylvania State University, University Park, PA 16802, USA}
\author[0000-0002-4436-6923]{Fan Zou}
\affiliation{Department of Astronomy and Astrophysics, The Pennsylvania State University, 525 Davey Lab, University Park, PA 16802, USA}
\affiliation{Institute for Gravitation and the Cosmos, The Pennsylvania State University, University Park, PA 16802, USA}
\author[0009-0001-8023-5701]{Ziyuan Zhu}
\affiliation{Department of Astronomy, School of Physics and Technology, Wuhan University, Wuhan 430072, People's Republic of China}
\author[0000-0002-8686-8737]{Franz E. Bauer}
\affiliation{Instituto de Astrof\'isica, Pontificia Universidad Cat\'olica de Chile, Casilla 306, Santiago 22, Chile}
\affiliation{Millennium Institute of Astrophysics (MAS), Nuncio Monse\~nor S\'otero Sanz 100, Providencia, Santiago, Chile}
\affiliation{Space Science Institute, 4750 Walnut Street, Suite 205, Boulder, CO 80301, USA}
\author[0000-0001-6317-8488]{Nathan Cristello}
\affiliation{Department of Astronomy and Astrophysics, The Pennsylvania State University, 525 Davey Lab, University Park, PA 16802, USA}
\author[0000-0002-9036-0063]{Bin Luo}
\affiliation{School of Astronomy and Space Science, Nanjing University, Nanjing 210093, People's Republic of China}
\affiliation{Key Laboratory of Modern Astronomy and Astrophysics (Nanjing University), Ministry of Education, Nanjing 210093, People's Republic of China}
\author[0000-0002-8577-2717]{Qingling Ni}
\affiliation{Max-Planck-Institut f\"ur extraterrestrische Physik (MPE), Gie\ss enbachstra\ss e 1, D-85748 Garching bei M\"unchen, Germany}
\author[0000-0003-0680-9305]{Fabio Vito}
\affiliation{INAF--Osservatorio di Astrofisica e Scienza dello Spazio di Bologna, Via Gobetti 93/3, I-40129 Bologna, Italy}
\author[0000-0002-1935-8104]{Yongquan Xue}
\affiliation{CAS Key Laboratory for Research in Galaxies and Cosmology, Department of Astronomy, University of Science and Technology of China, Hefei 230026, China}
\affiliation{School of Astronomy and Space Science, University of Science and Technology of China, Hefei 230026, People's Republic of China}

\begin{abstract}
Dust-obscured galaxies (DOGs) are enshrouded by dust, and many are believed to host accreting supermassive black holes (SMBHs), which makes them unique objects for probing the coevolution of galaxies and SMBHs. We select and characterize DOGs in the $13\,\deg^2$ XMM-Spitzer Extragalactic Representative Volume Survey (XMM-SERVS), leveraging the superb multiwavelength data from \mbox{X-rays} to radio. We select 3738 DOGs at $z\approx1.6-2.1$ in XMM-SERVS, while maintaining good data quality without introducing significant bias. This represents the largest DOG sample with thorough multiwavelength source characterization. Spectral energy distribution (SED) modeling shows DOGs are a heterogeneous population consisting of both normal galaxies and active galactic nuclei (AGNs). Our DOGs are massive ($\log\mstar/\msun\approx10.7-11.3$), 
174 are detected in \mbox{X-rays}, and they are generally radio-quiet systems. X-ray detected DOGs are luminous  
and are moderately to heavily obscured in \mbox{X-rays}. Stacking analyses for the \mbox{X-ray} undetected DOGs show highly significant average detections
. Critically, we compare DOGs with matched galaxy populations. DOGs have similar AGN fractions compared with typical galaxy populations. \mbox{X-ray} detected DOGs have higher \mstar\ and higher X-ray obscuration, but they are not more star-forming than typical \mbox{X-ray} AGNs. The results potentially challenge the relevance of the merger-driven galaxy-SMBH coevolution framework for \mbox{X-ray} detected DOGs.

\end{abstract}

\keywords{}

\section{Introduction}\label{sec:introduction}

Over the past couple of decades, astronomers have developed a coevolution framework between supermassive black holes (SMBHs) and galaxies \citep[e.g.,][]{Sanders+1988,DiMatteo+2005, Hopkins+2006}. As cold gas accumulates, for example, major mergers can trigger strong star formation (SF) in host galaxies; gas reservoirs also fuel the accretion of central SMBHs, allowing them to be observed as AGNs. Contemporaneously, gas and dust can enshroud the nucleus and cause severe obscuration. AGN feedback also further impacts host galaxies, in which AGN outflows and radiation may suppress SF activity. Such a coevolution framework provides a possible explanation for how the central AGN impacts its host galaxy.

Since the commissioning of wide-field infrared (IR) observatories like the Spitzer Space Telescope and Wide-field Infrared Survey Explorer (\wise), studies of dusty galaxies have been greatly advanced as dust emission can be directly traced by IR observations. \citet{Dey+2008} used \spitzer\ data to efficiently select a sample of dust-obscured galaxies (DOGs) at $z\approx1.5-3$ with $\fir>0.3\,\mJy$ and $(R-[24])_\mathrm{Vega}\geq14$, where $[24]$ is the magnitude at $24\,\mum$. \citet{Toba+2016} applied different selection criteria using \wise\ with $f_\mathrm{22\mu m}>3.8\,\mJy$ and $i-[22]\geq7$ to select the so-called IR-bright DOGs, which are a subpopulation of hyperluminous IR galaxies (HyLIRGs; $\lir>10^{13}\lsun$), where $[22]$ is the AB magnitude at $22\,\mum$. These galaxies constitute a substantial fraction of the IR luminosity density among ultraluminous IR galaxies ($\lir>10^{12}\lsun$) \citep[e.g.,][]{Toba+2015, Toba+2017}. \citet{Eisenhardt+2012} and \citet{Wu+2012} applied the WISE ``W1W2-dropout" technique to select an even more extreme subpopulation among HyLIRGs. 
These objects are similar to DOGs but have higher dust temperatures (up to hundreds of K versus $30-40\,\rm K$), and are therefore dubbed Hot DOGs \citep[e.g.,][]{Tsai+2015,Assef+2016,Vito+2018}. For them, the bolometric luminosity (\lbol) is dominated by IR emission, with the most extreme sources reaching to $\lbol\gtrsim10^{14}\,\lsun$ \citep[e.g.,][]{Tsai+2015,Vito+2018}. 

The strong IR emission from these types of dusty galaxies can be explained by intense SF activity, along with potential contributions from central AGNs that are buried beneath obscuring gas and dust \citep[e.g.,][]{Fiore+2008, Lanzuisi+2009,Eisenhardt+2012,Vito+2018,Toba+2020sofia}. Phenomologically, based upon the spectral energy distribution (SED) shape in the mid-infrared (MIR), DOGs can be classified as ``Power-Law" (PL) or ``Bump" DOGs. PL DOGs exhibit a fairly monotonic MIR SED, while the Bump DOGs show a distinct SED ``bump" at observed-frame $\approx3-10\,\mum$, possibly due to stellar continuum peaking at rest-frame $\approx1.6\,\mum$ \citep[e.g.,][]{Dey+2008,Melbourne+2012,Toba+2015}. The fraction of PL DOGs generally increases with IR flux density, and their SED shape appears to be more AGN-like \citep[e.g.,][]{Melbourne+2012}. Therefore, PL DOGs are generally thought to correspond to AGN-dominated sources, while Bump DOGs correspond to galaxies undergoing strong SF.  Both observations and simulations suggest that DOGs evolve from Bump to PL phase \citep[e.g.,][]{Dey+2009,Bussmann+2012,Yutani+2022}, indicating a possible link with the merger-driven galaxy-SMBH \mbox{coevolution} framework.



X-ray observations can provide further evidence of AGN activity due to the reduced absorption bias of \mbox{X-rays} and the large contrast between the \mbox{X-ray} emission of AGNs and stellar components \citep[e.g.,][]{Brandt&Yang+2022}. The typical fraction of \mbox{X-ray} detected DOGs is $\approx10-20\%$, depending on the \mbox{X-ray} depth of the fields \citep[e.g.,][]{Fiore+2008,Corral+2016,Riguccini+2019}. It is found that \mbox{X-ray} detected DOGs generally have moderate-to-high $\lx$ ($2-10\,\mathrm{keV}$ $\lx\approx10^{43.5}-10^{45}\,\lumcgs$) with a wide range of \nh\ from moderate to Compton-thick (CT)  levels \citep[e.g.,][]{Fiore+2008,Lanzuisi+2009,Stern+2014,Corral+2016,Riguccini+2019,Toba+2020nustar,Kayal+2024}. Studies have attempted to characterize and search for connections between AGN luminosity, obscuration, and host-galaxy properties for different types of dusty galaxies to understand how they are related to the merger-driven galaxy-SMBH coevolution framework. In particular, \mbox{X-ray} studies of the most extreme Hot DOGs show that they are significantly more obscured in \mbox{X-rays} and have comparable or slightly lower \mbox{X-ray} luminosity than optical type~1 quasars with similar \lbol, which indicates that Hot DOGs are caught during extreme SMBH accretion and are likely in the late stage of major mergers \citep{Vito+2018}. However, contrary results are found for \mbox{X-ray}-selected heavily obscured AGNs, which have less extreme optical-IR colors than DOGs. Systematic studies of the origins of their \mbox{X-ray} obscuration and its correlation with host-galaxy properties and morphologies indicate that they are more likely triggered by secular processes instead of mergers \citep{Li+2020}. As for DOGs, theoretical studies suggest that they are in the end stage of major mergers where they are at the peak of SF and starting to transition to the AGN-dominated phase \citep[e.g.,][]{Hopkins+2006,Narayanan+2010,Yutani+2022}; observational studies of the morphology and dust properties of DOGs give some evidence supporting the relevance of the merger-driven coevolution framework for DOGs \citep[e.g.,][]{Bussmann+2012}, but the results are limited by low-resolution images and small sample sizes \citep{Netzer+2015}. Recent results from the James Webb Space Telescope (JWST) on small samples of submillimeter galaxies (SMGs), which are dust-enshrouded like DOGs, show that most of them have non-disturbed disks, suggesting that they may grow via secular processes \citep[e.g.,][]{Cheng+2023,Gillman+2023,LeBail+2024}.

Although the widely used color-based criteria can efficiently select large samples of DOGs, source characterization is often poor due to, e.g., limited multiwavelength coverage, which hinders further detailed analysis to understand their properties. For instance, \citet{Dey+2008} selected $\approx2600$ DOGs in the $\approx8.1\,\deg^2$ NOAO Deep Wide-Field Survey Bo\"otes field, but they were only able to study the IR properties of 86 sources with available spectroscopic redshifts (\specz s). Similarly, \citet{Toba+2016} selected 5311 IR-bright DOGs using WISE across 14555$\,\deg^2$, but only 67 sources have reliable \specz s. Studies also searched for DOGs in deep and medium-deep fields (e.g., the Cosmic Evolution Survey; COSMOS) where source characterization may be more secure. \citet{Riguccini+2019} selected far-infrared (FIR) detected DOGs in COSMOS, but since its sky area is relatively small ($\approx2.2\,\deg^2$), there were only 108 sources in the final sample.

In this work, we select DOGs in the XMM-Spitzer Extragalactic Representative Volume Survey (XMM-SERVS) fields \citep{Chen+2018,Ni+2021} and investigate their nature by analyzing their multiwavelength properties. XMM-SERVS, with $13\deg^2$ of coverage, contains the prime parts of three Deep-Drilling Fields (DDFs) of the Legacy Survey of Space and Time (LSST): Wide Chandra Deep Field-South (W-CDF-S; $4.6\deg^2$), European Large Area Infrared Space Observatory Survey-S1 (ELAIS-S1; $3.2\deg^2$), and XMM-Newton \mbox{Large-Scale} Structure (XMM-LSS; $5.3\deg^2$). These fields provide a large search volume with superb, uniform multiwavelength coverage from \mbox{X-rays} to radio. 
Additionally, XMM-SERVS has excellent prospects for future development as it has been selected for further photometric and spectroscopic surveys, including LSST \citep[e.g.,][]{Ivezic+2019}, Euclid \citep[e.g.,][]{EuclidCollaboration+2024}, Large Millimeter Telescope (LMT) TolTEC \citep[e.g.,][]{Wilson+2020}, Multi-Object Optical and Near-IR Spectrograph \citep[MOONS; e.g.,][]{Cirasuolo+2020}, Subaru Prime Focus Spectrograph \citep[PFS; e.g.,][]{Takada+2014}, and 4-meter Multi-Object Spectroscopic Telescope Wide-Area VISTA Extragalactic Survey \citep[4MOST WAVES; e.g.,][]{Driver+2019}.

This work presents a multiwavelength study of a large sample of DOGs in the XMM-SERVS fields with an emphasis on their \mbox{X-ray} properties. We aim to explore the origin of their \mbox{X-ray} emission, and how the obscuration is related to the host-galaxy properties. Critically, we assess if DOGs fit into the merger-driven galaxy-SMBH coevolution framework by comparing their properties with a control sample of \mbox{X-ray} AGNs.

The structure of this paper is as follows. In Section~\ref{sec:sample}, we present our sample selection. Section~\ref{sec:results} presents our analyses and results on the multiwavelength properties of DOGs. We discuss the physical implications and test the relevance of the merger-driven galaxy-SMBH framework for DOGs in Section~\ref{sec:discussion}. Finally, we summarize our work in Section~\ref{sec:summary}. 

Throughout this paper, we adopt a flat $\Lambda$CDM cosmology with $H_0=70\,\rm km\,s^{-1}\,Mpc^{-1}$, $\Omega_\Lambda=0.70$, and $\Omega_\mathrm{M}=0.30$. Magnitudes are given in the AB system unless otherwise specified. We use the nonparametric $k$-sample Anderson-Darling (AD) test for hypothesis testing, and a significance level of $\alpha=0.001$ is adopted. The AD test is generally more effective than other similar nonparametric tests, such as the two-sample Kolmogorov-Smirnov (KS) test, because it provides more uniform sensitivity across the full ranges of the tested distributions \citep[e.g.,][]{Stephens+1974,Hou+2009,Feigelson&Babu+2012}. We have verified that the statistical results using KS tests are not significantly different from those obtained with AD tests in this paper.

\section{Data and Sample Selection}\label{sec:sample}

\subsection{Initial Selection}\label{subsec:initial_select}

Our DOGs are selected in the XMM-SERVS fields. As per Section~\ref{sec:introduction}, the XMM-SERVS fields are covered by superb multiwavelength surveys from \mbox{X-rays} to radio. These surveys include
\begin{enumerate}
    \item X-ray: XMM-SERVS \citep{Chen+2018,Ni+2021}.
    \item UV: the Galaxy Evolution Explorer \citep[GALEX;][]{Martin+2005}.
    \item Optical: the Hyper Suprime-Cam (HSC) Subaru Strategic Program \citep{Aihara+2018}, HSC imaging in the \mbox{W-CDF-S} \citep{Ni+2019}, the VST Optical Imaging of the CDF-S and ELAIS-S1 \citep[VOICE;][]{Vaccari+2016}, the ESO-Spitzer Imaging extragalactic Survey \citep[ESIS;][]{Berta+2006}, the Dark Energy Survey \citep[DES;][]{Abbott+2021}, and the Canada-France-Hawaii Telescope Legacy Survey \citep[CFHTLS;][]{Hudelot+2012}.
    \item Near-infrared (NIR): the VISTA Deep Extragalactic Observations \citep[VIDEO;][]{Jarvis+2013} survey.
    \item Mid-infrared (MIR) and far-infrared (FIR): the Spitzer DeepDrill survey \citep[DeepDrill;][]{Lacy+2021}, the Spitzer Wide-area Infrared Extragalactic survey \citep[SWIRE;][]{Lonsdale+2003}, and the Herschel Multi-tiered Extragalactic Survey \citep[HerMES;][]{Oliver+2012}.
    \item Radio: the Australia Telescope Large Area Survey \citep[ATLAS; e.g.,][]{Franzen+2015}, the VLA survey \citep[e.g.,][]{Heywood+2020}, and the MeerKAT International GHz Tiered Extragalactic Exploration \citep[MIGHTEE;][]{Jarvis+2013, Heywood+2022} survey.
\end{enumerate}
All of these surveys are utilized in our work. For detailed lists of covered surveys, see Table~1 of \citet{Zou+2022} and Table~1 of \citet{Zhu+2023}.


Among the aforementioned surveys, the \spitzer\ $24\,\rm\mu m$ coverage reaches $0.1\,\mJy$ at $5\sigma$ depth \citep{Lonsdale+2004}, which is sufficient to completely sample DOGs (defined as having $\fir>0.3\,\mJy$). The \mbox{X-ray} survey made by \xmm\ has a roughly uniform $50\,\rm ks$ exposure across the fields, reaching a flux limit of \mbox{$\approx10^{-15}-10^{-14}\,\fluxcgs$} at $0.5-10\,\kev$. The survey is currently the largest medium-depth \mbox{X-ray} survey and has provided over 10200 AGNs. Additionally, sensitive radio surveys at $1.4\,\rm GHz$ ($5\sigma$ flux limit $\approx28-85\,\rm \mu Jy$) allow us to perform systematic radio analyses of DOGs. Moreover, the $0.36-4.5\,\rm\mu m$ photometry has been refined via a forced-photometry technique to reduce source confusion and ensure consistency among different bands \citep{Nyland+2017,Nyland+2023, Zou+2021a}. The forced photometry utilizes deep fiducial images from the VIDEO survey. The \specz s are taken from several spectroscopic surveys, and the photometric redshifts (\photoz s) are compiled from \citet{Ni+2021} and \citet{Zou+2021b} for W-CDF-S and ELAIS-S1, and from \citet{Chen+2018} for XMM-LSS. The \photoz s are primarily calculated using the \photoz\ code \texttt{EAZY} \citep{Brammer+2008}, which estimates \photoz\ by fitting the observed photometry with various galaxy (and optionally, AGN) templates. 
Although the \photoz s in XMM-SERVS generally have high quality with a catastrophic outlier fraction of a few percent, we will show in Section~\ref{subsec:Qz_cut} that, for DOGs particularly, the \photoz\ quality is not optimal and requires additional quality cuts to ensure reliable source characterization.

Utilizing the \mbox{X-ray} to FIR coverage, \citet{Zou+2022} have measured the host-galaxy properties including stellar mass (\mstar) and star-formation rate (SFR) in XMM-SERVS via fitting the SED with \texttt{CIGALE} \citep{Boquien+2019,Yang+2022}, where the AGN emission has been properly considered. We will use the catalogs provided by \citet{Zou+2022} as our parent sample. Note that \texttt{CIGALE} is not used as a \photoz\ estimator.


We filter out stellar objects reported in \citet{Zou+2022} and apply the same criteria as in \citet{Dey+2008} to select our preliminary DOG sample, i.e., \mbox{$\fir>0.3\,\mJy$} and \mbox{$\fir/f_R\geq982$},\footnote{The color-selection criterion is equivalent to the originally defined $(R-[24])_\mathrm{Vega}\geq14$ in \citet{Dey+2008}. We also apply the corrections from Appendix~D of \citet{Zhu+2023} to the $24\,\rm \mu m$ flux. A $0.1\,\rm mag$ offset is applied to the $R$-band photometry in XMM-LSS to account for the difference between our forced-photometry catalog and the original catalogs.} where $\fir$ and $f_R$ are the observed-frame flux densities at $24\,\mu\rm m$ and in the $R$ band, respectively. The selection is done in three steps. First, we restrict our sample region to the intersection of the footprints cataloged by \citet{Zou+2022}, the $R$-band coverage, the $24\,\mum$ coverage, and the \mbox{X-ray} coverage. This results in $\approx 2.2$~million sources within a $\approx20\%$ smaller area than XMM-SERVS: $3.5\deg^2$ in W-CDF-S, $2.7\deg^2$ in ELAIS-S1, and $4.1\deg^2$ in XMM-LSS. Second, we apply \mbox{$\fir>0.3\,\mJy$} in our sample region and select 31853 sources. After that, we convert the $R$-band magnitude measured through different $R$ filters in XMM-SERVS \citep[see Table~1 in][]{Zou+2022} to the same $R$ filter used in \citet{Dey+2008} using their best-fit SED for consistency. The correction is generally small. For sources with non-positive $R$-band flux measured in forced photometry, we use the $R$-band flux estimated from their best-fit SED. Finally, we apply \mbox{$\fir/f_R\geq982$} and obtain 3738 sources in XMM-SERVS. The sky density of our selected DOGs ($363\,\deg^{-2}$) is similar to that in \citet{Dey+2008} ($\approx321\,\deg^{-2}$). There are 174 DOGs detected in \mbox{X-rays}. The median net source counts at $0.5-10\,\kev$ of the \mbox{X-ray} detected DOGs are 128 for all XMM-Newton EPIC cameras (PN, MOS1, and MOS2) combined, and the corresponding $25-75\%$ quantile range is $87-200$. We refer to these 3738 DOGs as our ``full sample".

It is worth noting that our forced photometry utilizes the reddest VIDEO band in which the source is detected as the fiducial band. We further examine the VIDEO $K_S$ band magnitude distributions for all sources with $f_{24\,\rm\mu m}>0.3\,\mJy$. These sources (median $K_S=19.4$) are generally much brighter than the $K_S$ magnitude limit ($K_S=24$) in our fields, and only $\approx0.6\%$ are fainter than the $K_S$ magnitude limit. The results indicate that we do not miss a significant fraction of sources with $f_{24\,\rm\mu m}>0.3\,\mJy$ in our fiducial images, and our forced photometry allows us to sample almost all the DOGs in our search volume.

\begin{figure}[tb!]
    \centering
    \includegraphics[width=\linewidth]{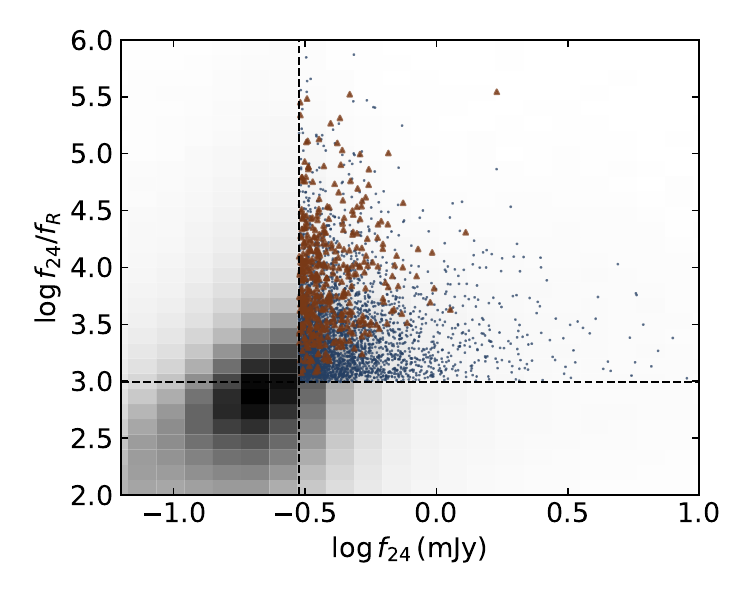}
    \caption{The $\fir/f_R$ versus $\fir$ distribution. The dashed lines represent our color-selection thresholds, and sources in the upper-right region are selected as DOGs. The blue points represent DOGs with positive measured $R$-band flux. The brown up-triangles represent DOGs with non-positive \mbox{$R$-band} flux measured via forced photometry, whose $R$-band flux is estimated from the best-fit SED. The grayscale cells are the 2-D histogram for all $24\,\rm\mu m$-detected galaxies in XMM-SERVS, with darker cells representing more galaxies.}\label{fig:selection}
\end{figure}
\begin{figure*}[tb!]
    \centering
    \includegraphics[width=\linewidth]{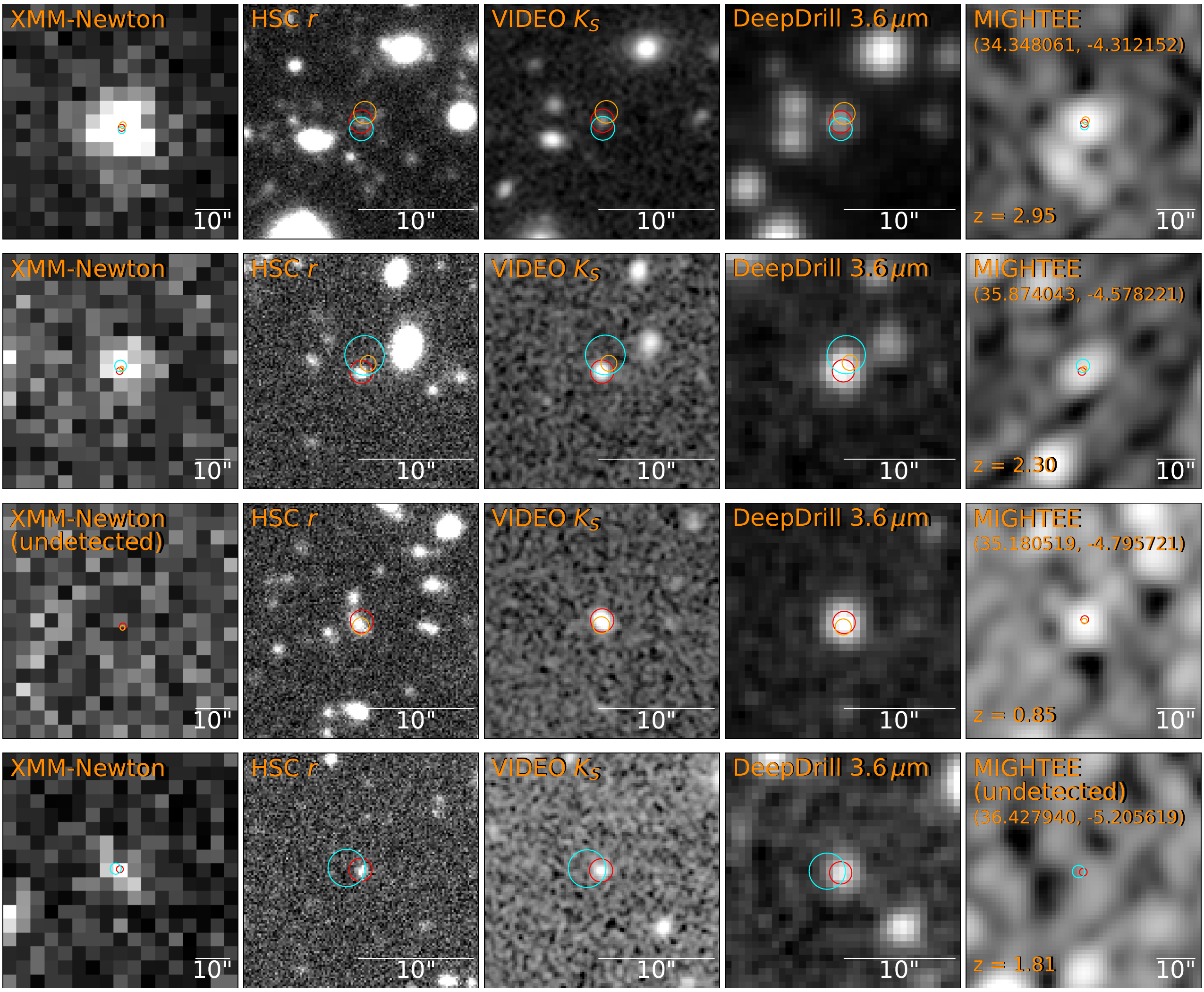}
    \caption{Illustrations of four DOGs in our full sample. For each source, we show the \xmm\ \mbox{0.5--10\,\kev} (first column; $60\arcsec\times60\arcsec$), HSC $r$-band (second column; $20\arcsec\times20\arcsec$), VIDEO $K_S$-band (third column; $20\arcsec\times20\arcsec$), DeepDrill IRAC $3.6\,\mum$ (fourth column; $20\arcsec\times20\arcsec$), and MIGHTEE $1.4\,\rm GHz$ (last column; $60\arcsec\times60\arcsec$) images. The first two sources are detected in all five bands. The third source is not detected in \mbox{X-rays}, and the fourth source is not detected in radio. The VIDEO position and the redshift are shown in the last column for each source. X-ray positions are marked as cyan circles with a 68\% error radius \citep{Chen+2018}; VIDEO positions are marked as red circles with a $1\arcsec$ radius; and MIGHTEE positions are marked as orange circles with a 68\% error radius \citep{Zhu+2023}.}\label{fig:postage}
\end{figure*}

We show the $\fir/f_R$ versus $\fir$ distribution for the 3738 selected DOGs and all $24\,\rm\mu m$-detected galaxies in XMM-SERVS in Figure~\ref{fig:selection}. By construction, DOGs have higher $\fir$ and are redder than typical galaxies. Among our DOGs, only $0.9\%$ (31) have available \specz s. Most of these sources lack detailed classification of galaxy/AGN type or publicly available spectra. We are able to identify one object as a type~1 AGN and one as a type~2 AGN, both observed by the Sloan Digital Sky Survey \citep{Abdurro'uf+2022}. The \specz\ fraction is much smaller than that generally for XMM-SERVS ($\approx4\%$), due to the faintness in optical bands imposed by our selection criteria. Thus, the majority of our sources only have \photoz s available. For illustration, we show a typical collection of four DOGs in Figure~\ref{fig:postage}. In the next subsection, we will further assess the reliability of the \photoz s.

\subsection{Additional Criterion for Reliable Redshifts}\label{subsec:Qz_cut}

Although the \photoz s in \citet{Chen+2018} and \citet{Zou+2021b} are generally reliable, they are expected to be less reliable for our DOGs. 
This is mainly because DOGs are extreme sources whose SEDs may be significantly faint in the optical bands and lack strong spectral features \citep[e.g.,][]{Perez-Gonzalez+2005,Polletta+2006,Dey+2008}, which makes it more difficult for \photoz\ determination using SED fitting. Among the 31 DOGs with \specz s, the outlier fraction\footnote{$f_\mathrm{outlier}$ is defined as the fraction of sources with \mbox{$|z_\mathrm{phot}-z_\mathrm{spec}|/(1+z_\mathrm{spec})>0.15$}, where $z_\mathrm{phot}$ and $z_\mathrm{spec}$ represent \photoz\ and \specz, respectively.} ($f_\mathrm{outlier}$) for their \photoz s is $14/31=45\%$.



\begin{figure}[htb!]
    \centering
    \includegraphics[width=\linewidth]{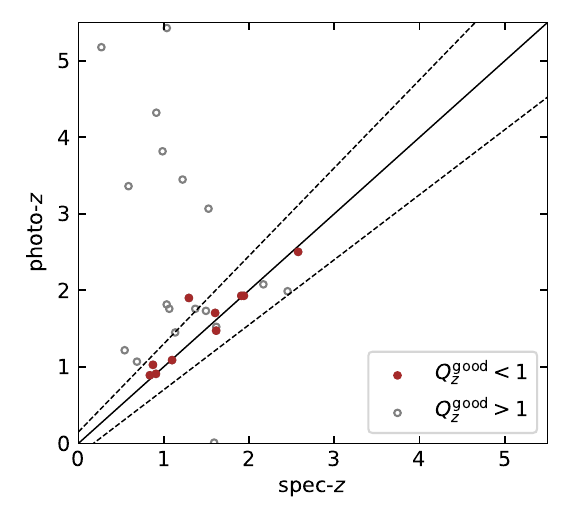}
    \caption{Comparison between the \photoz s and \specz s for sources with \specz\ measurements. The brown solid circles represent sources with $\qznew<1$. The grey empty circles are sources with $\qznew>1$. The $\qznew<1$ cut significantly reduces the outlier fraction. The solid line indicates the $z_\mathrm{phot} = z_\mathrm{spec}$ relation. The dashed lines represent the $|z_\mathrm{phot}-z_\mathrm{spec}|/(1+z_\mathrm{spec})>0.15$ relation.}\label{fig:photoz_compare}
\end{figure}

To minimize the impact of the above problem and reduce $f_\mathrm{outlier}$, we further select DOGs with more reliable \photoz s by employing the empirical \photoz\ quality indicator, $Q_z$, defined in Equation~8 of \citet{Brammer+2008}. It combines several pieces of information when deriving \photoz s: the best-fit statistic, the number of photometric bands used, and the total integrated probability of \photoz\ within $\pm0.2$ of the best-fit \photoz. Small $Q_z$ indicates high reliability. A general threshold for reliable \photoz\ is $Q_z<1$. However, this threshold may not be suitable for DOGs as they tend to be faint in the optical bands such that those bands do not necessarily provide useful constraints on the SED shape. Thus, we slightly modify the definition of $Q_z$: in Equation~8 of \citet{Brammer+2008}, instead of using the total number of bands, we only consider the number of ``good"  bands defined as having signal-to-noise ratio (SNR) greater than $3$. This new \photoz\ quality indicator ($Q_z^\mathrm{good}$) is more indicative of the quality of photometric measurements for sources with extreme colors similar to DOGs.\footnote{We have checked that for our DOGs, the bluest ``good" band is mostly VIDEO $Z$ band, $Y$ band, or $J$ band. At our median $z\approx1.8$, the ``good" bands cover rest-frame optical to $U$-band, which is acceptable for \photoz\ measurements.} We consider a threshold of $\qznew<1$ for high-quality \photoz s, which is more stringent than the nominal threshold of $\qz<1$. Among our DOGs with $\qznew<1$, the $f_\mathrm{outlier}$ for sources with \specz\ is reduced to $1/10=10\%$, indicating our $\qznew<1$ cut can greatly improve the \photoz\ quality for DOGs. The comparison between \photoz s and \specz s for DOGs with \specz\ is shown in Figure~\ref{fig:photoz_compare}.

\begin{figure}[tb!]
    \centering
    \includegraphics[width=\linewidth]{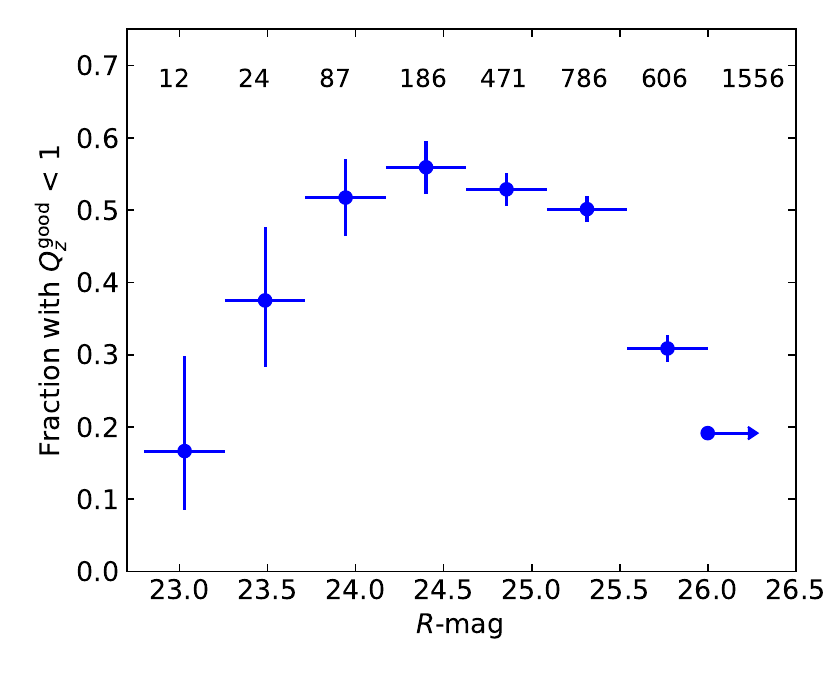}
    \caption{Fraction of sources in our full sample with $\qznew<1$ as a function of $R$-mag. Each magnitude bin has a width of 0.45 mag except for the last one showing all sources with $R\text{-mag}>26$. The number above each bin shows the number of sources in the bin. The $y$-axis error bar represents the $1\sigma$ binomial confidence interval.}\label{fig:Rcut}
\end{figure}

In Figure~\ref{fig:Rcut}, we plot the fraction of sources with high-quality \photoz s ($\qznew<1$) as a function of $R$-band magnitude. The highest fraction of sources with $\qznew<1$ is $\approx55\%$ at $R\approx24-25$. Toward the faint end, the fraction of sources with high-quality \photoz s decreases mainly due to the degradation of the photometric measurements. Toward the bright end, there is a decrease in the fraction. This arises from the increased fraction of sources hosting AGNs, and it has been found that larger fractional AGN contribution to the total flux will result in higher $Q_z$ values \citep[e.g.,][]{Zou+2023}.

Out of our full sample of 3738 DOGs, we refer to the 1309 sources with \specz s or $\qznew<1$ to be our ``core sample", among which 88 are detected in \mbox{X-rays}. The median net source counts at $0.5-10\,\kev$ of the \mbox{X-ray} detected DOGs are 126, and the corresponding $25-75\%$ quantile range is \mbox{$88-200$}. We also test that at the bright end of Figure~\ref{fig:Rcut} ($R<24.5$), the fraction of \mbox{X-ray} detected DOGs in the core sample (19.4\%) is similar to that in the full sample (20.3\%), to some extent indicating that the $\qznew$ cut does not preferentially exclude \mbox{X-ray} detected DOGs. Throughout the paper, we mainly report results for the core sample for a clear narrative flow, and we will show that the results for both samples are similar.\footnote{Our results are not highly sensitive to the $\rm SNR>3$ definition of a ``good" band, but the sample size can be affected by the definition. We verify that changing the definition of a ``good" band to having $\rm SNR>5$ reduces the core-sample size by $\approx1/4$, and the results throughout the paper remain largely unchanged.} A catalog containing all our selected DOGs is given in Table~\ref{tab:data_table}. We also summarize the subsamples used in this work in Table~\ref{tab:subsample} for readability.

\setlength\tabcolsep{2pt}
\begin{splitdeluxetable*}{cccccccccccccccBccccccccccccccBccccccccccccccc}
\tablecaption{Selected DOGs in XMM-SERVS. \label{tab:data_table}}
\tablewidth{0pt}
\tablehead{
\colhead{Field} & \colhead{RA} & \colhead{Dec} & \colhead{Tractor ID} & \colhead{$z$} & \colhead{$z_\mathrm{phot}^\mathrm{low}$} & \colhead{$z_\mathrm{phot}^\mathrm{up}$} & \colhead{$z$-type} & \colhead{$Q_z^\mathrm{good}$} & \colhead{Core sample} & \colhead{XID} & \colhead{X-ray AGN} & 
\colhead{\shortstack{Reliable SED \\ AGN }} & \colhead{\shortstack{SED AGN \\ candidate}} & \colhead{\shortstack{Radio AGN \\ via $\qfir$}}  & \colhead{\shortstack{Radio AGN \\ in \\ \citet{Zhu+2023}}} & \colhead{$f_\mathrm{24\mu\rm m}$} & \colhead{Err\{$f_\mathrm{24\mu\rm m}$\}} & \colhead{$f_\mathrm{24\mu\rm m}/f_R$} & \colhead{$\log\mstar$} & \colhead{Err\{$\log\mstar$\}} & \colhead{$\log \rm SFR$} & \colhead{Err\{$\log \rm SFR$\}} & \colhead{$\log\mathrm{SFR_{norm}}$} & \colhead{Err\{$\log\mathrm{SFR_{norm}}$\}} &
\colhead{$\log L_\mathrm{bol}$} & \colhead{$\log L_\mathrm{1.4\,GHz}$} & \colhead{Err\{$\log L_\mathrm{1.4\,GHz}$\}} & \colhead{$\alpha_r$} & \colhead{$\log L_\mathrm{IR}$} & \colhead{$\log\lsix$} & \colhead{Err\{$\log\lsix$\}} & \colhead{$f_\mathrm{6\mu\rm m,AGN}$} & \colhead{Err\{$f_\mathrm{6\mu\rm m,AGN}$\}} & \colhead{$\log\lxobs$} & \colhead{Err\{$\log\lxobs$\}} & \colhead{$\mathrm{HR}^\mathrm{med}$} & \colhead{$\mathrm{HR}^\mathrm{low}$} & \colhead{$\mathrm{HR}^\mathrm{up}$} & \colhead{$\log\nh^\mathrm{med}$} & \colhead{$\log\nh^\mathrm{low}$} & \colhead{$\log\nh^\mathrm{up}$} & \colhead{\shortstack{lower-luminosity \\ Hot DOG}}\\
\colhead{} & \colhead{[deg]} & \colhead{[deg]} & \colhead{} & \colhead{} & \colhead{} & \colhead{} & \colhead{} & \colhead{} & \colhead{} & \colhead{} &
\colhead{} & \colhead{} & \colhead{} & \colhead{} & \colhead{} & \colhead{[mJy]} & \colhead{[mJy]} & \colhead{} & \colhead{[\msun]} & \colhead{[\msun]} & \colhead{[\msun/yr]} & \colhead{[\msun/yr]} &
\colhead{} & \colhead{} & \colhead{[\lsun]} & \colhead{[W/Hz]} & \colhead{[W/Hz]} &
\colhead{} & \colhead{[W]} & \colhead{[\lumcgs]} & \colhead{[\lumcgs]} & \colhead{} & \colhead{} & \colhead{[\lumcgs]} & \colhead{[\lumcgs]} & \colhead{} & \colhead{} & \colhead{} & \colhead{[\nhcgs]} & \colhead{[\nhcgs]} & \colhead{[\nhcgs]}\\
\colhead{(1)} & \colhead{(2)} & \colhead{(3)} & \colhead{(4)} & \colhead{(5)} & \colhead{(6)} & \colhead{(7)} & \colhead{(8)} & \colhead{(9)} &
\colhead{(10)} & \colhead{(11)} & \colhead{(12)} & \colhead{(13)} & \colhead{(14)} & \colhead{(15)} & \colhead{(16)} & \colhead{(17)} & \colhead{(18)} &
\colhead{(19)} & \colhead{(20)} & \colhead{(21)} & \colhead{(22)} & \colhead{(23)} & \colhead{(24)} & \colhead{(25)} & \colhead{(26)} & \colhead{(27)} & \colhead{(28)} & \colhead{(29)}  & \colhead{(30)} & \colhead{(31)} & \colhead{(32)} & \colhead{(33)}  & \colhead{(34)} & \colhead{(35)} & \colhead{(36)} & \colhead{(37)} & \colhead{(38)} & \colhead{(39)} & \colhead{(40)} & \colhead{(42)} & \colhead{(42)} & \colhead{(43)}
}
\startdata
W-CDF-S&51.912735&--28.035887&467148&3.32 & 2.37 & 3.43 &\photoz &3.3652&0&WCDFS0256&1&1&1&1&0&0.535& 0.019 &2748&11.67 & 0.04 &2.36 & 0.20 &0.50 & 0.20 &13.13&    25.670 & 0.024 & --99 &39.60&45.70 & 0.03 &0.87& 0.06 &44.45& 0.08&--0.362& --0.511 & --0.215 & 22.92 & 22.40 & 23.24 & 0\\
W-CDF-S&52.992867&--27.844992&505262&1.90 & 1.77 & 2.21 &\photoz &0.3293&1&WCDFS2040&1&0&1&--1&--1&0.882& 0.017 &1679&11.22 & 0.11 & 1.98 & 0.27 &0.15&0.27&12.75&  --99& --99 & --99 &39.24&45.21& 0.13&0.76& 0.09&44.29& 0.10 &--0.299& --0.460 & --0.145 & 22.60 & 22.16 & 22.92 & 0\\
W-CDF-S&51.996666&--28.574844&328555&1.55& 1.52 & 2.11 & \photoz &1.1147&0&WCDFS0116&1&1&1&--1&--1&1.328& 0.017 &4799&11.34 & 0.21 &2.30  & 0.19&0.63&0.19&12.70&  --99 & --99 & --99 &39.16&45.40& 0.16&0.86& 0.06&43.81& 0.10 &--0.038& --0.378 & 0.256 & 22.92 & 22.28 & 23.28 & 0\\
W-CDF-S&52.026657&--28.955915&262032&1.65& 1.50 & 1.68 & \photoz &1.2292&0&WCDFS0149&1&1&1&--1&--1&0.797& 0.015 &1289&11.25 & 0.05 & 1.99 & 0.11&0.26&0.11&12.63&   --99& --99 & --99 &39.08&45.03& 0.07&0.81& 0.04&44.31& 0.15 &--0.209& --0.568 & 0.111 & 22.68 & 21.56 & 23.16  & 0\\
W-CDF-S&51.869560&--28.623209&456099&3.19& 2.56 & 3.35 & \photoz &4.8053&0&WCDFS0063&1&0&1&--1&--1&0.325& 0.020 &1550&10.89 & 0.12  &2.23 & 0.18 & 0.00&0.18&12.78&   --99 & --99 &  --99&39.18&45.46& 0.06&0.84& 0.08&45.16 & 0.11&--0.130& --0.274 & --0.008 & 23.36 & 23.12& 23.56  & 0\\
\enddata
\tablecomments{We only show five representative rows of our selected DOGs here. The full table is available as supplementary material. Column (1): field name. Columns (2) and (3): J2000 RA and Dec. Column (4): Tractor ID in \citet{Zou+2022}. Column (5): redshift. Columns (6) and (7): the 68\% lower and upper limits of \photoz. Sources with \specz\ are assigned --1. Column (8): redshift type. Column (9): new photo-$z$ quality indicator defined in Section~\ref{subsec:Qz_cut}. Column (10): flag for our core sample defined in Section~\ref{subsec:Qz_cut}. Column (11): \mbox{X-ray} source ID in \citet{Chen+2018} and \citet{Ni+2021}. Entries for sources not detected in \mbox{X-rays} are assigned --1. Column (12): flag for \mbox{X-ray} AGNs in \citet{Zou+2020}. Sources not detected in \mbox{X-rays} are assigned --1. Columns (13) and (14): flags for reliable SED AGNs, and SED AGN candidates in \citet{Zou+2022}. Column (15): flag for radio AGNs selected via \qfir\ in \mbox{Zhang et al. (submitted)}. Sources not detected in radio are assigned --1. Column (16): flag for radio AGNs selected via \qmir, morphology, or spectral index in \citet{Zhu+2023}. Sources not detected in radio are assigned --1. Columns (17) and (18): flux density at observed-frame $24\,\mu\rm m$ and its $1\sigma$ uncertainty. Column (19): $24\,\mu\rm m$-to-$R$ flux ratio. For the sources with non-positive $f_R$ via forced photometry, the $f_R$ is estimated from the best-SED, and the flux ratio is multiplied by --1. Columns (20) -- (23): logarithms of best-fit \mstar\ and SFR and their associated uncertainties in \citet{Zou+2022}. Columns (24) and (25): logarithms of \sfrnorm\ and its $1\sigma$ uncertainty. The uncertainty only considers the contribution from SFR (see Section~\ref{subsec:sed-fitting}). Column (26): logarithms of bolometric luminosity. Columns (27) and (28): logarithms of rest-frame 1.4\,GHz monochromatic luminosity and its $1\sigma$ uncertainty. Column (29): radio spectral slope calculated from measurements at 1.4\,GHz and higher/lower frequencies. For W-CDF-S and ELAIS-S1, ATLAS 2.3\,GHz is preferred over RACS. For XMM-LSS, LOFAR is preferred over RACS. Sources without multi-frequency measurements are assigned --99. Column (30): logarithms of total luminosity over rest-frame 8--1000\,$\mu$m. Sources not detected in radio are assigned --99. Columns (31) and (32): logarithms of rest-frame $6\,\mu\rm m$ luminosity contributed by the AGN component and its $1\sigma$ uncertainty. Sources with best-fit normal-galaxy models are assigned --99. Columns (33) and (34): fractional AGN flux contribution at rest-frame $6\,\mu\rm m$ and its $1\sigma$ uncertainty. Sources with best-fit normal-galaxy models are assigned --99. Columns (35) and (36): logarithm of the observed \mbox{X-ray} luminosity at rest-frame 2--10\,\kev\ and its associated uncertainty. Columns (37)--(39) the median, 68\% lower and upper limits of hardness ratio. Sources not detected in \mbox{X-rays} are assigned --99. Columns (40)--(42): the median, 68\% lower and upper limits of the intrinsic column density calculated via hardness ratio. Sources not detected in \mbox{X-rays} are assigned --99. Column (43): flag for lower-luminosity Hot DOG candidates selected in Appendix~\ref{appendix:b}}.
\end{splitdeluxetable*}

\begin{deluxetable*}{lcclc}
\centering
\tablecaption{Summary of the major subsamples of our DOGs. \label{tab:subsample}}
\tablewidth{0pt}
\tablehead{Subsample & \shortstack{Number in \\ Full sample} & \shortstack{Number in \\ Core sample} & Definition & \shortstack{First defined \\ in Section}}
\startdata
\multirow{2}{*}{Total} & \multirow{2}{*}{3738} & \multirow{2}{*}{1309} & \mbox{$\fir>0.3\,\mJy$} and \mbox{$\fir/f_R\geq982$} in our sample region & \multirow{2}{*}{\ref{subsec:initial_select}, \ref{subsec:Qz_cut}}\\
 & & & Core sample has $\qznew<1$ &\\
\hline
X-ray detected & 174 & 88 & Detected in \mbox{X-rays} & \ref{subsec:initial_select}\\
\multirow{2}{*}{X-ray AGN} & \multirow{2}{*}{174} & \multirow{2}{*}{88} & Identified as an \mbox{X-ray} AGN  & \multirow{2}{*}{\ref{subsec:sample_dist}}\\ [-3pt]
 & & & in \citet{Chen+2018} or \citet{Ni+2021} & \\
X-ray undetected & 3564 & 1221 & Not detected in \mbox{X-rays} & \ref{subsec:sed-fitting}\\
\hline
Radio detected & 745 (54) & 317 (27) & Detected in radio & \multirow{3}{*}{\ref{subsec:radio}}\\
\multirow{2}{*}{Radio AGN} & \multirow{2}{*}{172 (26)} & \multirow{2}{*}{73 (15)} & Identified as a radio AGN  & \\ [-3pt]
 & & & in \citet{Zhu+2023} or Zhang et al. (submitted) & \\
\hline
SED AGN candidate & 1887 (174) & 523 (88) & Identified as a SED AGN candidate in \citet{Zou+2022} & \multirow{3}{*}{\ref{subsec:sed-fitting}}\\
Reliable SED AGN & 412 (79) & 104 (37) & Identified as a reliable SED AGN in \citet{Zou+2022} &\\
Normal galaxy & 1851 (0) & 786 (0) & Not identified as a SED AGN candidate in \citet{Zou+2022} &\\
\hline
\multirow{4}{*}{``Safe"} & \multirow{4}{*}{2808 (81)} & \multirow{4}{*}{874 (42)} & $\mstar$ lower than the maximum $\mstar$ for reliable classification  & \multirow{4}{*}{\ref{subsec:host-galaxy}} \\ [-3pt]
 & & & of star-forming galaxies in Equation~\ref{eqn:max_mstar}, defined as & \\ [-3pt]
& & & the regions of the $z-\mstar$ plane where the fraction & \\ [-3pt]
& & &  of quiescent galaxies less than 0.5
\citep{Cristello+2024} & \\
\hline
\multirow{2}{*}{Lower-luminosity Hot DOG} & \multirow{2}{*}{62 (1)} & \multirow{2}{*}{7 (0)} & (1) $\rm W1-W4>9.7$ and $\rm W2-W4>8.2$ or  & \multirow{2}{*}{Appendix~\ref{appendix:b}} \\ [-3pt]
& & & (2) $\rm W1-W3>6.8$ and $\rm W2-W3>5.3$ & \\
\enddata
\tablecomments{The number of \mbox{X-ray} detected sources in the subsamples is shown in parentheses.}
\end{deluxetable*}

\subsection{SED Fitting}\label{subsec:sed-fitting}

In this subsection, we briefly explain the SED fitting and the SED-based classification in \citet{Zou+2022}. We will present our detailed analyses of the host-galaxy properties of our DOGs in Section~\ref{subsec:host-galaxy}. Interested readers can refer to \citet{Zou+2022} for more details on the SED fitting in XMM-SERVS.

\citet{Zou+2022} performed SED-fitting to classify non-stellar sources into AGN candidates and normal galaxies based upon calibrated Bayesian information criterion (BIC) and best-fit $\chi^2$ values; see their Section~3.2. They also classified a subset of AGN candidates as ``reliable SED AGNs" based upon further calibrations against the ultradeep Chandra Deep Field-South with $7\,\rm Ms$ Chandra observations \citep[CDF-S;][]{Luo+2017}. The reliable SED AGNs are expected to reach a $\gtrsim75\%$ purity, and the classification has been tested to be robust \citep[see Section~3.2.4 of][]{Zou+2022}. For each source, \citet{Zou+2022} provided the best-fit SED model for the statistically preferred category. They also included normal-galaxy fitting results for all sources in the catalog. The SEDs are generally of high quality as the median number of photometric bands with $\mathrm{SNR}>5$ is 9, and $\approx85\%$ of the sources have at least 7 photometric bands with $\mathrm{SNR}>5$ ranging from the UV to FIR.

Tables~\ref{tab:sed_gal} and \ref{tab:sed_agn} summarize the \texttt{CIGALE} parameter settings for normal galaxies and AGN candidates in \citet{Zou+2022}. We have verified that the SED-fitting parameter settings are suitable even for extreme sources like DOGs via several tests. 
First, we have confirmed that the best-fit $E(B-V)$ values for our sources are well below the maximum allowed value of $E(B-V)=1.5$ in the settings, indicating that the reddening is acceptably modeled. Second, we test a more complex dust-emission module adopted from \citet{Draine+2014} (\texttt{dl2014}), following the settings in \citet{Yang+2023}. In this test, the mass fraction of polycyclic aromatic hydrocarbons (PAHs) compared to total dust ($q_\mathrm{PAH}$) is allowed to be 0.47, 2.5, and 7.32, the minimum radiation field ($U_\mathrm{min}$) is allowed to vary from 0.1 to 50, the power-law slope ($\alpha$) is set to 2.0, and the fraction of the photodissociation region (PDR) is allowed to vary from 0.01 to 0.9. We find that the median differences for both \mstar\ and SFR are only $\approx0.1-0.2\,\rm dex$, and the normalized median absolute deviations ($\sigmaNMAD$) are $\approx0.2-0.3$, indicating general consistency between the two settings. 
Third, although there is no consensus on the exact star-formation history (SFH) for DOGs, \citet{Zou+2022} applied a delayed SFH, which has proven to be generally reliable even for AGN-host and/or bursty galaxies \citep[e.g.,][]{Carnall+2019, Lower+2020}. We further test a truncated delayed SFH (\texttt{sfhdelayedbq}) and a periodic SFH (\texttt{sfhperiodic}) following the settings in \mbox{Cristello et al. (submitted)} and \citet{Suleiman+2022}, which are dedicated to the SED analyses for DOGs. For the \texttt{sfhdelayedbq} module, we allow the $e$-folding time and the stellar age to vary in the ranges $0.5-5\,\rm Gyr$ and $1-10\,\rm Gyr$, the burst/quench age ($t_\mathrm{trun}$) is allowed in the range $10-800\,\rm Myr$, the factor for instantaneous change of SFR at $t_\mathrm{trun}$ is allowed in the range $0.1-50$, and the other parameters are set to their default values. For the \texttt{sfhperiodic} module, the types of individual SF episodes are allowed as ``exponential" or ``delayed", the period between each burst is allowed at 50 and 90$\,\rm Myr$, and the stellar age and the multiplicative factor for SFR are allowed in the ranges $0.1-10\,\rm Gyr$ and $1-4000$. The other parameters are set to their default values. For both types of SFHs, we do not find significant systematic differences in \mstar\ or SFR (median differences are $\lesssim0.2\,\rm dex$ and $\sigmaNMAD\approx0.2-0.3$).

To test further the reliability of our SFH, we fit the UV-to-NIR photometry of our normal-galaxy DOGs using the \texttt{Prospector-$\alpha$} model within \texttt{Prospector} \citep{Leja+2017, Johnson+2021}. This model is flexible and incorporates a six-component nonparametric SFH, which mitigates any systematic biases caused by the choice of a parametric SFH. We exclude AGN candidates because \texttt{Prospector} is not optimized for AGN-dominated sources. We find that the median difference for the \mstar\ is $\approx0.3\,\rm dex$, and the median difference for SFR is $\approx0.3\,\rm dex$. Furthermore, we verify that the best-fit SFHs generally do not exhibit recent starbursts (the median ratio of the SFR over the last $0-100\,\rm Myr$ compared to that over the last $100-300\,\rm Myr$ is only $\approx2.3$), indicating our delayed SFH should be suitable to model the SF for DOGs. It is worth noting that there are generally systematic ``factor-of-two" uncertainties among different SED-fitting results \citep[e.g.,][]{Leja+2019b}. This issue is inherent in SED-fitting methodologies, and solving it would require a more flexible SFH (e.g., a nonparametric SFH) at the cost of significantly higher computational requirements, which is beyond the scope of this work. Therefore, one should keep in mind possible systematic uncertainties depending on the adopted modules/parameter settings throughout this paper.

\begin{deluxetable*}{cccc}
\centering
\tablecaption{CIGALE parameter settings for normal galaxies.\label{tab:sed_gal}}
\tablewidth{0pt}
\tablehead{Module & Parameter & \shortstack{Name in the \texttt{CIGALE} \\ configuration file} & Possible values }
\startdata
\multirow{4}{*}{Delayed SFH} & \multirow{2}{*}{Stellar $e$-folding time} & \multirow{2}{*}{tau\_main} & 0.1, 0.2, 0.3, 0.4, 0.5, 0.6, 0.7, 0.8, 0.9, \\ 
 & & & 1, 2, 3, 4, 5, 6, 7, 8, 9, 10 Gyr \\
 & \multirow{2}{*}{Stellar age} & \multirow{2}{*}{age\_main} & 0.1, 0.2, 0.3, 0.4, 0.5, 0.6, 0.7, 0.8, 0.9, \\ 
 & & & 1, 2, 3, 4, 5, 6, 7, 8, 9, 10 Gyr \\
\hline
\multirow{2}{*}{\shortstack{Simple stellar population \\ \citet{BC+2003}}} & Initial mass function & imf & \citet{Chabrier+2003} \\ 
 & Metallicity & metallicity & 0.0001, 0.0004, 0.004, 0.008, 0.02, 0.05 \\ 
\hline
Nebular & -- & -- & -- \\
\hline
\multirow{3}{*}{\shortstack{Dust attenuation \\ \citet{Calzetti+2000}}} & \multirow{2}{*}{$E(B - V)_\mathrm{line}$} & \multirow{2}{*}{E\_BV\_lines} & 0, 0.05, 0.1, 0.15, 0.2, 0.25, 0.3, 0.4 \\ 
 & & & 0.5, 0.6, 0.7, 0.8, 0.9, 1, 1.2, 1.5 \\ 
                             & $E(B - V)_\mathrm{line} / E(B - V)_\mathrm{continuum}$ & E\_BV\_factor & 1 \\ 
\hline
\multirow{2}{*}{\shortstack{Dust emission \\ \citet{Dale+2014}}} & \multirow{2}{*}{Alpha slope} & \multirow{2}{*}{alpha} & \multirow{2}{*}{1.0, 1.25, 1.5, 1.75, 2.0, 2.25, 2.5, 2.75, 3.0} \\
& & & \\
\hline
X-ray & -- & -- & -- \\ 
\enddata
\tablecomments{Unlisted parameters are set to the default values. These are applied to all the sources.}
\end{deluxetable*}


\begin{deluxetable*}{cccc}
\centering
\tablecaption{CIGALE parameter settings for AGN candidates.\label{tab:sed_agn}}
\tablewidth{0pt}
\tablehead{Module & Parameter & \shortstack{Name in the \texttt{CIGALE} \\ configuration file} & Possible values}
\startdata
\multirow{2}{*}{Delayed SFH} & Stellar $e$-folding time & tau\_main & 0.1, 0.3, 0.5, 0.8, 1, 3, 5, 8, 10 Gyr\\ 
 & Stellar age & age\_main & 0.1, 0.3, 0.5, 0.8, 1, 3, 5, 8, 10 Gyr\\ 
\hline
\multirow{2}{*}{\shortstack{Simple stellar population \\ \citet{BC+2003}}} & Initial mass function & imf & \citet{Chabrier+2003} \\ 
 & Metallicity & metallicity & 0.02 \\ 
\hline
Nebular & -- & -- & -- \\
\hline
\multirow{3}{*}{\shortstack{Dust attenuation \\ \citet{Calzetti+2000}}} & \multirow{2}{*}{$E(B - V)_\mathrm{line}$} & \multirow{2}{*}{E\_BV\_lines} & 0, 0.05, 0.1, 0.15, 0.2, 0.25, 0.3, 0.4 \\ 
 & & & 0.5, 0.6, 0.7, 0.8, 0.9, 1, 1.2, 1.5 \\ 
                             & $E(B - V)_\mathrm{line} / E(B - V)_\mathrm{continuum}$ & E\_BV\_factor & 1 \\ 
\hline
\multirow{2}{*}{\shortstack{Dust emission \\ \citet{Dale+2014}}} & \multirow{2}{*}{Alpha slope} & \multirow{2}{*}{alpha} & \multirow{2}{*}{1.0, 1.25, 1.5, 1.75, 2.0, 2.25, 2.5, 2.75, 3.0} \\
& & & \\
\hline
\multirow{6}{*}{X-ray} & AGN photon index & gam & 1.8 \\
 & \multirow{2}{*}{AGN $\alpha_\mathrm{OX}$} & alpha\_ox & -1.9, -1.8, -1.7, -1.6, -1.5, \\
 & & & -1.4, -1.3, -1.2, -1.1 \\
 & \multirow{2}{*}{\shortstack{Maximum deviation of $\alpha_\mathrm{OX}$ \\ from the $\alpha_\mathrm{OX}-L_{\nu,2500}$ relation}} & \multirow{2}{*}{max\_dev\_alpha\_ox} & \multirow{2}{*}{0.2} \\
 & & & \\
 & AGN X-ray angle coefficients & angle\_coef & (0.5, 0)\\
\hline
\multirow{7}{*}{\shortstack{AGN \\ \citet{Stalevski+2012,Stalevski+2016}}} & Viewing angle & i & $0^{\circ}$, $10^{\circ}$, $30^{\circ}$, $50^{\circ}$, $70^{\circ}$, $90^{\circ}$ \\
 & Disk spectrum & disk\_type & \citet{Schartmann+2005} \\
 & \multirow{2}{*}{\shortstack{Modification of the optical \\ power-law index}} & \multirow{2}{*}{delta} & \multirow{2}{*}{--0.27} \\
 & & & \\
 & \multirow{2}{*}{AGN fraction} & \multirow{2}{*}{fracAGN} & 0, 0.05, 0.1, 0.2, 0.3, 0.4 \\
 & & & 0.5, 0.6, 0.7, 0.8, 0.9, 0.99 \\
 & $E(B-V)$ of the polar extinction & EBV & 0, 0.05, 0.1, 0.2, 0.3, 0.4, 0.5 \\
\enddata
\tablecomments{Unlisted parameters are set to the default values. These are only applied to AGN candidates.}
\end{deluxetable*}


Since \texttt{CIGALE} requires absorption-corrected \mbox{X-ray} flux, \citet{Zou+2022} applied a Bayesian approach to estimate the intrinsic \mbox{X-ray} luminosities for their \mbox{X-ray} detected AGNs directly from the \mbox{X-ray} count maps and adopted the \mbox{X-ray} luminosity function (XLF) as the prior. Their absorption correction is modest, and for our \mbox{X-ray} detected DOGs, the absorption-corrected flux from \citet{Zou+2022} is similar to ours using the \nh\ values derived from hardness ratios (HR) in Section~\ref{subsec:HR} (median difference $<0.01\,\rm dex$). They also showed that decreasing the uncertainty (i.e., increasing the weight) of the \mbox{X-ray} data points does not change the SED-fitting results materially, and thus the associated uncertainty for \nh\ will not significantly impact our results. For \mbox{X-ray} undetected sources, \mbox{X-ray} upper limits are used to constrain the AGN component if the sources are classified as AGN candidates. These upper limits are derived using the HB flux upper-limit maps and have been corrected for nominal intrinsic absorption based upon the XLF, as detailed in Section~2.2 of \citet{Zou+2022}. The correction is generally reliable for sources in XMM-SERVS. At the median redshift of our sources ($z\approx1.8$), the HB corresponds to rest-frame $\approx6-30\,\kev$, which is hardly affected by absorption with $\nh\lesssim10^{24}\,\nhcgs$. Besides, we have checked that the adopted \mbox{X-ray} upper limits are generally much higher than the predicted \mbox{X-ray} flux by \texttt{CIGALE}, indicating that our SED results are insensitive to the \mbox{X-ray} upper limits.

According to the classification of \citet{Zou+2022}, all of our \mbox{X-ray} detected DOGs are classified as AGN candidates, and 45\% (42\%) are classified as reliable SED AGNs in the full (core) sample. Among \mbox{X-ray} undetected DOGs, 9\% (6\%) are classified as reliable SED AGNs in the full (core) sample, 53\% (64\%) are classified as AGN candidates, and 38\% (30\%) are classified as normal galaxies. The total fraction of sources classified as reliable SED AGNs in the full (core) sample is 11\% (8\%). The values are consistent with the typical fraction of AGNs among general galaxy populations \citep[e.g.,][]{Xue+2010,Aird+2018,Zou+2024}. We have also checked the optical variability-selected AGN catalogs in W-CDF-S \citep{Falocco+2015,Poulain+2020}, but none of our DOGs is selected as a variable AGN candidate due to their faintness in the optical bands.

Since DOGs can also be classified into PL or Bump DOGs, and PL DOGs may preferentially host strong AGNs, we briefly compare the PL/Bump classification results with our SED-based classification results. We follow the method of \citet{Dey+2008} to classify our DOGs into PL/Bump DOGs, which is optimized for Spitzer IRAC photometry.\footnote{We do not use the ``$K_S$-excess" method to classify PL DOGs and Bump DOGs, because this method is optimized for WISE photometry, which is not included in our deeper photometric data \citep[e.g.,][]{Toba+2015,Noboriguchi+2019}.} We restrict our analysis to the 305 DOGs with all four IRAC bands having $\rm SNR>2$ in our core sample. We perform two power-law fits to the observed MIR photometry for each source with measurements in all four IRAC bands.  The first fit only includes the four IRAC bands (observed-frame 3.6--8.0\,\mum). The second fits the four IRAC bands along with MIPS 24\,\mum\ data. We then examine the power-law indices ($F_\nu\propto\lambda^\alpha_\mathrm{MIR}$) of the two fits following the selection criteria in \citet{Dey+2008}. These steps select PL DOGs with monotonic SEDs, and the rest are classified as Bump DOGs. We find that 41 sources are identified as PL DOGs, and all of these are classified as AGN candidates based upon their SEDs. The fraction of reliable SED AGNs (23/41=56\%) among PL DOGs is much higher than that for Bump DOGs, indicating PL DOGs indeed preferentially host AGNs, consistent with previous results \citep[e.g.,][]{Toba+2015}. However, there are still 55 reliable SED AGNs among the 264 Bump DOGs (21\%), which indicates the classification is rather phenomenological and incomplete for selecting AGN-dominated sources. The results for our full sample are similar.

\begin{figure*}[htp!]
    \centering
    \includegraphics[width=\linewidth]{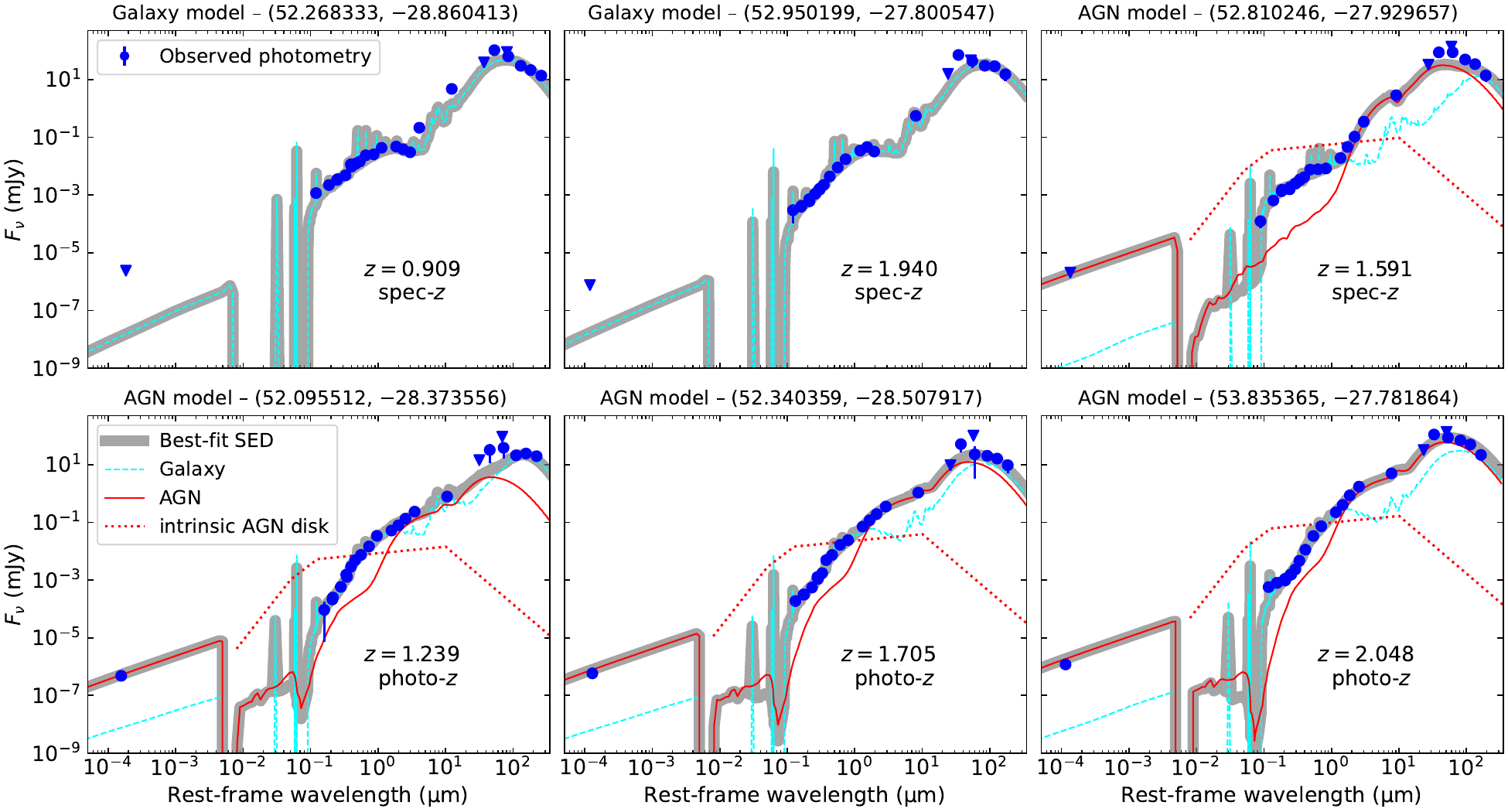}
    \caption{Example best-fit SED results for \mbox{X-ray} undetected (top row) and \mbox{X-ray} detected (bottom row) DOGs. All of the six sources have reliable redshifts (\specz s or \photoz s with $Q_z^\mathrm{good}<1$, as shown in each panel). The blue points and downward triangles are the observed photometry and upper limits, respectively. The thick-grey lines are the best-fit total models. The cyan-dashed lines and the red-solid lines represent galaxy and AGN components, respectively. For sources where the statistically preferred model is the AGN model, we also show the intrinsic AGN disk SEDs, which are only modeled at $\lambda\geq0.008\,\rm\mu m$ \citep{Yang+2022}. The sharp peaks at rest-frame $<0.1\,\mum$ are strong nebular emission lines, as modeled by the templates from \citet{Inoue+2011}. All sources shown with best-fit AGN models are classified as reliable SED AGNs. See Section~3 of \citet{Zou+2022} for more details about our source classification. We also show the coordinates (RA and Dec) of each source at the top of each panel.}\label{fig:SED}
\end{figure*}

Figure~\ref{fig:SED} shows examples of the best-fit SEDs for three \mbox{X-ray} undetected DOGs and three \mbox{X-ray} detected DOGs in the rest frame. All these sources have reliable redshifts (\specz s or \photoz s with $Q_z^\mathrm{good}<1$), and the SEDs are well characterized over a wide range of wavelengths. Generally, the \mbox{X-ray} emission is dominated by the AGN component. 
The galaxy component typically dominates the optical bands since the AGN continuum is heavily obscured in these bands. The intrinsic AGN disk SEDs \citep{Yang+2020,Yang+2022} are also shown in Figure~\ref{fig:SED}, and they are generally higher than the observed SED in the optical bands. In the MIR bands, sources classified as AGN candidates generally have a non-negligible AGN component, which contributes to their selection as DOGs.

The galaxy and AGN SEDs in Figure~\ref{fig:SED} and the diverse SED-based classification results indicate that DOGs are a heterogeneous population, which results from the color-selection criteria for DOGs. The MIR-to-optical color selection tends to identify both normal galaxies with significant optical obscuration and AGNs with strong MIR dust emission. DOGs with weaker galaxy IR emission are more likely to host AGNs with strong IR flux. Thus, our selection for reliable SED AGNs and/or \mbox{X-ray} detected DOGs may be biased toward sources with less galaxy FIR emission and lower SFR. This effect may contribute to our results in Section~\ref{subsec:host-galaxy}.

\subsection{Distributions of $z$, $L_\mathrm{bol}$, and $\lxobs$}\label{subsec:sample_dist}


\begin{figure}[tb!]
    \centering
    \includegraphics[width=\linewidth]{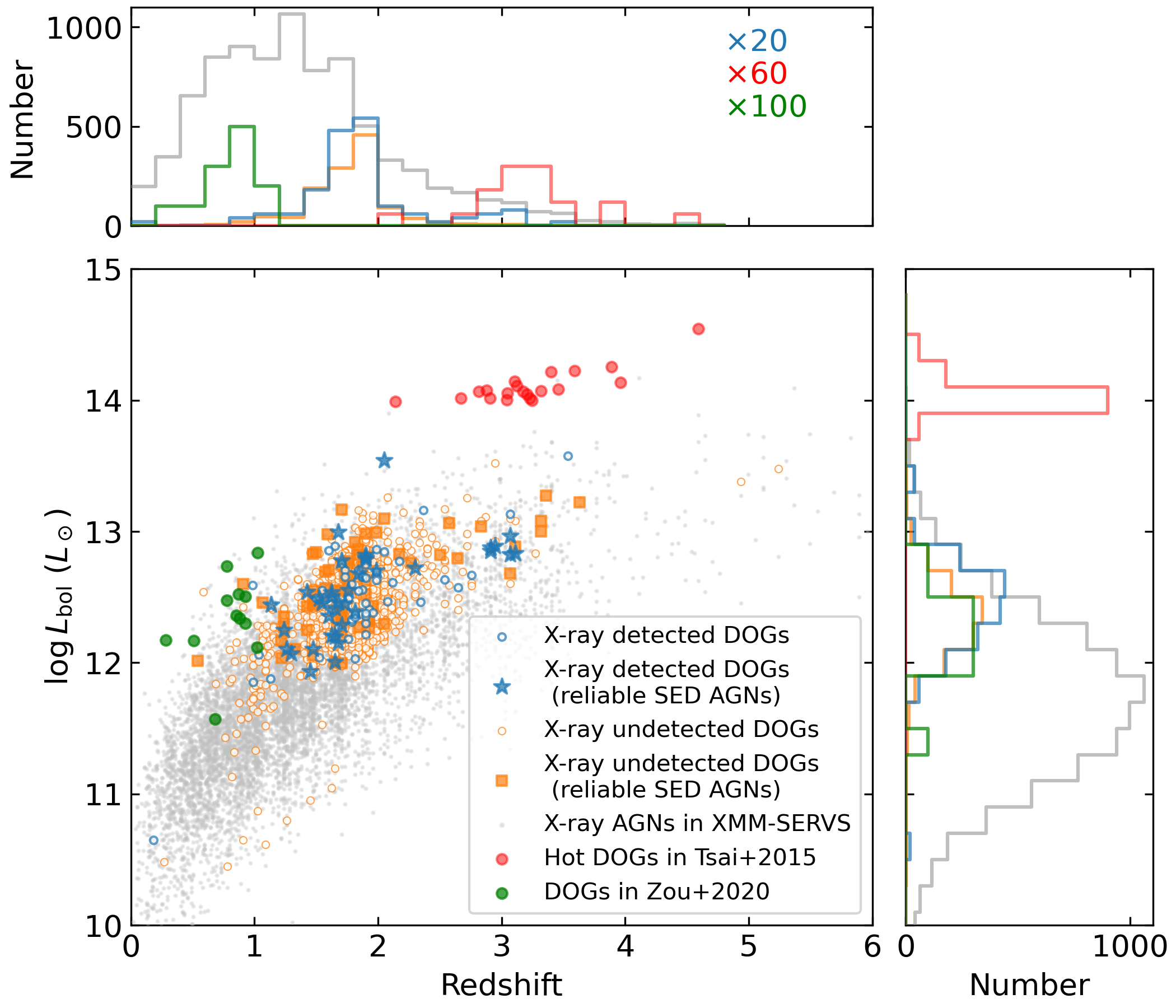}
    \caption{The \lbol\ versus $z$ distribution of our core sample. Our \mbox{X-ray} detected (undetected) DOGs are shown as blue (orange)-empty circles. Among them, sources classified as reliable SED AGNs in \citet{Zou+2022} (see text in Section~\ref{subsec:host-galaxy}) are shown as filled stars for \mbox{X-ray} detected DOGs and filled squares for \mbox{X-ray} undetected DOGs. For comparison, the \lbol\ versus $z$ distributions of \mbox{X-ray} AGNs in XMM-SERVS, Hot DOGs in \citet{Tsai+2015}, and DOGs in \citet{Zou+2020} are shown in grey, red, and green circles, respectively. The distributions of $z$ and \lbol\ for sources mentioned above are shown in the top and right subpanels, respectively, with the same colors as those in the legend (we do not further plot the distributions for our reliable SED AGNs). Note that for better visibility, in the top and right subpanels, the number of \mbox{X-ray} detected DOGs, Hot DOGs in \citet{Tsai+2015}, and DOGs in \citet{Zou+2020} are multiplied by factors of 20, 60, and 100, respectively.}\label{fig:z_lbol_dist}
\end{figure}

\begin{figure}[tb!]
    \centering
    \includegraphics[width=\linewidth]{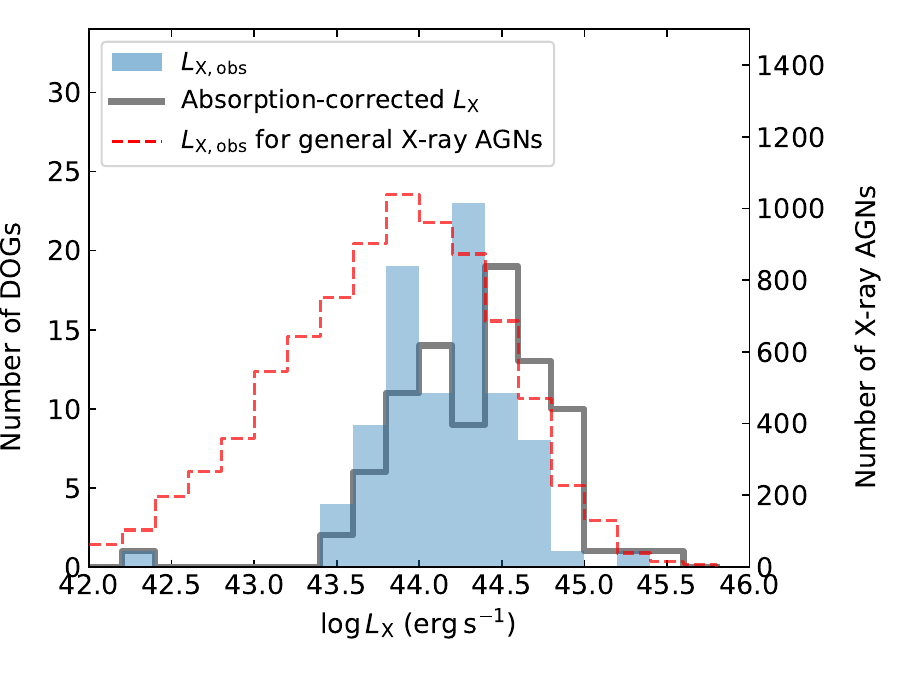}
    \caption{\lxobs\ (blue) and absorption-corrected \lx\ (gray) distributions for \mbox{X-ray} detected DOGs in the core sample. The red-dashed histogram represents the \lxobs\ distribution of general \mbox{X-ray} AGNs in XMM-SERVS. \mbox{X-ray} detected DOGs generally have high observed and absorption-corrected \mbox{X-ray} luminosity.}\label{fig:lxdist}
\end{figure}


We calculate the \lbol\ for our samples using their best-fit SED models reported by \citet{Zou+2022}, where we integrate the total SED models from X-rays to FIR (observed-frame $10^{-4}-1000\,\mum$). Figure~\ref{fig:z_lbol_dist} shows the \lbol\ versus $z$ distribution for our core sample. The median redshift is $z=1.82$, and the $25-75\%$ quantile range is $z=1.63-1.93$. An AD test on \mbox{X-ray} detected and \mbox{X-ray} undetected DOGs returns a $p$-value of 0.06, indicating their redshift distributions are not significantly different. The median $\log\lbol/\lsun\approx12.4$, and the $25-75\%$ quantile range is $\log\lbol/\lsun=12.3-12.6$. The results for our full sample are similar. For comparison, we also plot the distributions of general \mbox{X-ray} AGNs in XMM-SERVS \citep[selected via $\mathtt{flag\_Xrayagn}=1$ from the SED catalog of][]{Zou+2022}, Hot DOGs with $\log\lbol/\lsun>14$ in \citet{Tsai+2015}, and DOGs in \citet{Zou+2020}. DOGs selected by our criteria have a much narrower range of redshifts than typical \mbox{X-ray} AGNs. The narrow redshift distribution arises primarily because the $\fir/f_R>982$ criterion selects against sources with lower redshifts. At our median $z\approx1.8$, the observed $\fir/f_R$ approximately corresponds to the rest-frame $8\,\mum$-to-2200\,\AA\ flux ratio. At lower redshifts, the observed $R$ band corresponds to redder rest-frame optical bands, which mitigates dust obscuration. The \lbol\ for our DOGs is much higher than that for \mbox{X-ray} AGNs, but is significantly lower than that for the most-extreme Hot DOGs. Given the low sky density \citep[$\approx1$ per $30\deg^2$;][]{Wu+2012} and the much higher \lbol\ of Hot DOGs than that for our sample, it is unlikely that our sample contains any extreme Hot DOGs. However, we can select lower-luminosity analogs to Hot DOGs using the best-fit SED models, and we present our results in Appendix~\ref{appendix:b}.



We also show the observed \mbox{X-ray} luminosity at rest-frame \mbox{$2-10\,\rm keV$} (\lxobs) for \mbox{X-ray} detected DOGs in the core sample in Figure~\ref{fig:lxdist}. For AGNs, the hard X-ray spectrum ($>2\,\kev$) is generally characterized by a power-law: $N(E)\propto E^{-\Gamma}$, where $N(E)$ is the photon number flux as a function of photon energy, and $\Gamma$ is the ``intrinsic" power-law photon index. The intrinsic power-law photon index defines the spectral slope of the X-ray source, unaffected by any obscuring material (after correcting for Galactic absorption). For most AGNs, $\Gamma=1.7-2.2$ \citep[e.g.,][]{Scott+2011,Netzer+2015,Liu+2017}. The ``effective" power-law photon index ($\Gamma_\mathrm{eff}$), derived from a simple power-law fit, is a useful first-order descriptor of the spectral shape when the X-ray source is obscured (after correcting for Galactic absorption). For our DOGs, we assume a fixed effective power-law photon index of $\Gamma_\mathrm{eff}=1.4$ to allow for intrinsic absorption, and \lxobs\ is calculated from the observed flux (corrected for Galactic but not intrinsic absorption) in one band based upon the following priority order: \mbox{$2-10\,\rm keV$} (hard band; HB), \mbox{$0.5-10\,\rm keV$} (full band; FB), \mbox{$0.5-2\,\rm keV$} (soft band; SB) \citep{Chen+2018,Ni+2021}. This priority order is chosen to minimize absorption effects. An AD test on the full and core samples returns a $p$-value of 0.26, suggesting that the \lxobs\ distributions are not significantly different between the two samples. For comparison, in Figure~\ref{fig:lxdist} we show the absorption-corrected \lx\ values which are derived using the \nh\ values in Section~\ref{subsec:HR}, and we show the \lxobs\ for general \mbox{X-ray} AGNs in XMM-SERVS. Our DOGs are more luminous, with a median $\lxobs=10^{44.3}\,\lumcgs$. As our sources are at $z\approx2$, the HB coverage roughly corresponds to $6-30\,\kev$ in the rest frame, which substantially mitigates any intrinsic obscuration.

\section{Analyses and Results}\label{sec:results}

In this section, we investigate several properties of our full and core samples. Section~\ref{subsec:host-galaxy} presents the host-galaxy properties of our DOGs using the results from SED fitting. Section~\ref{subsec:HR} investigates \mbox{X-ray} hardness ratios (HRs) and the corresponding \nh. Section~\ref{subsec:X-ray_stacking} presents \mbox{X-ray} stacking for \mbox{X-ray} undetected DOGs to assess their typical \lxobs\ and \nh. Section~\ref{subsec:lx-l6} shows the \mbox{X-ray}$-$MIR relation. In Section~\ref{subsec:radio}, we present radio properties.


\subsection{Host-Galaxy Properties}\label{subsec:host-galaxy}

The host-galaxy properties are derived from SED fitting, which decomposes the galaxy component and, if present, the AGN component.


SED fitting returns host-galaxy properties, including \mstar\ and SFR. In general, \mstar\ measurements should be robust, as they are determined mainly by SEDs at rest-frame $\approx1\,\rm\mu m$ where the AGN component is often weaker than the galaxy component \citep[e.g.,][]{Ciesla+2015}. For luminous type~1 AGNs, \mstar\ is less reliable since the AGN component tends to dominate the emission from the UV to MIR. This should not impact our results significantly since there are only 3 reliable broad-line AGNs identified in \citet{Ni+2021} in our full sample. We further check that, among our sources with best-fit AGN models, only $\approx6\%$ and $\approx2\%$ in the full and core samples have fractional flux contributions by the AGN at rest-frame $1\,\mum$ greater than $0.2$, respectively. We also compare our best-fit \mstar\ with the \mstar\ estimated using normal-galaxy templates ($\mstar^\mathrm{gal}$) in Appendix~\ref{appendix:a}, and the results are generally consistent. Thus, we conclude that our \mstar\ measurements are not severely affected by AGN contributions. On the other hand, SFR measurements can incur more systematic uncertainties, but the inclusion of high-quality far-infrared (FIR) photometry can help obtain more reliable SFRs \citep[e.g.,][]{Netzer+2016}. Among the five Herschel bands in the FIR, the Herschel SPIRE $250\,\mum$ photometry gives the highest fractions of DOGs with high-SNR measurements. There are $\approx43\%$ of our \mbox{X-ray} detected DOGs and $\approx54\%$ of our \mbox{X-ray} undetected DOGs having $\mathrm{SNR}>5$ at Herschel SPIRE $250\,\mum$. These fractions are also much higher than the typical fractions in XMM-SERVS. We also show that excluding FIR photometry will not cause significant biases in Appendix~\ref{appendix:a}.


Figure~\ref{fig:mstar} shows the \mstar\ distribution of our core sample, and the results for our full sample are similar. Our DOGs are generally massive galaxies (median \mbox{$\log\mstar/\msun=11.0$}, and the \mbox{$25-75\%$} quantile range is $10.7-11.3$), which has also been noted by previous studies \citep[e.g.,][]{Bussmann+2012,Toba+2015,Riguccini+2019,Suleiman+2022}. Also, \mbox{X-ray} detected DOGs tend to have slightly higher \mstar\ than \mbox{X-ray} undetected ones, which is confirmed by an AD test with a $p$-value $<0.001$. Indeed, AGNs tend to reside in massive galaxies, and black-hole accretion rates traced by \mbox{X-ray} luminosity monotonically increase as \mstar\ increases \citep[e.g.,][]{Yang+2018,Zou+2024}. 

Our DOGs generally have high SFRs (median $\mathrm{SFR}=141\,\msun\,\rm yr^{-1}$ and the $25-75\%$ quantile range is \mbox{$65-277\,\msun\,\rm yr^{-1}$}). Instead of showing SFR distributions, we use the normalized SFR (\sfrnorm) to represent how ``starbursty" the source is compared with the SF main sequence (MS). \sfrnorm\ is defined as $\mathrm{SFR}/\mathrm{SFR_{MS}}$, where $\mathrm{SFR_{MS}}$ is the MS SFR. 
We do not directly adopt the MS results from other literature works because they may have systematic offsets due to different methods of deriving \mstar\ and SFR \citep[e.g.,][]{Mountrichas+2021}. Thus, we use the SED catalogs in XMM-SERVS to calibrate the MS directly for our sources. Following \citet{Cristello+2024}, for each DOG, we select all galaxies within $\pm0.1\,\rm dex$ in \mstar\ and $\pm0.075\times(1+z)$ in redshift. Among these matched galaxies, we select star-forming galaxies using their rest-frame $U-V$ and $V-J$ colors \citep[i.e., the $UVJ$ diagram; e.g.,][]{Williams+2009,Whitaker+2012,Lee+2018}, which constitute a reference star-forming galaxy sample for the DOGs. We utilize the $UVJ$ selection criteria for star-forming galaxies in Zhang et al. (submitted), which were calibrated specifically for XMM-SERVS using the methods in \citet{Williams+2009} and \citet{Whitaker+2015}. The adopted criteria are
\begin{equation}
    U-V<1.3
\end{equation}
as the horizontal cut, and
\begin{equation}
\begin{aligned}
    U-V &> 0.8\times(V-J) + 0.84\ (0.0<z<0.5)\\
    U-V &> 0.8\times(V-J) + 0.83\ (0.5<z<1.0)\\
    U-V &> 0.8\times(V-J) + 0.75\ (1.0<z<1.5)\\
    U-V &> 0.8\times(V-J) + 0.72\ (1.5<z<2.5)\\
    U-V &> 0.8\times(V-J) + 0.70\ (z>2.5)
\end{aligned}
\end{equation}
as the diagonal cut. Most of the sources in the catalog are identified as normal galaxies, and only $\approx3\%$ are identified as AGNs by \citet{Zou+2022}. Thus, our selection should not be materially impacted by AGNs. We use the median SFR of the selected star-forming galaxies as the $\rm SFR_{MS}$ of the corresponding DOG, and apply the above steps for all our DOGs. In principle, one must consider the mass-completeness limit, as the determination of the MS may be biased. However, since our sources generally have high \mstar\ with only $\approx3\%$ below the mass-completeness curves for XMM-SERVS \citep[see Section~2.4 of][]{Zou+2024}, we do not further apply a mass-completeness cut, and the results should not be significantly impacted. One caveat in the determination of the MS is that star-forming galaxies cannot be reliably distinguished at high-\mstar\ and/or low-$z$ due to the high fraction of quiescent or transitioning galaxies in those regimes, where the classification of star-forming galaxies (and, consequently, the determination of $\mathrm{SFR_{MS}}$) may become sensitive to the adopted methods \citep[e.g.,][]{Donnari+2019}. \citet{Cristello+2024} proposed a redshift-dependent maximum \mstar\ for reliable classifications using the following procedures. For each AGN in their sample, they selected all galaxies within $\pm0.1\,\rm dex$ in \mstar\ and $\pm0.075\times(1+z)$ in redshift. Among these reference galaxies, they classified star-forming and quiescent galaxies using the method in \citet{Tacchella+2022}. The regime in the $z-\mstar$ plane where the fraction of quiescent galaxies is less than 0.5 was determined to be the ``safe" regime. The maximum \mstar\ for the ``safe" regime can be well described at $z=0.1-4$ by the following equation:
\begin{equation}\label{eqn:max_mstar}
    \log\mstar=10.65+0.81\log z+0.38\log(1+z).
\end{equation}
The ``safe" regime is established to minimize MS offsets when probing the highest masses, ensuring that $\mathrm{SFR_{MS}}$ remains similar regardless of different MS definitions. We adopt Equation~\ref{eqn:max_mstar} to determine our ``safe" DOGs.

\sfrnorm\ has three sources of uncertainty: SFR, \mstar, and the determination of the MS. The systematic bias in how we determine the MS is generally not significant as long as we measure the SFRs for our DOGs and reference galaxies self-consistently and calibrate the MS. Furthermore, the MS uncertainty is also small for our ``safe" sample as it is constructed to minimize MS offsets. The relative uncertainty of \mstar\ is also generally smaller than that of SFR, so the \sfrnorm\ uncertainty is primarily driven by the SFR uncertainty of our DOGs. Our typical SFR uncertainty is $\lesssim 0.3\,\rm dex$, which is generally acceptable for SED fitting.

Figure~\ref{fig:sfr} shows the \sfrnorm\ distribution for our core sample, with the ``safe" sources in solid lines, as well as all sources, including the ``unsafe" ones, in dashed lines. An AD test comparing the distributions of the ``safe" and all sources returns a $p$-value of 0.06 for \mbox{X-ray} undetected DOGs, and 0.99 for \mbox{X-ray} detected DOGs, showing that the \sfrnorm\ distributions of the ``safe" and all sources are not significantly different. This implies that our MS definition is generally reliable even for high-\mstar\ and/or low-$z$ galaxies. Considering all the sources in the core sample, we find \mbox{X-ray} detected DOGs have a higher fraction of sources below the MS (32\% with $\log\sfrnorm<-0.4$), while the \mbox{X-ray} undetected ones are generally on or above the MS (only 8\% with $\log\sfrnorm<-0.4$). An AD test returns a $p$-value $<0.001$ when comparing the \mbox{X-ray} detected and undetected sources, confirming that they have different \sfrnorm\ distributions. The difference may be caused by selection effects for our color-based selection criteria, as discussed in \ref{subsec:sed-fitting}, where \mbox{X-ray} detected DOGs may be biased toward higher fractional FIR flux contributions from AGNs, which results in reduced levels of SFR from the galaxy component. We will further discuss how the \sfrnorm\ distributions of \mbox{X-ray} detected DOGs compare with matched typical \mbox{X-ray} AGNs in Section~\ref{subsec:compare}.

\begin{figure}[ht!]
    \centering
    \includegraphics[width=\linewidth]{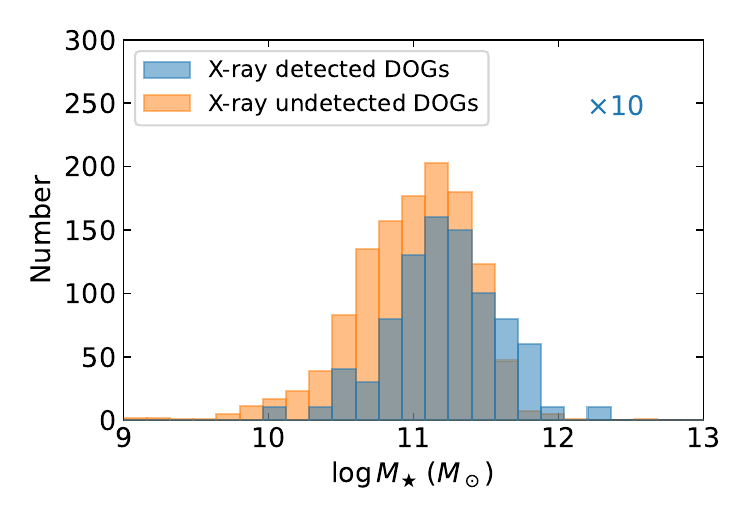}
    \caption{$\mstar$ distribution for the core sample. Blue and orange histograms represent \mbox{X-ray} detected and \mbox{X-ray} undetected DOGs, respectively. Note that the number of \mbox{X-ray} detected DOGs in each bin is multiplied by a factor of 10 for easier visibility. DOGs generally have high \mstar, and \mbox{X-ray} detected DOGs have slightly higher \mstar. The distributions for our full sample are similar.}\label{fig:mstar}
\end{figure}

\begin{figure}[ht!]
    \centering
    \includegraphics[width=\linewidth]{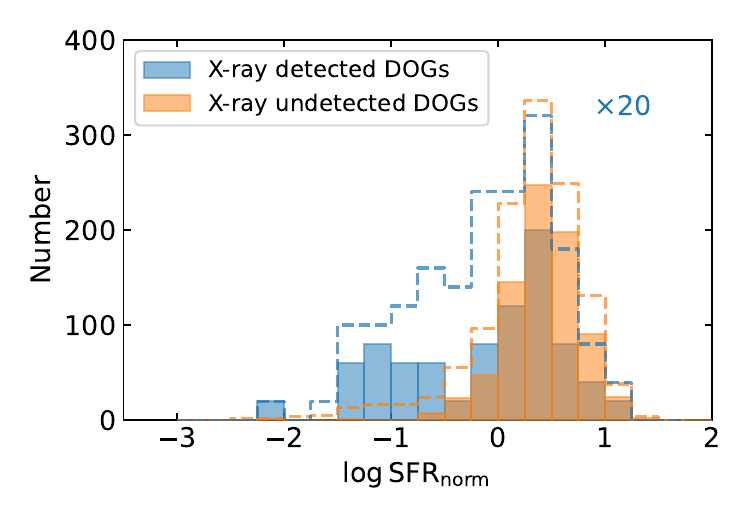}
    \caption{$\rm SFR/SFR_{MS}$ distribution for the core sample. Blue and orange histograms represent \mbox{X-ray} detected DOGs and \mbox{X-ray} undetected DOGs, respectively. Solid histograms represent the ``safe" sources where the corresponding star-forming MS can be reliably obtained. Dashed histograms represent all the sources in our sample. Note that the number of \mbox{X-ray} detected DOGs in each bin is multiplied by a factor of 20 for easier visibility. \mbox{X-ray} detected DOGs have a higher fraction of sources below the MS. The distributions for our full sample are similar.}\label{fig:sfr}
\end{figure}



\subsection{HRs and \nh}\label{subsec:HR}

\begin{figure*}[htb!]
    \centering
    \includegraphics[width=\linewidth]{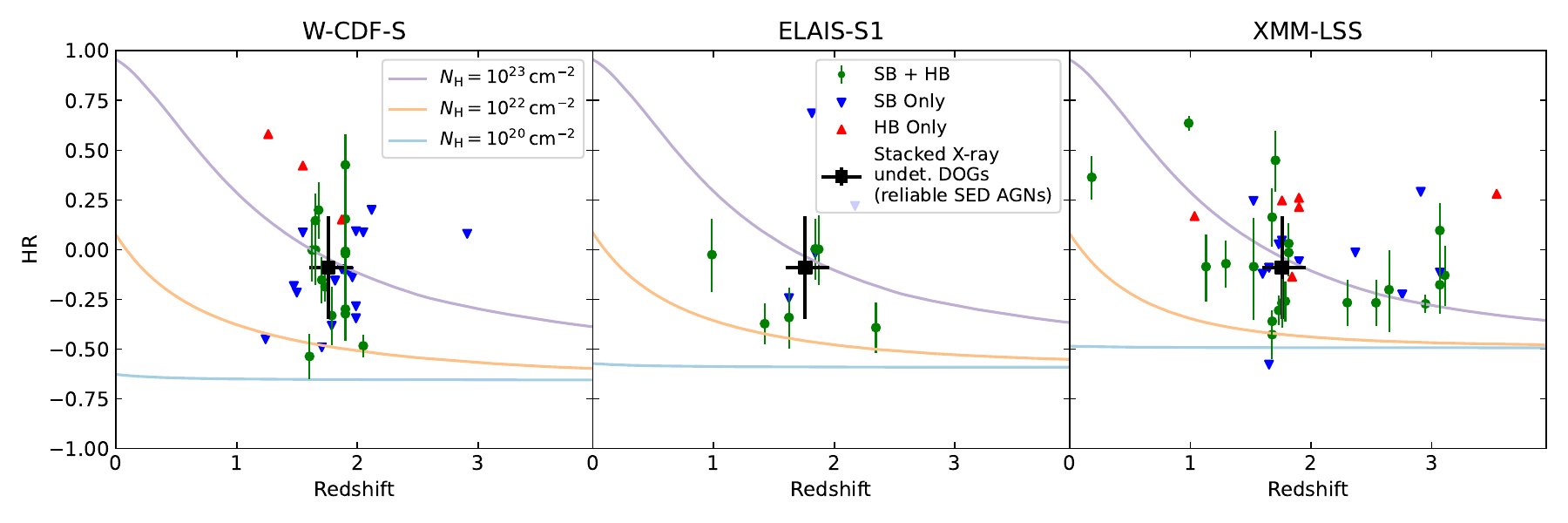}
    \caption{HR versus redshift in W-CDF-S (left), ELAIS-S1 (middle), and XMM-LSS (right) for the core sample. The colored points are our \mbox{X-ray} detected DOGs, and the legend in the middle panel indicates their detection status in the SB and HB. Green points with error bars represent the median HR values with $1\sigma$ uncertainties for sources detected in both bands. The blue downward triangles, which are detected in the SB but not the HB, represent $1\sigma$ HR upper limits. The red upward triangles, which are detected in the HB but not the SB, are $1\sigma$ HR lower limits. The black squares represent stacked \mbox{X-ray} undetected DOGs that are classified as reliable SED AGNs in the core sample (see Section~\ref{subsec:X-ray_stacking}), with the error bar in the $x$-axis direction showing the $25-75\%$ redshift quantile range. The solid lines are the expected relations for redshifted absorbed power-laws, whose intrinsic photon index $\Gamma=1.8$, with several different \nh\ values, as labeled in the legend.}\label{fig:NH}
\end{figure*}

In this subsection, we investigate the basic \mbox{X-ray} spectral properties of our \mbox{X-ray} detected DOGs. Given the limited counts (typically $90-200$ in the FB), we are not able to perform detailed \mbox{X-ray} spectral fitting for most of our DOGs detected in \mbox{X-rays}. We thus analyze the HRs for simplicity to probe their spectral properties. HR is defined as $(\hard-\soft)/(\hard+\soft)$, where \hard\ and \soft\ represent the HB and SB count rates, respectively. The cataloged HRs in XMM-SERVS are reliable mainly for sources detected in both the SB and HB, but many of our sources are detected only in one band. In particular, possible high \nh\ values present in DOGs can severely impact the detection in the SB while not affecting the HB too much. We thus apply a Bayesian method described in Appendix~A of \citet{Zou+2023} to calculate the HRs for our \mbox{X-ray} detected DOGs. We also calculate the expected \mbox{$z-\mathrm{HR}$} curves in each field, assuming an absorbed power-law with intrinsic photon index $\Gamma=1.8$. These curves are calculated using the Portable Interactive Multi-Mission Simulator (\texttt{PIMMS}). In brief, given a spectral model, we first obtain the net count rates for a given band (SB and HB) in a given instrument (PN, MOS1, and MOS2) as a function of redshift. Then, we weigh the count rates by the median exposure times of each instrument in each band across the field to calculate the HRs (assuming intrinsic $\Gamma=1.8$) at different redshifts. It is worth noting that using standard photoelectric absorption to calculate spectral shape is appropriate up to $\nh\approx10^{23.5}\,\nhcgs$; when $\nh\gtrsim10^{24}\,\nhcgs$ (i.e., CT absorption), the reflection component from the torus and other effects may become prominent. 

We present the results for our core sample in Figure~\ref{fig:NH}, along with the expected $z-\mathrm{HR}$ curves at different \nh\ values. Our \mbox{X-ray} detected DOGs generally have $\nh>10^{22}\,\nhcgs$ except for several sources, and a large fraction of them reach $\nh>10^{23}\,\nhcgs$. Among 174 \mbox{X-ray} detected sources in our full sample, 87 have $\nh>10^{23}\,\nhcgs$; 43 out of 88 \mbox{X-ray} detected sources in the core sample have $\nh>10^{23}\,\nhcgs$. These fractions are similar to \mbox{X-ray} spectral fitting results for DOGs in the ultradeep CDF-S field, where they found $\approx64\%$ of \mbox{X-ray} detected DOGs with $\nh>10^{23}\,\nhcgs$ \citep{Corral+2016}.

\begin{figure}[htb!]
    \centering
    \includegraphics[width=\linewidth]{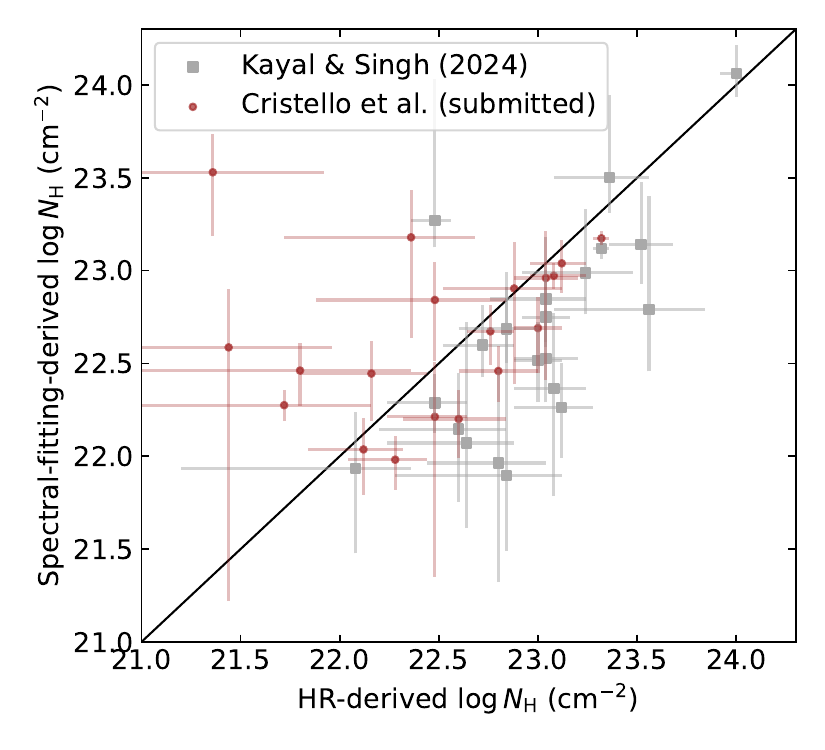}
    \caption{Comparison between our HR-derived $\nh$ values and those derived from \mbox{X-ray} spectral fitting. The grey squares represent the results from \citet{Kayal+2024}. The brown circles represent the results from Cristello et al. (submitted). The error bars represent $1\sigma$ uncertainties.}\label{fig:nh_compare}
\end{figure}
\begin{figure}[htb!]
    \centering
    \includegraphics[width=\linewidth]{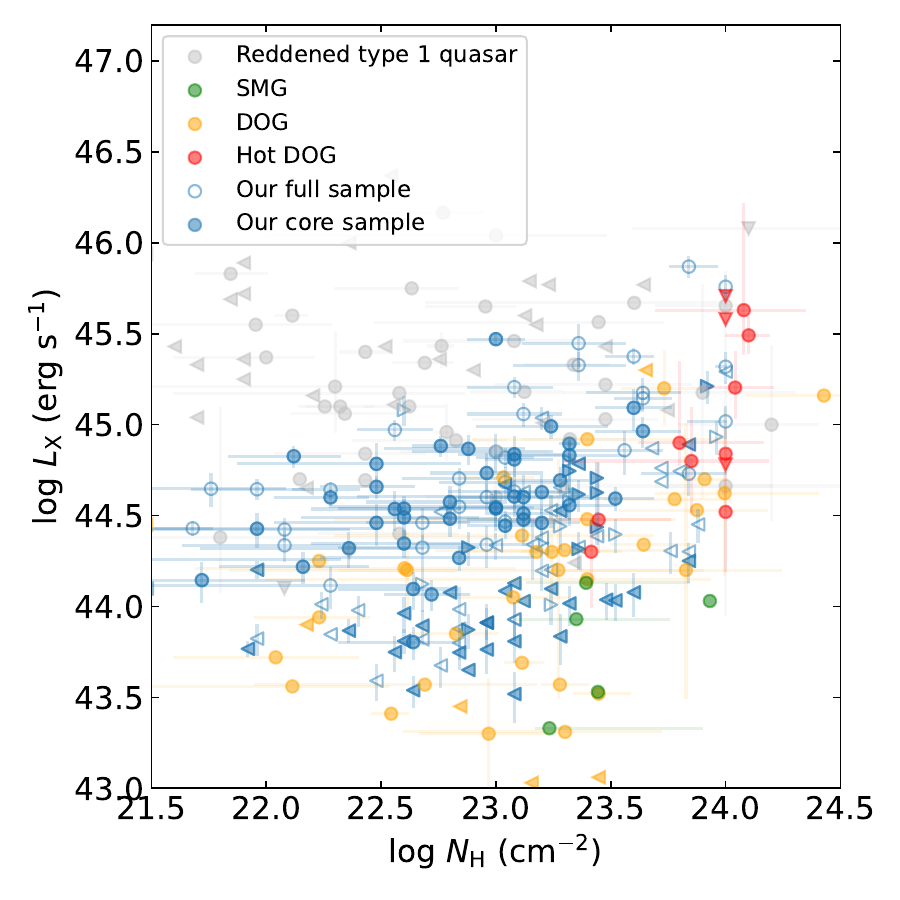}
    \caption{Comparison between our samples and previous studies in the $\nh-\lx$ plane. The \nh\ values of our DOGs are converted from their HRs. The blue points are our \mbox{X-ray} DOGs detected in both SB and HB with $1\sigma$ confidence error bars; the blue leftward and rightward triangles represent \mbox{X-ray} DOGs detected only in HB or SB ($1\sigma$ lower/upper limits for \nh), respectively. Our full and core samples are indicated by empty and filled symbols, respectively. The grey, green, yellow, and red points are reddened type~1 quasars, SMGs, DOGs, and Hot DOGs from previous studies, respectively (see the text for sample references). Our \mbox{X-ray} detected DOGs are luminous with a wide range of obscuration.}\label{fig:lx_nh}
\end{figure}

The calculated HRs can be further converted to \nh\ values by interpolating over the $z-\mathrm{HR}$ curves at different obscuration levels. The median $\nh$ for our core sample is $10^{22.8}\,\nhcgs$. Recently, \citet{Kayal+2024} and Cristello et al. (submitted) both performed \mbox{X-ray} spectral analyses for \mbox{X-ray} detected DOGs with sufficient counts in XMM-SERVS. The former covered XMM-LSS, while the latter covered all three XMM-SERVS fields. Note that \citet{Kayal+2024} used the HSC Subaru Strategic Program and the SWIRE band-merged catalogs as their parent sample to select DOGs, while Cristello et al. (submitted) utilized our DOG catalog directly. We compare our HR-derived $\nh$ with the results from \mbox{X-ray} spectral fitting in Figure~\ref{fig:nh_compare}. Our $\nh$ values appear systematically higher in general than those estimated via \mbox{X-ray} spectral analyses, which should be kept in mind for the following discussion. 
Figure~\ref{fig:lx_nh} shows the absorption-corrected rest-frame $2-10\,\kev$ \lx\ versus \nh\ for our \mbox{X-ray} detected DOGs as well as other AGN populations collected from the literature: reddened type~1 quasars \citep{Urrutia+2005,Martocchia+2017,Mountrichas+2017,Goulding+2018,Lansbury+2020}, SMGs \citep{Wang+2013}, DOGs \citep{Lanzuisi+2009,Corral+2016,Zou+2020}, and Hot DOGs \citep{Stern+2014,Assef+2016,Ricci+2017,Vito+2018}. We use the derived \nh\ from our HR results to correct for the intrinsic absorption using \texttt{sherpa}. The correction factor is generally modest (median value of $\approx1.9$) because, at the median redshift of our sources, the HB corresponds to rest-frame $\approx6-30\,\kev$, which is not significantly affected by absorption at the observed levels. Our \nh\ values span a wide range and are generally consistent with those for DOGs in previous studies. Our \mbox{X-ray} detected DOGs also show slightly higher $\lx$, which is mostly due to the wide homogeneous medium-deep \mbox{X-ray} coverage of XMM-SERVS that allows us to select more luminous sources compared with previous studies in small ultra-deep fields \citep[e.g.,][]{Corral+2016}. 

\subsection{X-ray Stacking}\label{subsec:X-ray_stacking}

\begin{figure*}[htb!]
    \centering
    \includegraphics[width=0.9\linewidth]{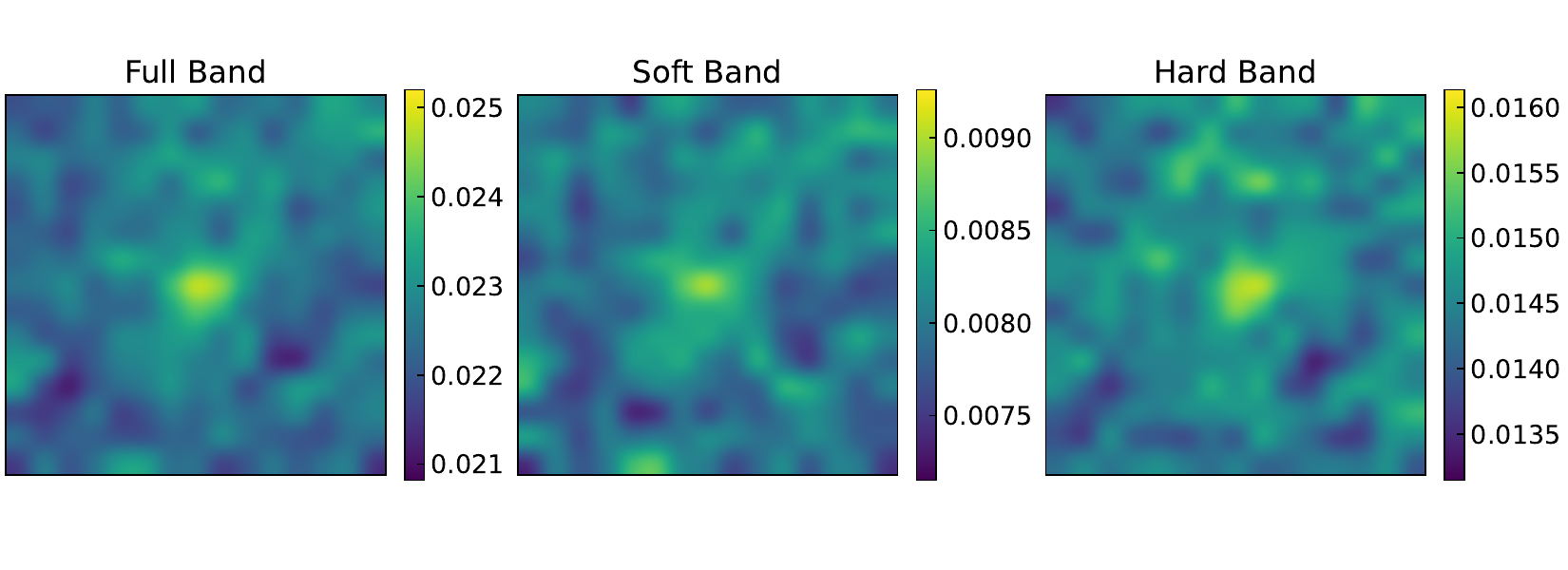}
    \caption{Smoothed stacked \mbox{X-ray} images combining all three EPIC cameras for our core-sample DOGs that are away from known \mbox{X-ray} sources (647 in total) in the FB, SB, and HB. Each image has a size of $60\arcsec\times 60\arcsec$. The numbers on the color bars are in units of $\rm count\,s^{-1}$.}\label{fig:stacking_img}
\end{figure*}

X-ray stacking allows for the detection, on average, of sources lying below the formal detection limits \citep[e.g.,][]{Vito+2016}. In this subsection, we stack the \mbox{X-ray} images of our \mbox{X-ray} undetected DOGs to study their \mbox{X-ray} properties further. We select DOGs at least 52$\arcsec$ away from all the \mbox{X-ray} sources to avoid contamination, which results in the selection of about 50\% of the sources. We restrict our stacking to regions where the total FB exposure from all three EPIC cameras $>14\,\rm ks$ in W-CDF-S, $>10\,\rm ks$ in ELAIS-S1, and $>25\,\rm ks$ in XMM-LSS, which constitute $\approx95\%$ of the area covered in \mbox{X-rays}. This step removes pixels with low exposures that may adversely affect the count-rate calculations. We also restrict the stacking for the other bands to the same regions. We end up with 1825 sources in the full sample and 647 in the core sample. We then extract the combined count-rate maps from all three EPIC cameras in all three \mbox{X-ray} bands within a $60\arcsec\times60\arcsec$ region around each selected source and sum the maps to obtain the stacked image. Figure~\ref{fig:stacking_img} shows the smoothed stacked images of \mbox{X-ray} undetected DOGs in the FB, SB, and HB for our core sample. The stacked signal is prominent in all three bands visually. To further assess the false-detection probabilities, we perform Monte Carlo stacking analyses \citep[e.g.,][]{Brandt+2001a}.

\begin{figure}[htb!]
    \centering
    \includegraphics[width=\linewidth]{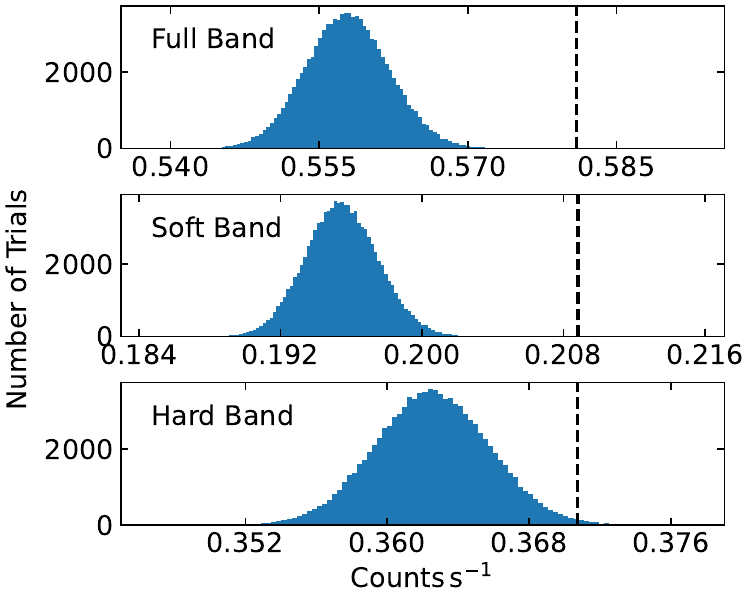}
    \caption{Results from our Monte Carlo \mbox{X-ray} stacking for the core sample. For each plot, we perform 100000 stacking trials at random positions around the undetected sources. The vertical dashed lines represent the stacked count rates at the positions of the undetected sources. In all three bands, we have significant detections. The results for our full sample are similar.}\label{fig:MC_stacking}
\end{figure}

The Monte Carlo stacking analyses are performed for the 1825/647 \mbox{X-ray} undetected DOGs in the full/core sample that are free from contamination and low-exposure regions. For our core sample, we stack 647 images using a $20\arcsec\times20\arcsec$ aperture ($5\times5$ pixels) at the position of each source to obtain the stacked source count rates. The aperture is chosen to be consistent with our calculations of count rates and flux in the next paragraph. We further perform local background stacking with 100000 trials, where we stack 647 random positions using the same aperture. These random positions are chosen to lie in an annular region (with an inner radius of $1\arcmin$ and an outer radius of $2\arcmin$) centered on each source (avoiding known \mbox{X-ray} sources) to reproduce the actual background distribution as closely as possible. The stacked exposure reaches $\approx55\,\rm Ms$, enabling the average detection below our survey sensitivity. We repeat the above procedure for all three bands and show the results for our core sample in Figure~\ref{fig:MC_stacking}. In all three bands, the resulting background distributions are nearly Gaussian. There are 0/0/573 out of 100000 trials in the FB/SB/HB with background count rates higher than the stacked source count rates. This corresponds to a false-detection level ($P_\mathrm{false}$) of $<10^{-5}$ for the FB and the SB, and $0.00573$ for the HB, indicating significant detections in all three bands. The results for our full sample are similar.

We further calculate the count rates within the \mbox{$5\times5$-pixel} aperture and net source fluxes using the single-camera exposure ($t$) maps, encircled-energy fraction (EEF) maps, and energy-conversion factors (ECFs) in \citet{Chen+2018} and \citet{Ni+2021}, where EEF is the expected fraction of source counts falling within the given aperture centered at the source position, and ECF is the expected ratio between the source flux and source counts. The $5\times5$-pixel aperture is chosen to be consistent with the EEF maps derived in \citet{Chen+2018} and \citet{Ni+2021}. We apply the same procedures as in our Monte Carlo count analyses to derive the stacked count rates and background count-rate distributions. The background count-rate distributions are also nearly Gaussian. Following \citet{Ruiz+2022}, we convert the stacked net count rates to fluxes in each band using
\begin{equation}
    f_\mathrm{X}=S\sum^{N}\sum_{i=1}^{3} t_i/\sum^{N}\sum_{i=1}^{3}t_i\,\mathrm{EEF}_i\,\mathrm{ECF}_i,
\end{equation}
where $S$ denotes the stacked net count rates within $5\times5$ pixels, $f_\mathrm{X}$ denotes the derived \mbox{X-ray} flux, $i$ denotes the cameras (PN, MOS1, and MOS2), and $N$ denotes the number of stacked sources. The stacking results are summarized in Table~\ref{tab:stacking}. For the 647 sources stacked in the core sample, after correction for Galactic absorption, the average observed HB net flux ($f_\mathrm{X}^\mathrm{HB}$) is $(3.3\pm1.3)\times10^{-16}\,\fluxcgs$. Assuming an effective power-law photon index of $\Gamma_\mathrm{eff}=1.4$, the median $\lxobs$ at rest-frame $2-10\,\kev$ is $(4.3\pm1.7)\times10^{42}\,\lumcgs$. The results for our full sample are similar. We also have verified that our \mbox{X-ray} stacking procedure does not have significant biases by stacking \mbox{X-ray} detected sources with known X-ray flux. 

We further estimate the non-AGN \mbox{X-ray} emission from host galaxies to assess if the stacked \mbox{X-ray} emission is sufficiently strong to indicate the presence of AGNs. \mbox{X-ray} emission from host galaxies is expected to primarily comes from \mbox{X-ray} binaries (XRBs) and hot gas. Following the same procedure as in Section~2.2 of \citet{Zou+2023}, we adopt the scaling relation in \citet{Lehmer+2016} to estimate the total HB flux from low-mass XRBs and high-mass XRBs assuming an intrinsic power-law photon index of 1.8, and the scaling relation in \citet{Kim&Fabbiano+2015} to estimate the \mbox{X-ray} emission from hot gas. We then apply K-corrections using \texttt{sherpa} \citep{Freeman+2001,Doe+2007} to convert the hot-gas \lx\ to the HB flux using the hot-gas spectra, with the gas temperature given by the scaling relation. For our core sample, the average $f_\mathrm{X}^\mathrm{HB}$ estimated from stacking is 3.8 times higher than the estimated average flux ($8.6\times10^{-17}\,\fluxcgs$) contributed by XRBs and hot gas. The results for our full sample are similar. The observed HB flux is much higher than the predicted values from non-AGN contributions, proving the presence of AGNs among our \mbox{X-ray} undetected DOGs.

Among \mbox{X-ray} undetected DOGs, objects can be classified into three categories following Section~3.5 of \citet{Zou+2022}: reliable SED AGNs, AGN candidates but not reliable SED AGNs, and normal galaxies (see Section~3.1). We further stack these three subsets separately to see if our SED-based classification truly reflects the contribution of AGNs. Among the 647 sources in the core sample, 37 are reliable SED AGNs, 191 are AGN candidates but not reliable SED AGNs, and 419 are normal galaxies. We use the same procedures as described earlier in this subsection to stack these subsets.

The stacking results are summarized in Table~\ref{tab:stacking}. We have checked that the stacked count rates are not dominated by any individual source. The results show that sources classified as AGN candidates indeed have more significant \mbox{X-ray} detections. We find the subsets for reliable SED AGNs produce the highest average \lxobs\ at rest-frame 2--10\,\kev, while no detections are found for normal galaxies in the HB. AGN candidates also show slightly elevated HR compared to \mbox{X-ray} undetected DOGs in general, which may be because normal galaxies contribute more to the SB count rates than to the HB. In fact, we have verified that both the predicted SB flux and HB flux for normal galaxies are consistent with the total XRB and hot-gas emission predicted by \citet{Lehmer+2016} and \citet{Kim&Fabbiano+2015}. Thus, the HR of our AGN candidates is more representative of the obscuration in the nuclear region. We also plot the HR for our stacked reliable SED AGNs in Figure~\ref{fig:NH}. On average, our \mbox{X-ray} undetected AGN candidates have $\nh\gtrsim10^{23}\,\nhcgs$, which is consistent with or slightly higher than the median value of $\nh=10^{22.8}\,\nhcgs$ for our \mbox{X-ray} detected DOGs.

We further calculate the total rest-frame \mbox{2--10\,\kev} \lxobs\ ($\lxobs^\mathrm{tot}$) of each stacked subset to probe their overall accretion power, assuming that each source contributes equally to the total $f_\mathrm{X}^\mathrm{HB}$. The total accretion power can be traced by the total intrinsic \lx. For \mbox{X-ray} undetected DOGs, we assume that every source is obscured at the average obscuration level of $\nh=10^{23}\,\nhcgs$ as shown in the previous paragraph. We then use \texttt{PIMMS} to convert the $\lxobs$ to intrinsic \lx. The correction is generally small (\mbox{$\approx5-10\%$}) as the HB corresponds to $\approx6-30\,\kev$ in the rest frame at the median redshift of our sources. We also use the intrinsic \lx\ derived in Section~\ref{subsec:HR} to assess the accretion power contributed by \mbox{X-ray} detected DOGs. We find the total accretion power for our DOGs is dominated by \mbox{X-ray} detected sources, which contribute $\approx{75\%}$ of the total intrinsic \lx\ for our core sample. Even in the extreme case where all the stacked \mbox{X-ray} undetected DOGs are heavily obscured at $\nh\approx10^{24}\,\nhcgs$, the correction factor for their total \lx\ would only be $\approx1.5$, which will not significantly impact our results. The results are consistent with previous measurements and/or simulations of SMBH growth, which conclude that most SMBH growth occurs in luminous AGNs \citep[e.g.,][]{Brandt&Alexander+2015,Vito+2016,Volonteri+2016,Zou+2024}.


\begin{deluxetable*}{lccccccccc}[ht!]
\tablecaption{X-ray stacking results for subsamples of \mbox{X-ray} undetected DOGs.\label{tab:stacking}}
\tablewidth{0pt}
\tablehead{
\colhead{Subsample} & \colhead{$N$} & \colhead{$P_\mathrm{false}^\mathrm{SB}$} & \colhead{\soft} & \colhead{$P_\mathrm{false}^\mathrm{HB}$} & \colhead{\hard} & \colhead{Hardness ratio} & \colhead{$f_\mathrm{X}^\mathrm{HB}$} & \colhead{$L_\mathrm{X,obs}$} & \colhead{$L_\mathrm{X,obs}^\mathrm{tot}$}\\
\colhead{(1)} & \colhead{(2)} & \colhead{(3)} & \colhead{(4)} & \colhead{(5)} & \colhead{(6)} & \colhead{(7)} & \colhead{(8)} & \colhead{(9)} & \colhead{(10)}
}
\startdata
\text{Full Sample} & & & & & & & & &\\
\phantom{p}All & 1825 & $10^{-5}$ & $2.11\pm0.21$ & $<10^{-5}$ & $1.70\pm0.31$ & $-0.11\pm0.10$ & $4.5\pm0.8$ & $0.62\pm0.11$ & $18.3\pm3.2$\\
\phantom{p}Reliable SED AGNs & 171 & 0.00158 & $4.06\pm0.76$ & $<10^{-5}$ & $4.50\pm1.12$ & $0.05\pm0.16$ & $12.4\pm3.1$ & $3.2\pm0.8$ & $\phantom{p}8.1\pm2.0$\\
\phantom{p}AGN candidates but not reliable & 694 & $<10^{-5}$ & $2.47\pm0.34$ & $<10^{-5}$ & $2.61\pm0.51$ & $0.03\pm0.12$ & $7.0\pm1.4$ & $1.12\pm0.22$ & $14.7\pm2.9$\\
\phantom{p}Normal galaxies & 960 & $<10^{-5}$ & $1.51\pm0.29$ & $0.12963$ & $0.54\pm0.41$ & ... & ... & ...  & ...\\
\hline
\text{Core Sample} & & & & & & & & &\\
\phantom{p}All & 647 & $<10^{-5}$ & $2.06\pm0.32$ & $0.00083$ & $1.27\pm0.50$ & $-0.24\pm0.18$ & $3.3\pm1.3$ & $0.43\pm0.17$ & $2.7\pm1.1$\\
\phantom{p}Reliable SED AGNs & 37 & $0.04796$ & $6.47\pm1.45$ & $0.00057$ & $5.34\pm2.60$ & $-0.09\pm0.26$ & $14.5\pm7.0$ & $1.7\pm0.8$ & $0.8\pm0.4$\\
\phantom{p}AGN candidates but not reliable & 191 & $0.00038$ & $2.10\pm0.57$ & $0.00121$ & $2.70\pm0.89$ & $0.13\pm0.22$ & $7.3\pm2.3$ & $0.83\pm0.26$ & $1.8\pm0.6$\\
\phantom{p}Normal galaxies & 419 & $0.00005$ & $1.65\pm0.40$ & $0.34431$ & $0.24\pm0.61$ & ... & ... & ... & ...\\
\enddata
\tablecomments{Our stacking utilizes \mbox{X-ray} images from all three EPIC cameras. We only show hardness ratio, flux, $\lxobs$, and $\lxobs^\mathrm{tot}$ when a detection of $>2\sigma$ ($P_\mathrm{false}<0.05$) is achieved. Column (1): Subsamples of \mbox{X-ray} undetected DOGs based upon the SED classification results. Column (2): Number of sources stacked. Column (3): Fraction of trials with stacked background count rates higher than the stacked count rate at the position of the source in the SB. Column (4): Average net count rate within a $20\arcsec\times20\arcsec$ aperture in the SB in units of $10^{-5}$ $\rm counts\,s^{-1}$. Column (5): Fraction of trials with stacked background count rates higher than the stacked count rate at the position of the source in the HB. Column (6): Average net count rate within a $20\arcsec\times20\arcsec$ aperture in the HB in units of $10^{-5}$ $\rm counts\,s^{-1}$. Column (7): Hardness ratio. Column (8): Galactic absorption-corrected average observed net flux in the HB in units of $10^{-16}\,\fluxcgs$. Column (9): Observed \mbox{X-ray} luminosity at rest-frame 2--10$\,$\kev\ in units of $10^{43}\,\lumcgs$ using the median redshift of each subset, calculated from the Galactic absorption-corrected average net HB flux assuming $\Gamma_\mathrm{eff}=1.4$. Column (10): The total rest-frame 2--10\,\kev\ \lxobs\ in units of $10^{45}\,\lumcgs$ contributed by each subset.}
\end{deluxetable*}

\subsection{$\lxobs$ versus $\lsix$}\label{subsec:lx-l6}

The MIR continuum luminosity is a robust indicator of the intrinsic AGN strength, as MIR emission is largely unaffected by obscuration except for the most extreme obscuration levels \citep[$A_V\approx30$; e.g.,][]{Stern+2015}, while our sources are much below such levels. Many studies have shown a tight relationship between the absorption-corrected \lx\ and the rest-frame $6\,\rm \mu m$ continuum luminosity contributed by AGNs \citep[$\nu L_\nu^\mathrm{AGN}$, written as \lsix\ hereafter; e.g.,][]{Fiore+2009, Lanzuisi+2009,Stern+2015,Chen+2017}. Since $\lxobs$ for AGNs will be significantly suppressed when $\nh$ is sufficiently large, comparing \lxobs\ versus \lsix\ with the nominal absorption-corrected $\lx-\lsix$ relation can be helpful to identify heavily-obscured and CT AGNs \citep[e.g.,][]{Rovilos+2014,Lanzuisi+2018,Guo+2021,Yan+2023}. 


The \lsix\ and its uncertainty are derived from the \texttt{CIGALE} SED-fitting output \texttt{agn.L\_6um}. It is worth noting that the rest-frame $6\,\mum$ luminosity utilized here should be solely contributed from AGNs.
At the redshift range of our DOGs ($z\approx1.5-2.5$), rest-frame $6\,\rm \mu m$ corresponds to $15-20\,\rm \mu m$ in the observed frame. Since our sources have high $24\,\rm\mu m$ fluxes by construction, and all sources have at least one photometric band with $\rm SNR>5$ at $4.5\,\mum$, $5.8\,\mum$, or $8\,\mum$, the measurements of \lsix\ should be reliable as long as the emission at rest-frame $6\,\mum$ is dominated by AGNs \citep[i.e., small galaxy contamination;][]{Yang+2020}. We calculate the AGN fractional flux contribution at rest-frame $6\,\mum$ using the \texttt{CIGALE} output \texttt{agn.fracAGN} with \texttt{lambda\_fracAGN} set to ``6/6" \citep{Yang+2022}. Around 90\% of AGN candidates among our DOGs have an AGN fractional contribution $>50\%$, indicating the \lsix\ measurements should be reliable.

\begin{figure}[t!]
    \centering
    \includegraphics[width=\linewidth]{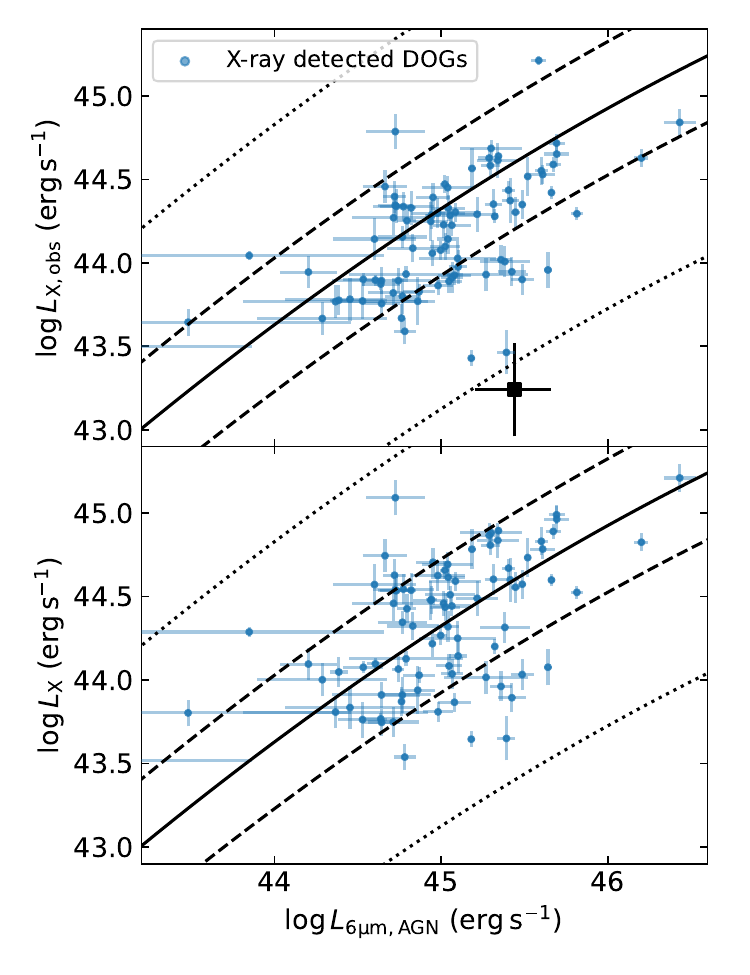}
    \caption{Top panel: the rest-frame $2-10\,\rm keV$ \lxobs\ versus \lsix\ for the core sample. Bottom Panel: the rest-frame $2-10\,\rm keV$ absorption-corrected \lx\ versus \lsix\ for the core sample. The blue points represent \mbox{X-ray} detected DOGs, and the associated error bars represent $1\sigma$ uncertainties. The black square represents the stacked average \lxobs\ versus the median \lsix\ for \mbox{X-ray} undetected reliable SED AGNs. The solid line shows the absorption-corrected $\lx-\lsix$ relation in \citet{Stern+2015}; the dashed and dotted lines are the $1\sigma$ and $3\sigma$ dispersions of their sample. The results for our full sample are similar.}\label{fig:Lx_6um}
\end{figure}

We plot \lxobs\ versus \lsix\ for our \mbox{X-ray} detected DOGs in the top panel of Figure~\ref{fig:Lx_6um}. 
We also show the absorption-corrected $\lx-\lsix$ relation in \citet{Stern+2015} along with the $1\sigma$ and $3\sigma$ dispersions of their sample. Most \mbox{X-ray} detected DOGs lie within the $1\sigma$ dispersion, although they generally show slightly suppressed \lxobs\ and some are below the $1\sigma$ dispersion range. Such deviations are likely due to obscuration as described in Section~\ref{subsec:HR}. We plot the absorption-corrected \lx\ versus \lsix\ in the bottom panel of Figure~\ref{fig:Lx_6um}, and most of our \mbox{X-ray} detected DOGs are now consistent with the \citet{Stern+2015} relation and do not show obvious systematic offsets. 
We further show the stacked average \lxobs\ versus the median \lsix\ for \mbox{X-ray} undetected reliable SED AGNs in our core sample (marked with a black square) in the top panel of Figure~\ref{fig:Lx_6um}. As all reliable SED AGNs have AGN fractional flux contribution $>50\%$ at rest-frame 6\mum, their median \lsix\ should be reliable. On average, reliable SED AGNs are $\approx3\sigma$ ($\approx1.5\,\rm dex$) below the $\lxobs-\lsix$ relation. We calculate the expected line-of-sight $\nh$ value corresponding to such a low $\lxobs/\lsix$ using photoelectric absorption with Compton-scattering losses, and obtain $\nh\approx10^{24}\,\nhcgs$. This value is much larger than the HR-derived $\nh$ of $\approx10^{23}\,\nhcgs$ in Section~\ref{subsec:X-ray_stacking} and Figure~\ref{fig:NH}. There are two main reasons for this discrepancy. First, our HR-derived $\nh$ does not consider a possible soft scattered component. A soft scattered component is often observed in the soft \mbox{X-ray} band of AGNs. For obscured AGNs, this component is likely a power-law scattered back into the line of sight, and it generally has $<10\%$ of the flux of the primary power-law \citep[e.g.,][]{Guainazzi&Bianchi+2007,Brightman+2012}. The presence of the soft scattered component will lead to an underestimated $\nh$ based upon HR. Second, $\lxobs/\lsix$ is not solely determined by $\nh$. Physical modeling shows that $\lxobs/\lsix$ may also be sensitive to AGN torus covering fraction and the incident \mbox{X-ray} continuum shape \citep{Yaqoob&Murphy+2011}; observations have also shown that AGNs with higher $\fir/f_R$ have lower $\lxobs/\lsix$ (up to $1\,\rm dex$ difference) when their $\nh$ values are similar to AGNs with lower $\fir/f_R$ \citep[e.g., see Figure~8 in][]{Li+2020}. Our DOGs indeed have high $\fir/f_R$ by construction, so the $\nh$ derived from the deviation from the absorption-corrected $\lx-\lsix$ relation may be overestimated. This effect may also partly explain why there are several \mbox{X-ray} detected DOGs slightly below the \citet{Stern+2015} relation after absorption correction. 

Apart from these two effects, it is also possible that galaxy contamination in the SB among our reliable SED AGNs results in a lower HR, as the purity of the reliable SED AGNs is $\gtrsim75\%$ (see Section~\ref{subsec:host-galaxy}). However, we have verified that galaxy contamination is generally small and should not cause significant differences in our HRs. Nevertheless, the significantly lower $\lxobs/\lsix$ for \mbox{X-ray} undetected reliable SED AGNs than those for \mbox{X-ray} detected DOGs implies that AGNs among \mbox{X-ray} undetected DOGs have heavier obscuration, and some may even reach CT levels.


%




\subsection{Radio Properties}\label{subsec:radio}

\begin{figure*}[htb!]
    \centering
    \includegraphics[width=0.9\linewidth]{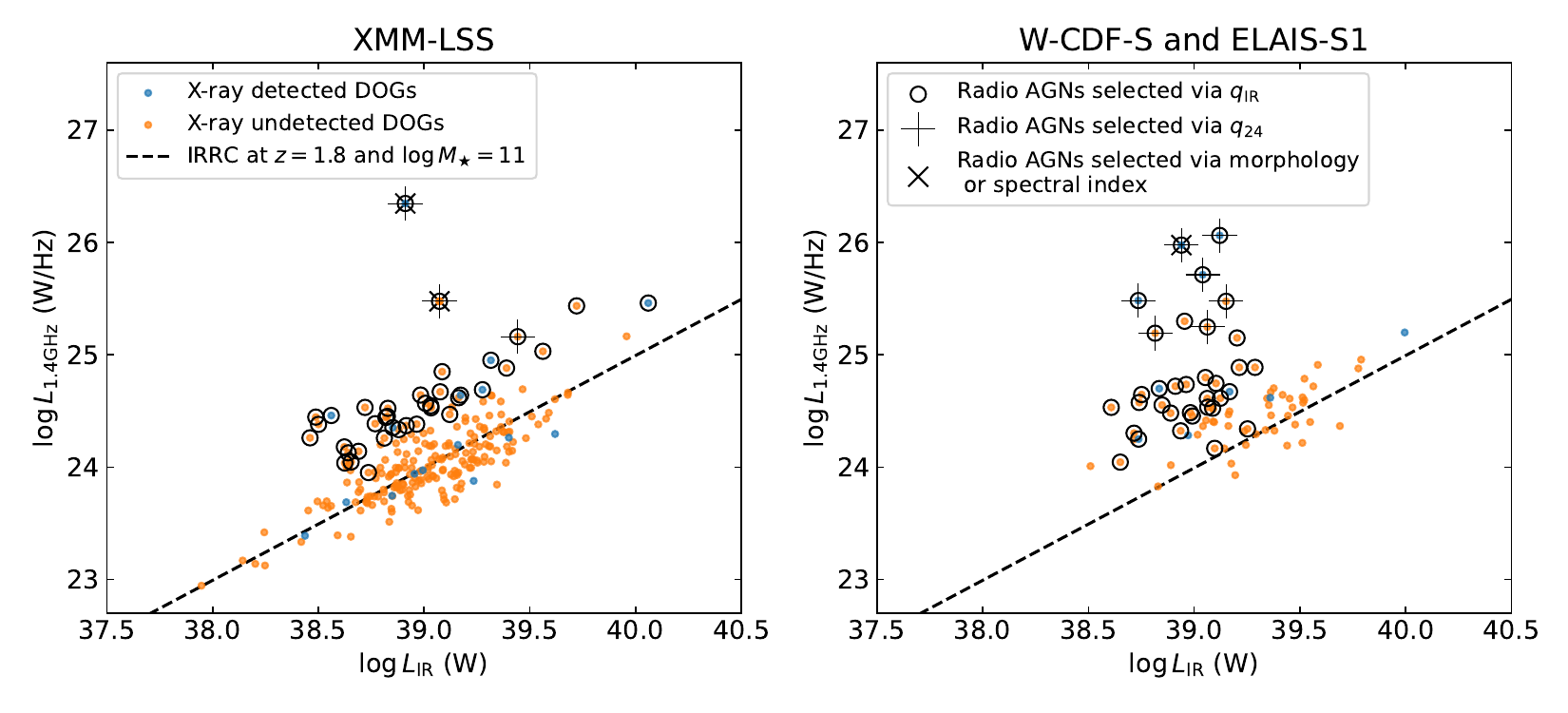}
    \caption{$L_\mathrm{1.4\,GHz}$ versus $L_\mathrm{IR}$ for our core sample. The left panel shows sources in XMM-LSS, and the right panel shows sources in W-CDF-S and ELAIS-S1. XMM-LSS has much higher radio sensitivity and thus has more radio-detected DOGs. \mbox{X-ray} detected and \mbox{X-ray} undetected DOGs are shown in blue and orange, respectively. The empty circles represent radio AGNs selected via \qfir\ in \mbox{Zhang et al. (submitted)}. The plus signs represent radio AGNs selected via \qmir\ in \citet{Zhu+2023}. The ``X"s represent radio AGNs selected via morphology or spectral index in \citet{Zhu+2023}. The dashed lines represent the IRRC of \citet{Delvecchio+2021} at $z=1.8$ and $\log\mstar/\msun=11$ with an additional 0.3\,dex offset.}\label{fig:radio}
\end{figure*}

The XMM-SERVS fields are also covered by sensitive radio surveys at $1.4\,\rm GHz$, including the ATLAS in W-CDF-S and ELAIS-S1 \citep{Norris+2006,Hales+2014,Franzen+2015}, and a VLA survey and the MIGHTEE survey in XMM-LSS \citep{Heywood+2020, Heywood+2022}. MIGHTEE covers $3.5\deg^2$ in XMM-LSS and reaches a superb $5\sigma$ sensitivity of $28\,\uJy$; the other surveys are relatively shallower ($5\sigma$ sensitivity $\approx85\,\uJy$) but cover wider areas. These radio data have been extracted and analyzed by \citet{Zhu+2023} and matched to our DOGs. There are 745 (20.0\%) and 317 (24.4\%) DOGs with $1.4\,\rm GHz$ detections in the full and core samples, respectively.

Both SF processes and AGN processes (e.g., jets) can produce radio emission from extragalactic sources. However, SF-related radio emission generally follows a tight correlation with the IR emission, which is known as the IR-radio correlation \citep[IRRC; e.g.,][]{Condon+1992,Tabatabaei+2017,Delvecchio+2017,Delvecchio+2021}. To identify radio AGNs, one can look for radio emission that exceeds the levels predicted by the IR emission from SF. Two parameters are often used to identify radio excess: one is the observed $24\,\mu$m-to-1.4\,GHz flux ratio \citep[e.g.,][]{Appleton+2004}: $\qmir=\log(f_\mathrm{24\,\rm \mu m}/f_\mathrm{1.4\,\rm GHz})$, where $f_\mathrm{1.4\,\rm GHz}$ is the flux density at observed-frame $1.4\,\rm GHz$; the other is the rest-frame FIR-to-radio flux ratio \citep[e.g.,][]{Sargent+2010}: $\qfir=\log(\frac{L_\mathrm{IR}[\mathrm{W}]}{3.75\times10^{12}[\mathrm{Hz}]})-\log(L_\mathrm{1.4GHz}[\mathrm{W/Hz}])$, where $L_\mathrm{IR}$ is the rest-frame \mbox{8--1000\,$\mu$m} total luminosity, and $L_\mathrm{1.4GHz}$ is the monochromatic luminosity at rest-frame 1.4\,GHz.\footnote{$L_\mathrm{IR}$ is calculated by integrating the best-fit SED models over rest-frame 8--1000\,$\mu$m. The rest-frame $L_\mathrm{1.4\,GHz}$ is converted from the observed-frame $f_\mathrm{1.4\,GHz}$ assuming a power-law radio spectral shape $f_\nu\propto\nu^{\alpha_r}$, where $\alpha_r=-0.7$.} Sources with radio excess (i.e., low \qmir\ or low \qfir) can be identified as radio AGNs. \citet{Zhu+2023} employed the criterion for \qmir\ in \citet{Appleton+2004} in XMM-SERVS and identified 1763 radio AGNs; \mbox{Zhang et al. (submitted)} employed the \qfir\ criterion in \citet{Delvecchio+2021} and identified 6766 radio AGNs in XMM-SERVS. The \qfir\ criterion is more complete than the \qmir\ criterion, leading to a sample size $\approx4$ times that of \citet{Zhu+2023} while maintaining a satisfactory purity of $\approx95.2\%$. In fact, the \qmir\ criterion may not be very applicable to our DOGs with luminous AGNs since there is significant AGN contamination in the MIR. As we have shown in Section~\ref{subsec:lx-l6}, most of our AGN candidates are dominated by the AGN component at rest-frame $6\,\mu\rm m$, which approximately corresponds to $12-24\,\mu\rm m$ in the observed-frame. On the other hand, $L_\mathrm{IR}$ over rest-frame \mbox{8--1000} \mum\ is less affected, where only 31\% of DOGs have fractional AGN contributions of $>50\%$. $L_\mathrm{IR}$ is also not primarily driven by the $24\,\mum$ photometry since approximately half of our DOGs have at least one Herschel FIR band with $\mathrm{SNR}>5$. Note that we do not use the conventional radio-loudness parameter for luminous quasars \citep[e.g.,][]{Kellermann+1994} since it assumes that the optical emission is dominated by AGNs, which is not the case for our sources. We also do not rely on the \texttt{radio} module in \texttt{CIGALE} to calculate $\qfir$ because \texttt{CIGALE} only considers the host-galaxy contribution to $\qfir$ and the AGN contribution is controlled by the radio-loudness parameter \citep{Yang+2022}, while we compare the total $\qfir$ with the IRRC to identify radio-excess AGNs.

Figure~\ref{fig:radio} presents $L_\mathrm{1.4\,\rm GHz}$ versus $L_\mathrm{IR}$ for our core sample. We plot sources in XMM-LSS separately, as MIGHTEE provides a much deeper radio depth than those in the other two fields, which results in more radio-detected DOGs in XMM-LSS (237) than in the other two fields (80). For comparison, we also show the IRRC of \citet{Delvecchio+2021}, assuming \mbox{$z=1.8$} and $\log\mstar/\msun=11$, with an additional 0.3\,dex offset that accounts for the systematic difference in \lir\ following Zhang et al. (submitted). Most DOGs follow a strong correlation between $L_\mathrm{1.4\,\rm GHz}$ and $L_\mathrm{IR}$, as predicted by the IRRC. We mark radio AGNs selected via \qfir\ and \qmir, as well as those selected via morphology or spectral index in \citet{Zhu+2023}, which are generally independent indicators of radio excess. All these radio AGNs show elevated $L_\mathrm{1.4\,\rm GHz}$. 39 (34) radio AGNs are identified via \qfir\ in XMM-LSS (W-CDF-S and ELAIS-S1), constituting 16.3\% (42.5\%) of the radio-detected DOGs. 
Fewer radio AGNs are identified via \qmir\ (3 in XMM-LSS and 7 in W-CDF-S and ELAIS-S1) and they generally have the strongest radio emission among DOGs. This can be explained by the fact that the AGN component generally is dominant at rest-frame 6\,$\mu$m, so sources require much stronger radio emission to be selected via \qmir.

Among the 73 radio AGNs selected via \qfir, only 16 of them are identified as reliable SED AGNs. Only one of the three radio AGNs selected by morphology or spectral index is identified as a reliable SED AGN. There are 15 radio AGNs detected in \mbox{X-rays}, and they are all identified as \mbox{X-ray} AGNs. We stack the \mbox{X-ray} images of the 30 \mbox{X-ray} undetected radio AGNs away from known \mbox{X-ray} sources, and we do not obtain detections at $>2\sigma$ significance in any \mbox{X-ray} band. Previous work on the VLA/FIRST $1.4\,\rm GHz$ detected radio-excess DOG J1406+0102 does not show an \mbox{X-ray} detection either \citep{Fukuchi+2023}. Despite the relatively small sample size of radio AGNs among DOGs, the AGN selection results based upon radio, SED, and \mbox{X-rays} show minimal overlap. This indicates that radio selection can identify AGNs that can hardly be selected via other methods among DOGs, which aligns with the general results for radio AGN selection in XMM-SERVS \citep{Zhu+2023}. 

\citet{Zhu+2023} also compiled counterparts of their $1.4\,\rm GHz$ radio sources at lower and higher radio frequencies in the three XMM-SERVS fields. The utilized radio surveys include the LOw Frequency ARray \citep[LOFAR;][]{Hale+2019} observations at $144\,\rm MHz$ of XMM-LSS, the Rapid ASKAP Continuum Survey \citep[RACS;][]{McConnell+2020} at $887.5\,\rm MHz$ of all three fields, and the $2.3\,\rm GHz$ ATLAS observations of W-CDF-S and ELAIS-S1 \citep{Zinn+2012}. Among our 237 (80) core-sample radio-detected DOGs in XMM-LSS (\mbox{W-CDF-S} and \mbox{ELAIS-S1}), only 9 (7) of them are detected by LOFAR or RACS (RACS or ATLAS $2.3\,\rm GHz$). We further calculate their radio spectral slopes between $1.4\,\rm GHz$ and lower/higher frequencies. Considering their higher detection rates compared to RACS, we use LOFAR measurements in XMM-LSS when possible, and we use ATLAS $2.3\,\rm GHz$ measurements for \mbox{W-CDF-S} and ELAIS-S1 when available. Among the 16 radio-detected DOGs, the median radio spectral slope ($\alpha_r$) is --0.65, and only 5 of them are identified as flat-spectrum radio sources (defined as $\alpha_r>-0.5$). There is one steep-spectrum radio source ($\alpha_r=-0.94$) showing extended double-lobe radio emission, which is consistent with the radio-AGN unification model where steep-spectrum radio sources tend to have lobe-dominated radio morphology \citep[e.g.,][]{Tadhunter+2008,Pyrzas+2015}.

\section{Discussion}\label{sec:discussion}

\subsection{AGN Fractions}\label{subsec:agn_fraction}


In this subsection, we investigate the fraction of our DOGs hosting accreting SMBHs above a certain accretion-rate threshold ($\lambda_\mathrm{thres}$), i.e., the AGN fraction (\fagn). Following \citet{Aird+2018}, we define specific black-hole accretion rate ($\lambda$) in dimensionless units, such that
\begin{equation}\label{eqn:lambda}
    \lambda=k\,\lx/\mstar,
\end{equation}
where 
\begin{equation}
    k=\frac{\kbol}{1.3\times10^{38}\,\lumcgs\times0.002\,\msun^{-1}}.
\end{equation}
\lx\ is the absorption-corrected rest-frame $2-10\,\kev$ luminosity, and \kbol\ is the bolometric correction factor \citep[we adopt $\kbol=25$ in this paper following][]{Aird+2018}. We choose the additional factor $k$ so that $\lambda$ is approximately the Eddington ratio (\lambedd). We further assume a constant $\lambda$ at its typical value $\lambda_0$ for sources accreting above a certain threshold ($\lambda_0>\lambda_\mathrm{thres}$; i.e., \fagn\ of our sources accrete at $\lambda_0$, and the others accrete below $\lambda_\mathrm{thres}$). \citet{Aird+2018} have shown that for AGNs accreting at $\lambda>0.01$, $\langle\lambda\rangle\approx0.1-1$, where $\langle\lambda\rangle$ is the average specific accretion rate\footnote{These values do not consider to low-excitation radio galaxies (LERGs), which generally accrete with $\lambda<0.01$ \citep{Best&Heckman+2012}.}. In fact, the similarity between $\lambda$ and $\lambedd$ already provides a reasonable range of typical $\lambda$, as we would expect most sources are at sub-Eddington levels.

Considering the contribution from all sources, Equation~\ref{eqn:lambda} becomes
\begin{equation}\label{eqn:total}
    \fagn=k\,\lambda_0^{-1}\,\sum \lx/\sum \mstar,
\end{equation}
where the summations run over all sources in the sample. A simple, intuitive physical interpretation of Equation~\ref{eqn:total} is that $\lambda_0$ and $\fagn$ are degenerate: for a given total intrinsic \mbox{X-ray} luminosity from all sources, the more powerful the central engines are, the lower is the required AGN fraction.

The total intrinsic \mbox{X-ray} luminosity is contributed by both the \mbox{X-ray} detected sources and the undetected ones, and our $\fagn$ have considered both. We use the absorption-corrected \lx\ in Sections~\ref{subsec:HR} and \ref{subsec:X-ray_stacking} for \mbox{X-ray} detected and \mbox{X-ray} undetected DOGs, respectively. Assuming a constant $\lambda_0=0.1$, we obtain $\fagn=15\%$ for our full sample; if $\lambda_0=1$, $\fagn=1.5\%$. As for the core sample, when $\lambda_0=0.1-1$, $\fagn=20-2.0\%$. These values are generally consistent with the typical AGN fraction of $10-20\%$ at $z\approx1.8$ and $\log\mstar\approx11$ \citep[i.e., our median $z$ and $\log\mstar$,][]{Xue+2010,Aird+2018,Zou+2024}. This indicates that DOGs do not appear to host a distinctively higher fraction of AGNs. The AGN fraction for DOGs is also similar to that of $17^{+16}_{-6}\%$ for SMGs at $z\approx2-3$ \citep{Wang+2013} which are also strongly star-forming and dusty \citep[e.g.,][]{Alexander+2005}. However, Hot DOGs, a more extreme subset of HyLIRGs with extreme MIR colors, have been found to host stronger AGN activity than reddened quasars \citep{Vito+2018}. They are thought to be at the peak of SMBH accretion when the feedback has not yet swept away the surrounding gas and dust in the merger-driven coevolution framework. Our results show that DOGs, in general, do not present significantly different SMBH accretion compared with AGNs among a matched typical galaxy population, at least for those selected via the criteria of \citet{Dey+2008}.

\subsection{Comparison between \mbox{X-ray} Detected DOGs and non-DOG \mbox{X-ray} AGNs}\label{subsec:compare}


\begin{figure*}[htbp!]
    \centering
    \includegraphics[width=\linewidth]{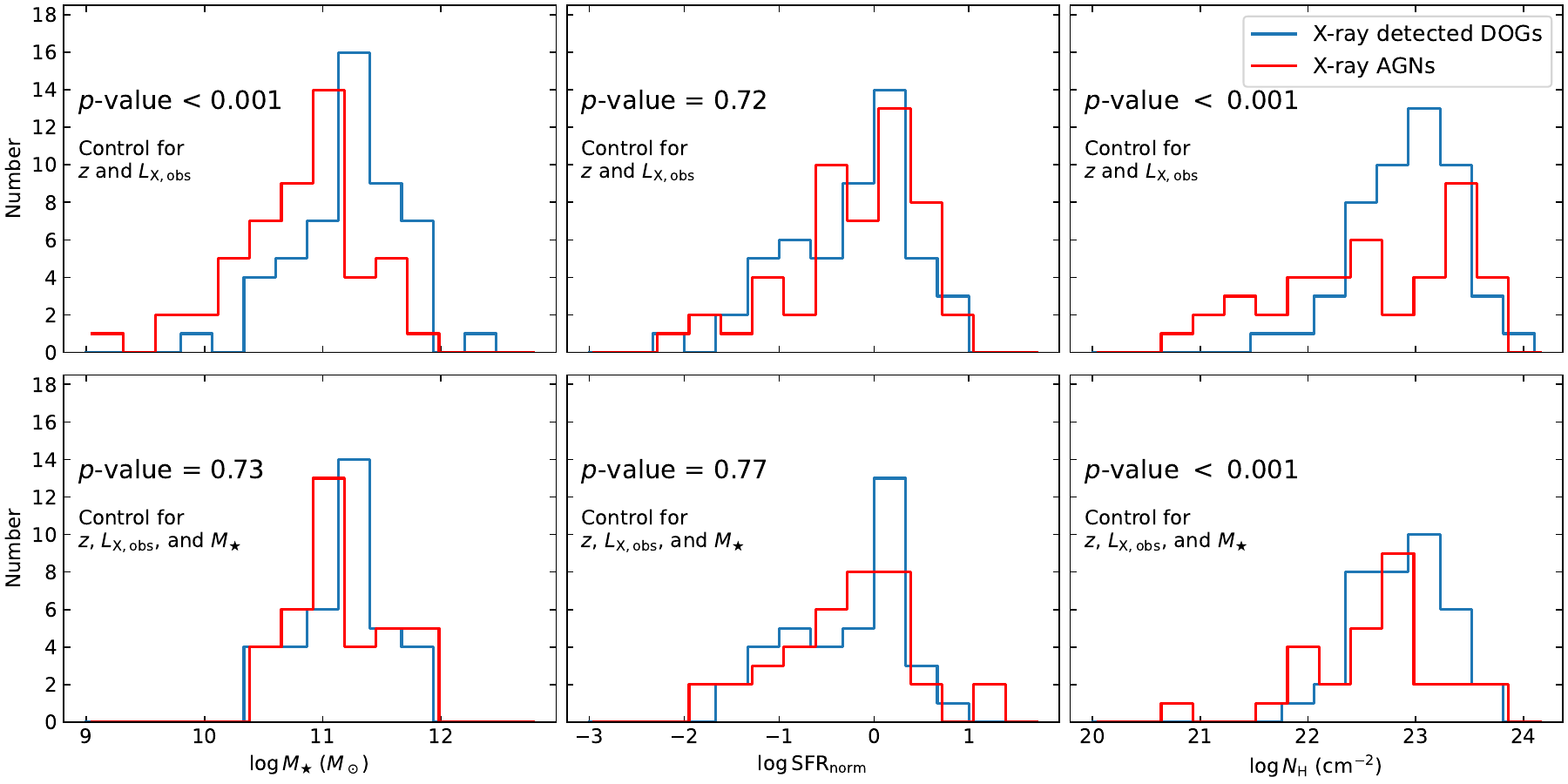}
    \caption{Example distributions of \mstar\ (left column), \sfrnorm\ (middle column), and \nh\ (right column) for our \mbox{X-ray} detected DOGs in the core sample (blue) compared with those for non-DOG \mbox{X-ray} AGNs (red). We control for $z$ and \lxobs\ in the top row, and we control for $z$, \lxobs, and \mstar\ in the bottom row. When we only control for $z$ and \lxobs, \mstar\ and \nh\ both show distinct differences between \mbox{X-ray} detected DOGs and \mbox{X-ray} AGNs, while the difference in \sfrnorm\ is not significant. After we further control for \mstar, the difference in \sfrnorm\ remains insignificant, and the difference in \nh\ remains significant. Note that further controlling for \mstar\ makes building consistent comparison samples more difficult, which results in fewer sources in the bottom panels.}\label{fig:compare_mstar_sfr_nh}
\end{figure*}

\begin{deluxetable*}{ccccccccc}
\tablewidth{\linewidth} 
\tablecaption{Differences between \mbox{X-ray} detected DOGs and non-DOG \mbox{X-ray} AGNs after 1000 trials.}\label{tab:table}
\tablehead{
Subsample & \shortstack{Controlled\\ parameters}  & \shortstack{$p$-values\\ for \mstar} & \shortstack{$p$-values\\ for \sfrnorm} & \shortstack{$p$-values\\ for \nh} & $\Delta\overline{\log\mstar}$ & $\Delta\overline{\log\sfrnorm}$ & $\Delta\overline{\log\nh}$ & \shortstack{Number of sources\\ compared}\\
\colhead{(1)} & \colhead{(2)} & \colhead{(3)} & \colhead{(4)} & \colhead{(5)} & \colhead{(6)} & \colhead{(7)} & \colhead{(8)} & \colhead{(9)}
}
\startdata 
\multirow{2}{*}{Full} & $z$ and $\lxobs$  & $<0.001$ & $0.8_{-0.3}^{+0.1}$ & $<0.001$ & $0.38_{-0.07}^{+0.04}$ & $0.05_{-0.07}^{+0.06}$ & $0.40_{-0.04}^{+0.09}$ &  87\\
 & $z$, $\lxobs$, and \mstar  & $>0.97$ & $0.5_{-0.3}^{+0.3}$ & $<0.001$ & $-0.01_{-0.02}^{+0.03}$ & $0.18_{-0.08}^{+0.09}$ & $0.61_{-0.23}^{+0.15}$ & 70\\
\multirow{2}{*}{Core} & $z$ and $\lxobs$ & $<0.001$ & $0.7_{-0.3}^{+0.2}$ & $<0.001$ & $0.35_{-0.06}^{+0.07}$ & $-0.01_{-0.07}^{+0.10}$ & $0.52_{-0.20}^{+0.20}$ & 51 \\
 & $z$, $\lxobs$, and \mstar & $>0.94$ & $0.4_{-0.2}^{+0.3}$ & $<0.001$ & $0.06_{-0.04}^{+0.05}$ & $0.20_{-0.09}^{+0.10}$ & $0.44_{-0.12}^{+0.20}$ & 39\\
\enddata
\tablecomments{Column (1): our full sample and core sample. Column (2): parameters controlled when making the comparison \mbox{X-ray} AGN sample. Columns (3)--(5): median $p$-values and their associated 68\% confidence intervals when performing AD tests on \mstar, \sfrnorm, and \nh\ between \mbox{X-ray} detected DOGs and non-DOG \mbox{X-ray} AGNs. If the median $p$-values are less than 0.001 (greater than 0.97), we only show ``$<0.001$" (``$>0.97$"). Columns (6)--(7): differences in the median values of $\log\mstar$, $\log\sfrnorm$, and $\log\nh$. We show the median differences and the associated 68\% confidence intervals. Column (8): Number of sources in each comparison sample.}
\end{deluxetable*}

Under the merger-driven galaxy-SMBH coevolution framework, DOGs also represent the peak phase of SF before the fast SMBH growth. In this subsection, we discuss how \mbox{X-ray} detected DOGs differ from \mbox{X-ray} AGNs not selected as DOGs. 
We focus on their host-galaxy properties and \mbox{X-ray} obscuration. All our \mbox{X-ray} detected DOGs are identified as AGN candidates using the SED-based method of \citet{Zou+2022}, and they are all identified as \mbox{X-ray} AGNs in \citet{Chen+2018} and \citet{Ni+2021}.  Together with the high \lxobs\ shown in Figure~\ref{fig:lxdist}, the results indicate good classification purity.

Since host-galaxy properties and \mbox{X-ray} obscuration correlate with $z$ and $\lx$, we need to control for these factors. For instance, high-luminosity AGNs generally have enhanced SFRs compared to star-forming galaxies at similar $z$ and \mstar\ \citep[e.g.,][]{Mountrichas+2024}. Also, \mbox{X-ray} obscuration in AGNs shows significant cosmic evolution, increasing strongly from $z\approx0$ to $z\approx2$ \citep[e.g.,][]{Buchner+2015,Georgakakis+2017,Liu+2017}; at a given redshift, AGNs tend to have less \mbox{X-ray} obscuration at higher luminosities \citep[e.g.,][]{Merloni+2014,Liu+2017}. 
We consider \mbox{X-ray} detected sources within $z=0-4$ and $\log\lxobs=43-46$, where 87 sources in the full sample and 51 sources in the core sample are included. We divide the $z-\log\lxobs$ plane into a grid with $\Delta z=0.2$ and $\Delta\log\lxobs=0.4\,\rm dex$. In each cell, we denote the number of \mbox{X-ray} detected DOGs in field $S$ ($S=1,2,\text{ or }3$, denoting one of the three fields covered by XMM-SERVS) as $N_{S,1}$, and the number of non-DOG \mbox{X-ray} AGNs as $N_{S,2}$. We randomly select $\min\{N_{S,1}, N_{S,2}\}$ \mbox{X-ray} detected DOGs and the same number of non-DOG \mbox{X-ray} AGNs in field $S$. We then combine all the selected sources across the three fields. The above steps conserve the number of sources from different fields in our comparison sample to mitigate any possible effects across different fields. After repeating the above steps in each cell and each field, we can construct new \mbox{X-ray} detected DOG and non-DOG \mbox{X-ray} AGN samples with similar distributions of $z$ and $\lxobs$. Since there are many more general \mbox{X-ray} AGNs than \mbox{X-ray} detected DOGs, in most grids the number of selected objects is controlled by the number of \mbox{X-ray} detected DOGs. 
For a typical example of those randomly chosen sources based upon our core sample (51 \mbox{X-ray} detected DOGs and 51 non-DOG \mbox{X-ray} AGNs), we use a two-sample AD test to examine the consistency of $z$ and $\lxobs$, and the $p$-values of $z$ and $\lxobs$ are 0.97 and 0.78, respectively, indicating that these two parameters have been controlled acceptably.

 In the top row of Figure~\ref{fig:compare_mstar_sfr_nh}, we compare the distributions of \mstar, \sfrnorm, and \nh\ for a typical example of randomly chosen \mbox{X-ray} DOGs and \mbox{X-ray} AGNs using the above procedure. We do not exclude ``unsafe" sources (i.e., whose MS cannot be reliably obtained) from the comparison of \sfrnorm\ because the \sfrnorm\ distribution is not significantly affected by them as shown in Section~\ref{subsec:host-galaxy}. We find that \mbox{X-ray} detected DOGs generally have higher \mstar\ and \nh\ than typical \mbox{X-ray} AGNs, while the SFR does not have significant differences. We further run AD tests on these distributions, finding that the difference is statistically significant for \mstar\ and \nh\ with both $p$-values $<0.001$. As for \sfrnorm, the difference is insignificant with a $p$-value of 0.72. The result that \mbox{X-ray} detected DOGs appear to have higher obscuration levels than typical \mbox{X-ray} AGNs could be at least partly caused by their higher \mstar. There have been findings of a positive correlation between \mstar\ and \mbox{X-ray} obscuration level \citep[e.g.,][]{Lanzuisi+2017}. It is also possible that galaxy-scale gas and dust can contribute to the obscuration of AGNs \citep{Buchner&Bauer+2017,Gilli+2022}, and the \nh\ of galaxy-scale gas also follows a positive correlation with \mstar\ \citep[e.g.,][]{Buchner+2017}. To eliminate the impact of \mstar\ on \nh, we further control for \mstar\, in addition to $z$ and \lxobs\ and test if the difference in \nh\ remains. We show the new results for \mstar, \sfrnorm, and \nh\ in the bottom row of Figure~\ref{fig:compare_mstar_sfr_nh}. As expected, the difference in \mstar\ is no longer significant with $p$-$\mathrm{value} = 0.73$. However, \nh\ is still significantly different between \mbox{X-ray} detected DOGs and \mbox{X-ray} AGNs with $p$-$\mathrm{value} < 0.001$. The difference in \sfrnorm\ remains insignificant with \mbox{$p$-$\mathrm{value}=0.77$}.

We further use a Monte Carlo method to check the robustness of our statistical results. We repeat the above test 1000 times for both the full and core samples, and each time the randomly chosen samples are different. We also calculate the difference in the median values of $\log\mstar$ ($\Delta\overline{\log\mstar}$), $\log\sfrnorm$ ($\Delta\overline{\log\sfrnorm}$), and $\log\nh$ ($\Delta\overline{\log\nh}$) between our \mbox{X-ray} detected DOGs and matched \mbox{X-ray} AGNs. We show the median \mbox{$p$-values} and the median differences and their associated 68\% confidence intervals in Table~\ref{tab:table}. 
Overall, the results show that \mbox{X-ray} detected DOGs generally have higher \mstar\ and higher \nh\ than typical \mbox{X-ray} AGNs across different samples when $z$ and $\lxobs$ are controlled. Also, \mbox{X-ray} detected DOGs do not appear to be more actively star-forming than typical \mbox{X-ray} AGNs. 
After we further control for \mstar, the difference in \nh\ is still significant; \mbox{X-ray} detected DOGs appear to have slightly higher \sfrnorm, but the difference is still insignificant given the large $p$-values. Note that we do not apply any selection correction to obtain the intrinsic \nh\ distribution; rather, we present the observed distributions and compare two sub-populations within the same parent sample (i.e., \mbox{X-ray} sources in XMM-SERVS). 

Apart from \mstar, \sfrnorm, and \nh, we further examine the IR flux densities of DOGs. We find that the reliable SED AGN fraction among our core sample increases significantly with $f_\mathrm{24\mum}$, which is consistent with previous works finding that the fraction of PL sources among DOGs increases with IR flux density \cite[e.g.,][]{Melbourne+2012,Toba+2015}. The results are similar for our full sample. We then compare $f_\mathrm{24\mum}$ between matched \mbox{X-ray} detected DOGs and non-DOG \mbox{X-ray} AGNs. These samples are constructed following the same procedure outlined in this Section. We find that $f_\mathrm{24\mum}$ for our \mbox{X-ray} detected DOGs is significantly higher than for matched \mbox{X-ray} AGNs. This is expected, as our DOGs are constructed to have $f_\mathrm{24\mum}>0.3\,\mJy$, while many matched typical X-ray AGNs have $f_\mathrm{24\mum}$ below this threshold.

The similar \sfrnorm\ and higher \mstar\ for \mbox{X-ray} detected DOGs compared with typical \mbox{X-ray} AGNs cast doubt on the relevance of the merger-driven coevolution framework for DOGs, which postulates that DOGs should be on the peak phase of SF \citep[e.g.,][]{Hopkins+2006,Hopkins+2008,Narayanan+2010,Yutani+2022}. 
We argue that a more natural interpretation of our results is that \mbox{X-ray} detected DOGs can be regarded as analogs to extreme type~2 AGNs. Type~2 AGNs show strong dust emission and are heavily obscured in the optical bands, which contributes to the selection of \mbox{X-ray} detected DOGs. Traditionally, AGNs can be classified into type~1 and type~2 objects, where type~1 AGNs show broad optical emission lines (specifically, the Balmer lines), while type~2 AGNs only have narrow lines. Both DOGs and type~2 AGNs are obscured in the optical band, although broad-line DOGs \citep[e.g.,][]{Toba+2017,Zou+2020} and blue-excess IR-bright DOGs \citep[BluDOGs; e.g.,][]{Noboriguchi+2019, Noboriguchi+2022} are also observed. These may be attributed to strong starburst or leaked UV emission from the central AGN. These sources may be explained by leaked emission via reflection or from partially covered Broad-Line Regions (BLRs), especially in cases where the AGN component is strong \citep[e.g.,][]{Assef+2016}. Recent studies have shown that type~2 AGNs tend to have higher \mstar\ than type~1 AGNs, but both types have similar SFR distributions \citep[e.g.,][]{Zou+2019,Mountrichas+2021}. The higher \mstar\ for type~2 AGNs can be explained if galaxy-wide gas and dust contribute to the obscuration of AGNs, and thus type~2 AGNs tend to reside in more massive galaxies because more massive galaxies have more dust \citep[e.g.,][]{Whitaker+2017}. These results are consistent with our findings if \mbox{X-ray} detected DOGs are considered extreme type~2 AGNs. In addition, observations have shown that PL DOGs tend to have slightly higher \mstar\ than Bump DOGs \citep[e.g.,][]{Bussmann+2012}. As we confirmed in Section~\ref{subsec:sed-fitting}, PL DOGs are more AGN-dominated, and their higher \mstar\ can be explained if PL DOGs are more similar to type~2 AGNs than Bump DOGs due to their higher AGN purity. 
Since DOGs are more obscured in the optical bands due to galaxy-wide dust, the higher \mstar\ values of our \mbox{X-ray} detected DOGs also indicate that the galaxy-wide dust is indeed connected to \mstar. 
However, the consistently higher \nh\ for \mbox{X-ray} detected DOGs even when we control for \mstar\ also shows that their AGN obscuration is not solely determined by the galaxy-wide obscuration and should be primarily contributed by the higher nuclear obscuration among \mbox{X-ray} detected DOGs.

Our results are in agreement with those in \citet{Li+2020} who found that secular processes, instead of mergers, are most probable to trigger \mbox{X-ray}-selected, heavily obscured AGNs that show less-extreme optical-IR colors. However, we cannot simply rule out the relevance of mergers. For our \mbox{X-ray} detected DOGs, the similar \sfrnorm\ compared with matched typical \mbox{X-ray} AGNs could be explained in the merger-driven SMBH-galaxy coevolution framework if \mbox{X-ray} detected DOGs are in a slightly later evolutionary phase than \mbox{X-ray} undetected ones, in which the star-formation has been reduced a bit due to feedback. Observational constraints on the host morphology and/or stellar and gas dynamics can provide more direct evidence for/against the relevance of mergers for DOGs. The relevance may also differ among DOGs with different extreme levels, and further division of DOGs into different subsets may help us understand such differences. For instance, Hot DOGs show larger \nh\ than those derived for type~1 quasars with similar luminosities \citep{Vito+2018}, consistent with a post-merger stage;
high-\lambedd\ DOGs are found to be in similar evolutionary stages as Hot DOGs, and both SMBH accretion and host-galaxy SF are reaching the highest level \citep{Zou+2020}.

\section{Summary and Future Work}\label{sec:summary}

In this work, we select 3738 DOGs in XMM-SERVS as a full sample using the same selection criteria as \citet{Dey+2008}, among which 174 are detected in \mbox{X-rays}. To ensure reliable redshifts, we further select 1309 DOGs with \specz s or reliable \photoz s as a core sample, among which 88 are detected in \mbox{X-rays}. The large survey volume and deep multiwavelength coverage of XMM-SERVS provide high-quality characterization of DOGs and make our sample size substantially exceed those of previous comparable studies. We analyze DOG properties based upon SED-fitting, \mbox{X-ray}, and radio observations. Critically, we assess if DOGs represent the key evolutionary phase in the merger-driven galaxy-SMBH coevolution framework. The main results are the following:

\begin{enumerate}
    \item Our core sources are at $z=1.63-1.93$ ($25-75$\% quantiles) with a median $\lbol=10^{12.4}\,\lsun$. The median \lbol\ is much higher than that for typical \mbox{X-ray} AGNs but is much lower than for the most-extreme Hot DOGs. There are $\approx10\%$ of DOGs identified as reliable SED AGNs, and we confirm that the phenomenologically defined PL DOGs indeed preferentially host AGNs. The \mbox{X-ray} detected DOGs are generally luminous (median $\lxobs=10^{44.3}\,\lumcgs$). The results for our full sample are similar. See Sections~\ref{subsec:sed-fitting} and \ref{subsec:sample_dist}.
    \item Our DOGs are massive with median $\log\mstar\approx11$. \mbox{X-ray} detected DOGs have slightly higher \mstar\ than \mbox{X-ray} undetected DOGs, which can be explained by the connection between SMBH accretion and \mstar. As for \sfrnorm, \mbox{X-ray} undetected DOGs generally lie on or above the star-formation MS relation calibrated in XMM-SERVS, reaching up to starburst galaxies. In contrast, a significant fraction of \mbox{X-ray} detected DOGs lie below the MS. See Section~\ref{subsec:host-galaxy}. 
    \item We perform \mbox{X-ray} stacking analyses of \mbox{X-ray} undetected DOGs and find significant average detections in all three bands for both the full and core samples. The median average \lxobs\ is \mbox{$(4.3\pm1.7)\times10^{42}\,\lumcgs$} for our core sample. Stacking several subsets also reveals that the total accretion power of \mbox{X-ray} undetected DOGs is primarily contributed by AGN candidates identified via their SEDs. See Section~\ref{subsec:X-ray_stacking}.
    \item We find about half of our \mbox{X-ray} detected DOGs are identified as heavily obscured \mbox{($\nh>10^{23}\,\nhcgs$)} AGNs based upon their HRs. \mbox{X-ray} stacking also indicates that \mbox{X-ray} undetected DOGs identified as AGN candidates have \mbox{$\nh\gtrsim10^{23}\,\nhcgs$} on average. The results on $\lxobs$ versus $\lsix$ confirm their heavily obscured nature. Overall, \mbox{X-ray} undetected DOGs identified as reliable SED AGNs are more obscured in \mbox{X-rays} than \mbox{X-ray} detected ones, and some likely reach CT levels. See Sections~\ref{subsec:HR} and \ref{subsec:lx-l6}.
    \item Most radio-detected DOGs show a strong correlation between $L_\mathrm{1.4\,\rm GHz}$ and $L_\mathrm{IR}$, and DOGs selected as radio AGNs show much elevated $L_\mathrm{1.4\,\rm GHz}$. We find that radio selection can select AGNs among DOGs that can hardly be selected via other methods, consistent with the results on general radio AGN selection in, e.g., \citet{Zhu+2023}. See Section~\ref{subsec:radio}.
    \item Combining the individual fluxes of \mbox{X-ray} detected DOGs and the stacked \mbox{X-ray} images for \mbox{X-ray} undetected DOGs, we estimate the AGN fractions to be \mbox{1.5--15\%} and 2.0--20\% for our DOG full sample and core sample, respectively, considering a constant $\lambda_0=1-0.1$. The values are consistent with the AGN fraction of 10--20\% for typical galaxy populations at $z\approx2$ with $\log\mstar\approx11$, indicating that DOGs do not present significantly different SMBH accretion compared with AGNs in typical galaxy populations. See Section~\ref{subsec:agn_fraction}.
    \item We control for $z$ and \lxobs\ and find that \mbox{X-ray} detected DOGs have higher \mstar\ and \nh\ than non-DOG \mbox{X-ray} AGNs in XMM-SERVS, while their \sfrnorm\ distributions are similar even after we further control for \mstar. 
    The results challenge the relevance of the merger-driven galaxy-SMBH coevolution framework for \mbox{X-ray} detected DOGs and suggest that they may be analogs to extreme type~2 AGNs that show similar behaviors. See Section~\ref{subsec:compare}.
\end{enumerate}

This work presents the largest sample of DOGs in sensitive multiwavelength surveys. Critical advances for this sample will be made once future deep spectroscopic and photometric data from, e.g., MOONS, PFS, 4MOST WAVES, LSST, Euclid, and LMT TolTEC are gathered in XMM-SERVS. In particular, deep spectroscopic surveys can not only provide a larger sample with better completeness of reliable redshifts, but also provide rich diagnostic information from spectroscopy, which can help better characterize the sample and even break it down into physically relevant subsets \citep[e.g., \mbox{high-\lambedd} DOGs;][]{Zou+2020}. LMT TolTEC will provide sensitive submillimeter data, which will help better characterize the star-formation and dust properties of these DOGs. Spatially resolved imaging and/or spectroscopy from JWST for a representative subset of DOGs can probe the host morphology and dynamics, providing a direct test of the relevance of mergers for DOGs.

\begin{acknowledgments}
We thank the anonymous referee for constructive comments and suggestions. We thank Joel Leja and Michael Eracleous for constructive comments. ZY, FZ, and WNB acknowledge financial support from NSF grants AST-2106990 and AST-2407089, CXC grant AR4-25008X, and the Penn State Eberly Endowment. ZZ acknowledges financial support from Wuhan University. FEB acknowledges support from ANID-Chile BASAL CATA FB210003, FONDECYT Regular 1241005, and Millennium Science Initiative Program ICN12\_009. BL acknowledges financial support from the National Natural Science
Foundation of China grant 11991053. YQX acknowledges financial support from NSFC~12025303.

\end{acknowledgments}

\appendix

\section{The Reliability of \mstar\ and SFR Measurements \label{appendix:a}}

\begin{figure}[htbp!]
    \centering
    \includegraphics[width=\linewidth]{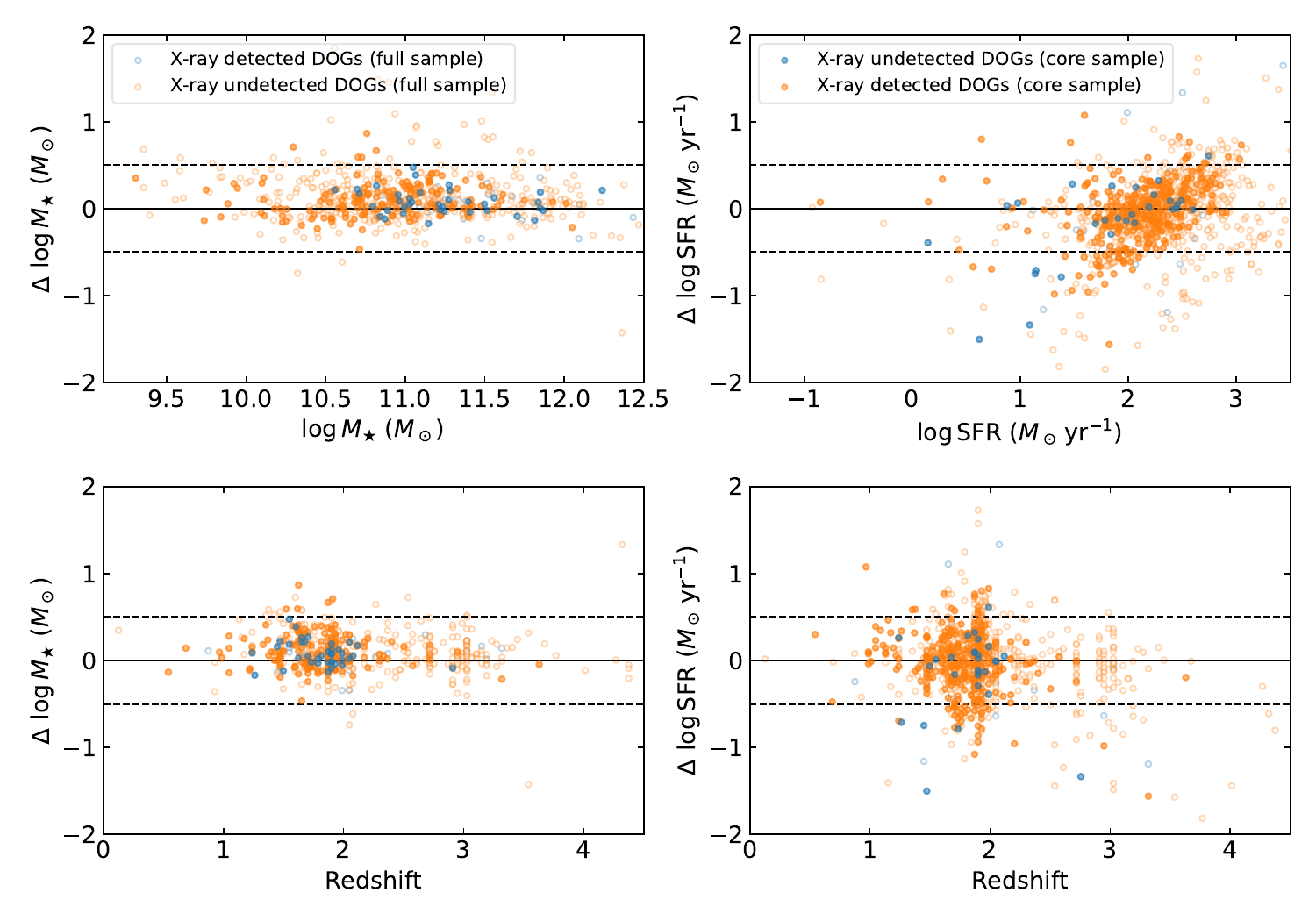}
    \caption{Top-left panel: Comparison between $\mstar^\mathrm{gal}$ and the adopted AGN-template-based \mstar\ values ($\Delta\log\mstar=\log\mstar^\mathrm{gal} - \log\mstar$) for sources with preferred AGN models in W-CDF-S. Bottom-left panel: $\Delta\log\mstar$ versus redshift. Top-right panel: Comparison between SFR measurements with and without FIR data in W-CDF-S. Bottom-right panel: $\Delta\log\mathrm{SFR}$ versus redshifts. The full and core samples are shown with empty and filled circles, respectively. \mbox{X-ray} detected and \mbox{X-ray} undetected DOGs are shown in blue and orange, respectively. The differences are generally within 0.5~dex and do not evolve with redshift. The median differences and $\sigmaNMAD$ are small, indicating that our \mstar\ and SFR measurements do not have significant biases.}\label{fig:mstar_sfr_diff}
\end{figure}

In Section~\ref{subsec:host-galaxy}, we measured \mstar\ and SFR of our DOGs via SED-fitting. In this Appendix, we further check the reliability of our measurements. For sources hosting strong AGN components, the rest-frame NIR SED may be dominated by the AGN component, making \mstar\ usually less reliable. We denote $\mstar^\mathrm{gal}$ as the \mstar\ fitted using normal-galaxy templates (i.e., without the AGN component) in \citet{Zou+2022}. Since the contribution from the AGN component is then assigned to the galaxy component, $\mstar^\mathrm{gal}$ can be considered a soft upper limit for the true \mstar. We use our samples in W-CDF-S as an illustration. We plot the difference in $\log\mstar$ ($\Delta\log\mstar$) in the top-left panel of Figure~\ref{fig:mstar_sfr_diff}. The median $\Delta\log\mstar=0.08$ and $\sigmaNMAD=0.17$ for our core sample. The results are similar for the other two fields, and indicate that the AGN component does not impact our measurements of \mstar\ significantly. We also plot $\Delta\log\mstar$ versus redshift in the bottom-left panel. The results show that $\Delta\log\mstar$ is not redshift dependent.

The inclusion of FIR photometry helps accurately measure SFR. About 55\% of our sources have $250\,\mum$ Herschel SPIRE photometry with $\mathrm{SNR}>5$, which is much higher than the typical fraction in general galaxy samples in XMM-SERVS, and we expect our SFR measurements to be reliable for these sources. However, we still test if the measurements for the rest of the sources are reliable. We exclude the FIR photometry for sources in W-CDF-S and measure their SFRs using the same method as in \citet{Zou+2022}, and we show the difference in SFR ($\Delta\log\mathrm{SFR}$) in the top-right panel of Figure~\ref{fig:mstar_sfr_diff}. Visually, there is a weak trend such that $\Delta\log\mathrm{SFR}$ slightly increases toward higher SFR. However, the small median $\Delta\log\mathrm{SFR}$ of 0.02 and $\sigmaNMAD$ of 0.26 for our core sample are similar to or even better than the typical comparison results for XMM-SERVS AGNs in general \citep[Table~7 and Figure~29 of][]{Zou+2022}. The results are similar for the other two fields. These indicate that the SFR measurements are not significantly biased for sources without high-quality FIR photometry. We also show $\Delta\log\mathrm{SFR}$ versus redshift in the bottom-right panel. The results show that $\Delta\log\mathrm{SFR}$ does not evolve with redshift.

\section{Selection of lower-luminosity Hot DOG Candidates \label{appendix:b}}

In Section~\ref{sec:sample}, we presented the selection of DOGs and showed that, as expected, they generally have lower \lbol\ than the most extreme Hot DOGs with $\lbol>10^{14}\,\lsun$. In this Appendix, we further describe how we select lower-luminosity analogs to those extreme Hot DOGs using the best-fit SED model and selection criteria similar to those in \citet{Eisenhardt+2012}. The adopted selection criteria in \citet{Eisenhardt+2012} are $\mathrm{W1}>17.4$, and either (1) $\mathrm{W4}<7.7$ and $\mathrm{W2}-\mathrm{W4}>8.2$ or (2) $\mathrm{W3}<10.6$ and $\mathrm{W2-W3}>5.3$. Such criteria select luminous \citep[$\lbol>10^{13}\,\lsun$, some even exceed $10^{14}\,\lsun$;][]{Tsai+2015,Li+2024} and rare ($\approx1$ per 40 $\deg^2$) Hot DOGs that are characterized by hot dust temperatures and extreme MIR colors.

Our SEDs fits do not directly provide dust temperature, and thus we select lower-luminosity Hot DOG candidates among our DOG full sample using slightly modified color criteria compared to \citet{Eisenhardt+2012}: (1) $\mathrm{W1}-\mathrm{W4}>9.7$ and $\mathrm{W2}-\mathrm{W4}>8.2$ or (2) $\mathrm{W1}-\mathrm{W3}>6.8$ and $\mathrm{W2-W3}>5.3$, i.e., we lift the magnitude cuts on W1 and W3/W4 in \citet{Eisenhardt+2012} and convert them to color cuts. This modification allows the selection of sources with lower luminosity but with similarly extreme MIR colors. The synthetic WISE photometry in \mbox{W1--W4} is calculated from the best-fit SEDs, where we calculate the expected WISE flux in \mbox{W1--W4} by convolving the best-fit SED with the WISE filter response. The flux is then converted to WISE magnitude using the zero-point values in \citet{Wright+2010}. We have checked that our SED-based WISE magnitude is generally consistent with the AllWISE catalog \citep{Cutri+2021} for sources that are detected in the corresponding WISE band.

\begin{figure}[htbp!]
    \centering
    \includegraphics[width=0.7\linewidth]{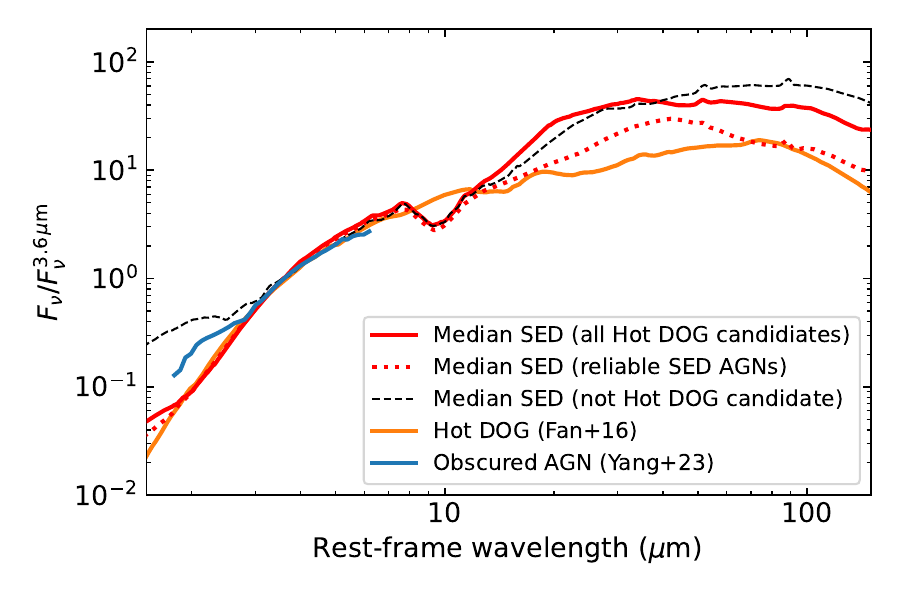}
    \caption{Comparison between different types of median SEDs (normalized at rest-frame 3.6\,\mum). The red solid and dotted curves represent all our lower-luminosity Hot DOG candidates, and those selected as reliable SED AGNs, respectively. The black dashed curve represents DOGs that are not selected as lower-luminosity Hot DOG candidates. The orange and blue solid curves represent Hot DOGs \citep{Fan+2016} and obscured AGNs \citep{Yang+2023}.}\label{fig:hotdog}
\end{figure}

We end up with 62 sources out of 3738 DOGs in the full sample selected as lower-luminosity Hot DOG candidates using our modified selection criteria. These candidates are marked in Table~\ref{tab:data_table}. Among these candidates, 26 are identified as reliable SED AGNs. The median \lbol\ of the 62 candidates and the 26 reliable SED AGNs are $10^{12.5}\,\lsun$ and $10^{12.8}\,\lsun$, both of which are much lower than the typical \lbol\ of $\approx10^{13}-10^{14}\,\lsun$ for Hot DOGs selected via ``W1W2-dropout". The rest-frame median SEDs for all our Hot DOG candidates and those selected as reliable SED AGNs are shown in Figure~\ref{fig:hotdog}. For comparison, we also plot the median SED for DOGs that are AGN candidates but are not selected as lower-luminosity Hot DOG candidates, the median SED for Hot DOGs in \citet{Fan+2016}, and the median SED for high-redshift obscured AGNs in \citet{Yang+2023}, all of which are normalized at rest-frame 3.6\,\mum. The median SED for our lower-luminosity Hot DOG candidates is similar to that for Hot DOGs in \citet{Fan+2016} in the MIR at rest-frame $\lesssim15\mum$. This wavelength range fully covers the WISE bands assuming the median $z\approx1.8$ of our DOGs. Our Hot DOG candidates show much redder MIR colors than those that are not selected, indicating their relatively stronger hot-dust emission. At longer wavelengths that the WISE bands do not cover, our Hot DOG candidates generally show similar FIR SEDs compared with DOGs that are not selected as Hot DOG candidates. However, Hot DOG candidates selected as reliable SED AGNs still have similar FIR SEDs compared with Hot DOGs in \citet{Fan+2016}. This is due to the fact that those Hot DOG candidates selected as reliable SED AGNs are more similar to Hot DOGs selected via ``W1W2-dropout", which have higher \lbol\ and a stronger AGN component in the MIR. These results indicate our criteria indeed select candidates for lower-luminosity version of Hot DOGs. We have also tried the selection criteria in Section~2 of \citet{Ricci+2017} for low-redshift Hot DOG candidates. Their criteria are similar to ours except that they have an additional requirement of $\mathrm{W1}<17.4$ to preferentially select bright low-redshift Hot DOGs. We do not find any such low-redshift Hot DOG candidates among our full sample DOGs.

One of our selected Hot DOG candidates is detected in \mbox{X-rays} (XID: WCDFS0530) at $z=1.79$. XMM-Newton observed this source two times, and the total cleaned exposure time is 49.9\,ks. It is only marginally detected in the FB, with 76 net counts in total from all three EPIC cameras \citep{Ni+2021}. We reduce the observations and extract the spectra using the XMM-Newton Science Analysis System (SAS; v21.0). We use \texttt{XSPEC} \citep{Arnaud+1996} v12.12.1 to jointly fit all individual \mbox{X-ray} spectra, in which the \textit{C}-statistic is adopted. We limit our spectral fitting to \mbox{0.5--1.4\,\kev} and \mbox{1.6--7.5\,\kev} due to the high background above $7.8\,\kev$ and the Al K$\alpha$ instrumental background lines at $\approx1.5\,\kev$. We fit the spectra with a simple power-law with Galactic absorption. The best-fit effective power-law photon index is $\Gamma_\mathrm{eff}=0.9_{-0.8}^{+0.9}$, and the best-fit observed-frame 2--10\,\kev\ flux is $5_{-5}^{+10}\times10^{-15}\,\fluxcgs$. Assuming $\Gamma_\mathrm{eff}$ at its best-fit value, its best-fit rest-frame 2--10\,\kev\ \lxobs\ is $4_{-4}^{+8}\times10^{43}\,\lumcgs$. The \lsix\ of this source is $1.3\times10^{46}\,\lumcgs$, which places this source $\approx3\sigma$ below the \citet{Stern+2015} relation in the $\lxobs-\lsix$ plane (see Figure~\ref{fig:Lx_6um}). The hard best-fit $\Gamma_\mathrm{eff}$ and low $\lxobs/\lsix$ both indicate the lower-luminosity Hot DOG candidate WCDFS0530 is heavily obscured, consistent with the \mbox{X-ray} study on Hot DOGs in \citet{Vito+2018}.

We also stack the \mbox{X-ray} images for all our \mbox{X-ray} undetected Hot DOG candidates and those selected as reliable SED AGNs (following Section~\ref{subsec:X-ray_stacking}, stacked sources are away from known \mbox{X-ray} sources), but none of the \mbox{X-ray} bands shows detections at $>2\sigma$ for both samples.


\newpage

\bibliography{sample631}{}

\begin{thebibliography}{}
\expandafter\ifx\csname natexlab\endcsname\relax\def\natexlab#1{#1}\fi
\providecommand{\url}[1]{\href{#1}{#1}}
\providecommand{\dodoi}[1]{doi:~\href{http://doi.org/#1}{\nolinkurl{#1}}}
\providecommand{\doeprint}[1]{\href{http://ascl.net/#1}{\nolinkurl{http://ascl.net/#1}}}
\providecommand{\doarXiv}[1]{\href{https://arxiv.org/abs/#1}{\nolinkurl{https://arxiv.org/abs/#1}}}

\bibitem[{{Abbott} {et~al.}(2021){Abbott}, {Adam{\'o}w}, {Aguena}, {Allam}, {Amon}, {Annis}, {Avila}, {Bacon}, {Banerji}, {Bechtol}, {Becker}, {Bernstein}, {Bertin}, {Bhargava}, {Bridle}, {Brooks}, {Burke}, {Carnero Rosell}, {Carrasco Kind}, {Carretero}, {Castander}, {Cawthon}, {Chang}, {Choi}, {Conselice}, {Costanzi}, {Crocce}, {da Costa}, {Davis}, {De Vicente}, {DeRose}, {Desai}, {Diehl}, {Dietrich}, {Drlica-Wagner}, {Eckert}, {Elvin-Poole}, {Everett}, {Evrard}, {Ferrero}, {Fert{\'e}}, {Flaugher}, {Fosalba}, {Friedel}, {Frieman}, {Garc{\'\i}a-Bellido}, {Gaztanaga}, {Gelman}, {Gerdes}, {Giannantonio}, {Gill}, {Gruen}, {Gruendl}, {Gschwend}, {Gutierrez}, {Hartley}, {Hinton}, {Hollowood}, {Honscheid}, {Huterer}, {James}, {Jeltema}, {Johnson}, {Kent}, {Kron}, {Kuehn}, {Kuropatkin}, {Lahav}, {Li}, {Lidman}, {Lin}, {MacCrann}, {Maia}, {Manning}, {Maloney}, {March}, {Marshall}, {Martini}, {Melchior}, {Menanteau}, {Miquel}, {Morgan}, {Myles}, {Neilsen}, {Ogando}, {Palmese}, {Paz-Chinch{\'o}n}, {Petravick},
  {Pieres}, {Plazas}, {Pond}, {Rodriguez-Monroy}, {Romer}, {Roodman}, {Rykoff}, {Sako}, {Sanchez}, {Santiago}, {Scarpine}, {Serrano}, {Sevilla-Noarbe}, {Smith}, {Smith}, {Soares-Santos}, {Suchyta}, {Swanson}, {Tarle}, {Thomas}, {To}, {Tremblay}, {Troxel}, {Tucker}, {Turner}, {Varga}, {Walker}, {Wechsler}, {Weller}, {Wester}, {Wilkinson}, {Yanny}, {Zhang}, {Nikutta}, {Fitzpatrick}, {Jacques}, {Scott}, {Olsen}, {Huang}, {Herrera}, {Juneau}, {Nidever}, {Weaver}, {Adean}, {Correia}, {de Freitas}, {Freitas}, {Singulani}, {Vila-Verde}, \& {Linea Science Server}}]{Abbott+2021}
{Abbott}, T.~M.~C., {Adam{\'o}w}, M., {Aguena}, M., {et~al.} 2021, \apjs, 255, 20, \dodoi{10.3847/1538-4365/ac00b3}

\bibitem[{{Abdurro'uf} {et~al.}(2022){Abdurro'uf}, {Accetta}, {Aerts}, {Silva Aguirre}, {Ahumada}, {Ajgaonkar}, {Filiz Ak}, {Alam}, {Allende Prieto}, {Almeida}, {Anders}, {Anderson}, {Andrews}, {Anguiano}, {Aquino-Ort{\'\i}z}, {Arag{\'o}n-Salamanca}, {Argudo-Fern{\'a}ndez}, {Ata}, {Aubert}, {Avila-Reese}, {Badenes}, {Barb{\'a}}, {Barger}, {Barrera-Ballesteros}, {Beaton}, {Beers}, {Belfiore}, {Bender}, {Bernardi}, {Bershady}, {Beutler}, {Bidin}, {Bird}, {Bizyaev}, {Blanc}, {Blanton}, {Boardman}, {Bolton}, {Boquien}, {Borissova}, {Bovy}, {Brandt}, {Brown}, {Brownstein}, {Brusa}, {Buchner}, {Bundy}, {Burchett}, {Bureau}, {Burgasser}, {Cabang}, {Campbell}, {Cappellari}, {Carlberg}, {Wanderley}, {Carrera}, {Cash}, {Chen}, {Chen}, {Cherinka}, {Chiappini}, {Choi}, {Chojnowski}, {Chung}, {Clerc}, {Cohen}, {Comerford}, {Comparat}, {da Costa}, {Covey}, {Crane}, {Cruz-Gonzalez}, {Culhane}, {Cunha}, {Dai}, {Damke}, {Darling}, {Davidson}, {Davies}, {Dawson}, {De Lee}, {Diamond-Stanic}, {Cano-D{\'\i}az}, {S{\'a}nchez},
  {Donor}, {Duckworth}, {Dwelly}, {Eisenstein}, {Elsworth}, {Emsellem}, {Eracleous}, {Escoffier}, {Fan}, {Farr}, {Feng}, {Fern{\'a}ndez-Trincado}, {Feuillet}, {Filipp}, {Fillingham}, {Frinchaboy}, {Fromenteau}, {Galbany}, {Garc{\'\i}a}, {Garc{\'\i}a-Hern{\'a}ndez}, {Ge}, {Geisler}, {Gelfand}, {G{\'e}ron}, {Gibson}, {Goddy}, {Godoy-Rivera}, {Grabowski}, {Green}, {Greener}, {Grier}, {Griffith}, {Guo}, {Guy}, {Hadjara}, {Harding}, {Hasselquist}, {Hayes}, {Hearty}, {Hern{\'a}ndez}, {Hill}, {Hogg}, {Holtzman}, {Horta}, {Hsieh}, {Hsu}, {Hsu}, {Huber}, {Huertas-Company}, {Hutchinson}, {Hwang}, {Ibarra-Medel}, {Chitham}, {Ilha}, {Imig}, {Jaekle}, {Jayasinghe}, {Ji}, {Johnson}, {Jones}, {J{\"o}nsson}, {Katkov}, {Khalatyan}, {Kinemuchi}, {Kisku}, {Knapen}, {Kneib}, {Kollmeier}, {Kong}, {Kounkel}, {Kreckel}, {Krishnarao}, {Lacerna}, {Lane}, {Langgin}, {Lavender}, {Law}, {Lazarz}, {Leung}, {Leung}, {Lewis}, {Li}, {Li}, {Lian}, {Liang}, {Lin}, {Lin}, {Lin}, {Lintott}, {Long}, {Longa-Pe{\~n}a}, {L{\'o}pez-Cob{\'a}}, {Lu},
  {Lundgren}, {Luo}, {Mackereth}, {de la Macorra}, {Mahadevan}, {Majewski}, {Manchado}, {Mandeville}, {Maraston}, {Margalef-Bentabol}, {Masseron}, {Masters}, {Mathur}, {McDermid}, {Mckay}, {Merloni}, {Merrifield}, {Meszaros}, {Miglio}, {Di Mille}, {Minniti}, {Minsley}, {Monachesi}, {Moon}, {Mosser}, {Mulchaey}, {Muna}, {Mu{\~n}oz}, {Myers}, {Myers}, {Nadathur}, {Nair}, {Nandra}, {Neumann}, {Newman}, {Nidever}, {Nikakhtar}, {Nitschelm}, {O'Connell}, {Garma-Oehmichen}, {Luan Souza de Oliveira}, {Olney}, {Oravetz}, {Ortigoza-Urdaneta}, {Osorio}, {Otter}, {Pace}, {Padilla}, {Pan}, {Pan}, {Parikh}, {Parker}, {Peirani}, {Pe{\~n}a Ram{\'\i}rez}, {Penny}, {Percival}, {Perez-Fournon}, {Pinsonneault}, {Poidevin}, {Poovelil}, {Price-Whelan}, {B{\'a}rbara de Andrade Queiroz}, {Raddick}, {Ray}, {Rembold}, {Riddle}, {Riffel}, {Riffel}, {Rix}, {Robin}, {Rodr{\'\i}guez-Puebla}, {Roman-Lopes}, {Rom{\'a}n-Z{\'u}{\~n}iga}, {Rose}, {Ross}, {Rossi}, {Rubin}, {Salvato}, {S{\'a}nchez}, {S{\'a}nchez-Gallego}, {Sanderson}, {Santana
  Rojas}, {Sarceno}, {Sarmiento}, {Sayres}, {Sazonova}, {Schaefer}, {Schiavon}, {Schlegel}, {Schneider}, {Schultheis}, {Schwope}, {Serenelli}, {Serna}, {Shao}, {Shapiro}, {Sharma}, {Shen}, {Shetrone}, {Shu}, {Simon}, {Skrutskie}, {Smethurst}, {Smith}, {Sobeck}, {Spoo}, {Sprague}, {Stark}, {Stassun}, {Steinmetz}, {Stello}, {Stone-Martinez}, {Storchi-Bergmann}, {Stringfellow}, {Stutz}, {Su}, {Taghizadeh-Popp}, {Talbot}, {Tayar}, {Telles}, {Teske}, {Thakar}, {Theissen}, {Tkachenko}, {Thomas}, {Tojeiro}, {Hernandez Toledo}, {Troup}, {Trump}, {Trussler}, {Turner}, {Tuttle}, {Unda-Sanzana}, {V{\'a}zquez-Mata}, {Valentini}, {Valenzuela}, {Vargas-Gonz{\'a}lez}, {Vargas-Maga{\~n}a}, {Alfaro}, {Villanova}, {Vincenzo}, {Wake}, {Warfield}, {Washington}, {Weaver}, {Weijmans}, {Weinberg}, {Weiss}, {Westfall}, {Wild}, {Wilde}, {Wilson}, {Wilson}, {Wilson}, {Wolf}, {Wood-Vasey}, {Yan}, {Zamora}, {Zasowski}, {Zhang}, {Zhao}, {Zheng}, {Zheng}, \& {Zhu}}]{Abdurro'uf+2022}
{Abdurro'uf}, {Accetta}, K., {Aerts}, C., {et~al.} 2022, \apjs, 259, 35, \dodoi{10.3847/1538-4365/ac4414}

\bibitem[{{Aihara} {et~al.}(2018){Aihara}, {Armstrong}, {Bickerton}, {Bosch}, {Coupon}, {Furusawa}, {Hayashi}, {Ikeda}, {Kamata}, {Karoji}, {Kawanomoto}, {Koike}, {Komiyama}, {Lang}, {Lupton}, {Mineo}, {Miyatake}, {Miyazaki}, {Morokuma}, {Obuchi}, {Oishi}, {Okura}, {Price}, {Takata}, {Tanaka}, {Tanaka}, {Tanaka}, {Uchida}, {Uraguchi}, {Utsumi}, {Wang}, {Yamada}, {Yamanoi}, {Yasuda}, {Arimoto}, {Chiba}, {Finet}, {Fujimori}, {Fujimoto}, {Furusawa}, {Goto}, {Goulding}, {Gunn}, {Harikane}, {Hattori}, {Hayashi}, {He{\l}miniak}, {Higuchi}, {Hikage}, {Ho}, {Hsieh}, {Huang}, {Huang}, {Imanishi}, {Iwata}, {Jaelani}, {Jian}, {Kashikawa}, {Katayama}, {Kojima}, {Konno}, {Koshida}, {Kusakabe}, {Leauthaud}, {Lee}, {Lin}, {Lin}, {Mandelbaum}, {Matsuoka}, {Medezinski}, {Miyama}, {Momose}, {More}, {More}, {Mukae}, {Murata}, {Murayama}, {Nagao}, {Nakata}, {Niida}, {Niikura}, {Nishizawa}, {Oguri}, {Okabe}, {Ono}, {Onodera}, {Onoue}, {Ouchi}, {Pyo}, {Shibuya}, {Shimasaku}, {Simet}, {Speagle}, {Spergel}, {Strauss}, {Sugahara},
  {Sugiyama}, {Suto}, {Suzuki}, {Tait}, {Takada}, {Terai}, {Toba}, {Turner}, {Uchiyama}, {Umetsu}, {Urata}, {Usuda}, {Yeh}, \& {Yuma}}]{Aihara+2018}
{Aihara}, H., {Armstrong}, R., {Bickerton}, S., {et~al.} 2018, \pasj, 70, S8, \dodoi{10.1093/pasj/psx081}

\bibitem[{{Aird} {et~al.}(2018){Aird}, {Coil}, \& {Georgakakis}}]{Aird+2018}
{Aird}, J., {Coil}, A.~L., \& {Georgakakis}, A. 2018, \mnras, 474, 1225, \dodoi{10.1093/mnras/stx2700}

\bibitem[{{Alexander} {et~al.}(2005){Alexander}, {Bauer}, {Chapman}, {Smail}, {Blain}, {Brandt}, \& {Ivison}}]{Alexander+2005}
{Alexander}, D.~M., {Bauer}, F.~E., {Chapman}, S.~C., {et~al.} 2005, \apj, 632, 736, \dodoi{10.1086/444342}

\bibitem[{{Appleton} {et~al.}(2004){Appleton}, {Fadda}, {Marleau}, {Frayer}, {Helou}, {Condon}, {Choi}, {Yan}, {Lacy}, {Wilson}, {Armus}, {Chapman}, {Fang}, {Heinrichson}, {Im}, {Jannuzi}, {Storrie-Lombardi}, {Shupe}, {Soifer}, {Squires}, \& {Teplitz}}]{Appleton+2004}
{Appleton}, P.~N., {Fadda}, D.~T., {Marleau}, F.~R., {et~al.} 2004, \apjs, 154, 147, \dodoi{10.1086/422425}

\bibitem[{{Arnaud}(1996)}]{Arnaud+1996}
{Arnaud}, K.~A. 1996, in Astronomical Society of the Pacific Conference Series, Vol. 101, Astronomical Data Analysis Software and Systems V, ed. G.~H. {Jacoby} \& J.~{Barnes}, 17

\bibitem[{{Assef} {et~al.}(2016){Assef}, {Walton}, {Brightman}, {Stern}, {Alexander}, {Bauer}, {Blain}, {Diaz-Santos}, {Eisenhardt}, {Finkelstein}, {Hickox}, {Tsai}, \& {Wu}}]{Assef+2016}
{Assef}, R.~J., {Walton}, D.~J., {Brightman}, M., {et~al.} 2016, \apj, 819, 111, \dodoi{10.3847/0004-637X/819/2/111}

\bibitem[{{Berta} {et~al.}(2006){Berta}, {Rubele}, {Franceschini}, {Held}, {Rizzi}, {Lonsdale}, {Jarrett}, {Rodighiero}, {Oliver}, {Dias}, {Buttery}, {Fiore}, {La Franca}, {Puccetti}, {Fang}, {Shupe}, {Surace}, \& {Gruppioni}}]{Berta+2006}
{Berta}, S., {Rubele}, S., {Franceschini}, A., {et~al.} 2006, \aap, 451, 881, \dodoi{10.1051/0004-6361:20054548}

\bibitem[{{Best} \& {Heckman}(2012)}]{Best&Heckman+2012}
{Best}, P.~N., \& {Heckman}, T.~M. 2012, \mnras, 421, 1569, \dodoi{10.1111/j.1365-2966.2012.20414.x}

\bibitem[{{Boquien} {et~al.}(2019){Boquien}, {Burgarella}, {Roehlly}, {Buat}, {Ciesla}, {Corre}, {Inoue}, \& {Salas}}]{Boquien+2019}
{Boquien}, M., {Burgarella}, D., {Roehlly}, Y., {et~al.} 2019, \aap, 622, A103, \dodoi{10.1051/0004-6361/201834156}

\bibitem[{{Brammer} {et~al.}(2008){Brammer}, {van Dokkum}, \& {Coppi}}]{Brammer+2008}
{Brammer}, G.~B., {van Dokkum}, P.~G., \& {Coppi}, P. 2008, \apj, 686, 1503, \dodoi{10.1086/591786}

\bibitem[{{Brandt} \& {Alexander}(2015)}]{Brandt&Alexander+2015}
{Brandt}, W.~N., \& {Alexander}, D.~M. 2015, \aapr, 23, 1, \dodoi{10.1007/s00159-014-0081-z}

\bibitem[{{Brandt} \& {Yang}(2022)}]{Brandt&Yang+2022}
{Brandt}, W.~N., \& {Yang}, G. 2022, in Handbook of X-ray and Gamma-ray Astrophysics, 78, \dodoi{10.1007/978-981-16-4544-0_130-1}

\bibitem[{{Brandt} {et~al.}(2001){Brandt}, {Hornschemeier}, {Alexander}, {Garmire}, {Schneider}, {Broos}, {Townsley}, {Bautz}, {Feigelson}, \& {Griffiths}}]{Brandt+2001a}
{Brandt}, W.~N., {Hornschemeier}, A.~E., {Alexander}, D.~M., {et~al.} 2001, \aj, 122, 1, \dodoi{10.1086/321135}

\bibitem[{{Brightman} \& {Ueda}(2012)}]{Brightman+2012}
{Brightman}, M., \& {Ueda}, Y. 2012, \mnras, 423, 702, \dodoi{10.1111/j.1365-2966.2012.20908.x}

\bibitem[{{Bruzual} \& {Charlot}(2003)}]{BC+2003}
{Bruzual}, G., \& {Charlot}, S. 2003, \mnras, 344, 1000, \dodoi{10.1046/j.1365-8711.2003.06897.x}

\bibitem[{{Buchner} \& {Bauer}(2017)}]{Buchner&Bauer+2017}
{Buchner}, J., \& {Bauer}, F.~E. 2017, \mnras, 465, 4348, \dodoi{10.1093/mnras/stw2955}

\bibitem[{{Buchner} {et~al.}(2017){Buchner}, {Schulze}, \& {Bauer}}]{Buchner+2017}
{Buchner}, J., {Schulze}, S., \& {Bauer}, F.~E. 2017, \mnras, 464, 4545, \dodoi{10.1093/mnras/stw2423}

\bibitem[{{Buchner} {et~al.}(2015){Buchner}, {Georgakakis}, {Nandra}, {Brightman}, {Menzel}, {Liu}, {Hsu}, {Salvato}, {Rangel}, {Aird}, {Merloni}, \& {Ross}}]{Buchner+2015}
{Buchner}, J., {Georgakakis}, A., {Nandra}, K., {et~al.} 2015, \apj, 802, 89, \dodoi{10.1088/0004-637X/802/2/89}

\bibitem[{{Bussmann} {et~al.}(2012){Bussmann}, {Dey}, {Armus}, {Brown}, {Desai}, {Gonzalez}, {Jannuzi}, {Melbourne}, \& {Soifer}}]{Bussmann+2012}
{Bussmann}, R.~S., {Dey}, A., {Armus}, L., {et~al.} 2012, \apj, 744, 150, \dodoi{10.1088/0004-637X/744/2/150}

\bibitem[{{Calzetti} {et~al.}(2000){Calzetti}, {Armus}, {Bohlin}, {Kinney}, {Koornneef}, \& {Storchi-Bergmann}}]{Calzetti+2000}
{Calzetti}, D., {Armus}, L., {Bohlin}, R.~C., {et~al.} 2000, \apj, 533, 682, \dodoi{10.1086/308692}

\bibitem[{{Carnall} {et~al.}(2019){Carnall}, {Leja}, {Johnson}, {McLure}, {Dunlop}, \& {Conroy}}]{Carnall+2019}
{Carnall}, A.~C., {Leja}, J., {Johnson}, B.~D., {et~al.} 2019, \apj, 873, 44, \dodoi{10.3847/1538-4357/ab04a2}

\bibitem[{{Chabrier}(2003)}]{Chabrier+2003}
{Chabrier}, G. 2003, \pasp, 115, 763, \dodoi{10.1086/376392}

\bibitem[{Chen {et~al.}(2017)Chen, Hickox, Goulding, Stern, Assef, Kochanek, Brown, Harrison, Hainline, Alberts, Alexander, Brodwin, Moro, Forman, Gorjian, Jones, Murray, Pope, \& Rovilos}]{Chen+2017}
Chen, C.-T.~J., Hickox, R.~C., Goulding, A.~D., {et~al.} 2017, The Astrophysical Journal, 837, 145, \dodoi{10.3847/1538-4357/837/2/145}

\bibitem[{Chen {et~al.}(2018)Chen, Brandt, Luo, Ranalli, Yang, Alexander, Bauer, Kelson, Lacy, Nyland, Tozzi, Vito, Cirasuolo, Gilli, Jarvis, Lehmer, Paolillo, Schneider, Shemmer, Smail, Sun, Tanaka, Vaccari, Vignali, Xue, Banerji, Chow, Häußler, Norris, Silverman, \& Trump}]{Chen+2018}
Chen, C.-T.~J., Brandt, W.~N., Luo, B., {et~al.} 2018, \mnras, 478, 2132, \dodoi{10.1093/mnras/sty1036}

\bibitem[{{Cheng} {et~al.}(2023){Cheng}, {Huang}, {Smail}, {Yan}, {Cohen}, {Jansen}, {Windhorst}, {Ma}, {Koekemoer}, {Willmer}, {Willner}, {Diego}, {Frye}, {Conselice}, {Ferreira}, {Petric}, {Yun}, {Gim}, {Polletta}, {Duncan}, {Holwerda}, {R{\"o}ttgering}, {Honor}, {Hathi}, {Kamieneski}, {Adams}, {Coe}, {Broadhurst}, {Summers}, {Tompkins}, {Driver}, {Grogin}, {Marshall}, {Pirzkal}, {Robotham}, \& {Ryan}}]{Cheng+2023}
{Cheng}, C., {Huang}, J.-S., {Smail}, I., {et~al.} 2023, \apjl, 942, L19, \dodoi{10.3847/2041-8213/aca9d0}

\bibitem[{{Ciesla} {et~al.}(2015){Ciesla}, {Charmandaris}, {Georgakakis}, {Bernhard}, {Mitchell}, {Buat}, {Elbaz}, {LeFloc'h}, {Lacey}, {Magdis}, \& {Xilouris}}]{Ciesla+2015}
{Ciesla}, L., {Charmandaris}, V., {Georgakakis}, A., {et~al.} 2015, \aap, 576, A10, \dodoi{10.1051/0004-6361/201425252}

\bibitem[{{Cirasuolo} {et~al.}(2020){Cirasuolo}, {Fairley}, {Rees}, {Gonzalez}, {Taylor}, {Maiolino}, {Afonso}, {Evans}, {Flores}, {Lilly}, {Oliva}, {Paltani}, {Vanzi}, {Abreu}, {Accardo}, {Adams}, {{\'A}lvarez M{\'e}ndez}, {Amans}, {Amarantidis}, {Atek}, {Atkinson}, {Banerji}, {Barrett}, {Barrientos}, {Bauer}, {Beard}, {B{\'e}chet}, {Belfiore}, {Bellazzini}, {Benoist}, {Best}, {Biazzo}, {Black}, {Boettger}, {Bonifacio}, {Bowler}, {Bragaglia}, {Brierley}, {Brinchmann}, {Brinkmann}, {Buat}, {Buitrago}, {Burgarella}, {Burningham}, {Buscher}, {Cabral}, {Caffau}, {Cardoso}, {Carnall}, {Carollo}, {Castillo}, {Castignani}, {Catelan}, {Cicone}, {Cimatti}, {Cioni}, {Clementini}, {Cochrane}, {Coelho}, {Colling}, {Contini}, {Contreras}, {Conzelmann}, {Cresci}, {Cropper}, {Cucciati}, {Cullen}, {Cumani}, {Curti}, {Da Silva}, {Daddi}, {Dalessandro}, {Dalessio}, {Dauvin}, {Davidson}, {de Laverny}, {Delplancke-Str{\"o}bele}, {De Lucia}, {Del Vecchio}, {Dessauges-Zavadsky}, {Di Matteo}, {Dole}, {Drass}, {Dunlop},
  {D{\"u}nner}, {Eales}, {Ellis}, {Enriques}, {Fasola}, {Ferguson}, {Ferruzzi}, {Fisher}, {Flores}, {Fontana}, {Forchi}, {Francois}, {Franzetti}, {Gargiulo}, {Garilli}, {Gaudemard}, {Gieles}, {Gilmore}, {Ginolfi}, {Gomes}, {Guinouard}, {Gutierrez}, {Haigron}, {Hammer}, {Hammersley}, {Haniff}, {Harrison}, {Haywood}, {Hill}, {Hubin}, {Humphrey}, {Ibata}, {Infante}, {Ives}, {Ivison}, {Iwert}, {Jablonka}, {Jakob}, {Jarvis}, {King}, {Kneib}, {Laporte}, {Lawrence}, {Lee}, {Li Causi}, {Lorenzoni}, {Lucatello}, {Luco}, {Macleod}, {Magliocchetti}, {Magrini}, {Mainieri}, {Maire}, {Mannucci}, {Martin}, {Matute}, {Maurogordato}, {McGee}, {Mcleod}, {McLure}, {McMahon}, {Melse}, {Messias}, {Mucciarelli}, {Nisini}, {Nix}, {Norberg}, {Oesch}, {Oliveira}, {Origlia}, {Padilla}, {Palsa}, {Pancino}, {Papaderos}, {Pappalardo}, {Parry}, {Pasquini}, {Peacock}, {Pedichini}, {Pello}, {Peng}, {Pentericci}, {Pfuhl}, {Piazzesi}, {Popovic}, {Pozzetti}, {Puech}, {Puzia}, {Raichoor}, {Randich}, {Recio-Blanco}, {Reis}, {Reix}, {Renzini},
  {Rodrigues}, {Rojas}, {Rojas-Arriagada}, {Rota}, {Royer}, {Sacco}, {Sanchez-Janssen}, {Sanna}, {Santos}, {Sarzi}, {Schaerer}, {Schiavon}, {Schnell}, {Schultheis}, {Scodeggio}, {Serjeant}, {Shen}, {Simmonds}, {Smoker}, {Sobral}, {Sordet}, {Sp{\'e}rone}, {Strachan}, {Sun}, {Swinbank}, {Tait}, {Tereno}, {Tojeiro}, {Torres}, {Tosi}, {Tozzi}, {Tresiter}, {Valenti}, {Valenzuela Navarro}, {Vanzella}, {Vergani}, {Verhamme}, {Vernet}, {Vignali}, {Vinther}, {Von Dran}, {Waring}, {Watson}, {Wild}, {Willesme}, {Woodward}, {Wuyts}, {Yang}, {Zamorani}, {Zoccali}, {Bluck}, \& {Trussler}}]{Cirasuolo+2020}
{Cirasuolo}, M., {Fairley}, A., {Rees}, P., {et~al.} 2020, The Messenger, 180, 10, \dodoi{10.18727/0722-6691/5195}

\bibitem[{{Condon}(1992)}]{Condon+1992}
{Condon}, J.~J. 1992, \araa, 30, 575, \dodoi{10.1146/annurev.aa.30.090192.003043}

\bibitem[{{Corral} {et~al.}(2016){Corral}, {Georgantopoulos, I.}, {Comastri, A.}, {Ranalli, P.}, {Akylas, A.}, {Salvato, M.}, {Lanzuisi, G.}, {Vignali, C.}, \& {Koutoulidis, L.}}]{Corral+2016}
{Corral}, {Georgantopoulos, I.}, {Comastri, A.}, {et~al.} 2016, A\&A, 592, A109, \dodoi{10.1051/0004-6361/201527624}

\bibitem[{{Cristello} {et~al.}(2024){Cristello}, {Zou}, {Brandt}, {Chen}, {Leja}, {Ni}, \& {Yang}}]{Cristello+2024}
{Cristello}, N., {Zou}, F., {Brandt}, W.~N., {et~al.} 2024, \apj, 962, 156, \dodoi{10.3847/1538-4357/ad2177}

\bibitem[{{Cutri} {et~al.}(2021){Cutri}, {Wright}, {Conrow}, {Fowler}, {Eisenhardt}, {Grillmair}, {Kirkpatrick}, {Masci}, {McCallon}, {Wheelock}, {Fajardo-Acosta}, {Yan}, {Benford}, {Harbut}, {Jarrett}, {Lake}, {Leisawitz}, {Ressler}, {Stanford}, {Tsai}, {Liu}, {Helou}, {Mainzer}, {Gettngs}, {Gonzalez}, {Hoffman}, {Marsh}, {Padgett}, {Skrutskie}, {Beck}, {Papin}, \& {Wittman}}]{Cutri+2021}
{Cutri}, R.~M., {Wright}, E.~L., {Conrow}, T., {et~al.} 2021, {VizieR Online Data Catalog: AllWISE Data Release (Cutri+ 2013)}, VizieR On-line Data Catalog: II/328. Originally published in: IPAC/Caltech (2013)

\bibitem[{{Dale} {et~al.}(2014){Dale}, {Helou}, {Magdis}, {Armus}, {D{\'\i}az-Santos}, \& {Shi}}]{Dale+2014}
{Dale}, D.~A., {Helou}, G., {Magdis}, G.~E., {et~al.} 2014, \apj, 784, 83, \dodoi{10.1088/0004-637X/784/1/83}

\bibitem[{{Delvecchio} {et~al.}(2017){Delvecchio}, {Smol{\v{c}}i{\'c}}, {Zamorani}, {Lagos}, {Berta}, {Delhaize}, {Baran}, {Alexander}, {Rosario}, {Gonzalez-Perez}, {Ilbert}, {Lacey}, {Le F{\`e}vre}, {Miettinen}, {Aravena}, {Bondi}, {Carilli}, {Ciliegi}, {Mooley}, {Novak}, {Schinnerer}, {Capak}, {Civano}, {Fanidakis}, {Herrera Ruiz}, {Karim}, {Laigle}, {Marchesi}, {McCracken}, {Middleberg}, {Salvato}, \& {Tasca}}]{Delvecchio+2017}
{Delvecchio}, I., {Smol{\v{c}}i{\'c}}, V., {Zamorani}, G., {et~al.} 2017, \aap, 602, A3, \dodoi{10.1051/0004-6361/201629367}

\bibitem[{{Delvecchio} {et~al.}(2021){Delvecchio}, {Daddi}, {Sargent}, {Jarvis}, {Elbaz}, {Jin}, {Liu}, {Whittam}, {Algera}, {Carraro}, {D'Eugenio}, {Delhaize}, {Kalita}, {Leslie}, {Moln{\'a}r}, {Novak}, {Prandoni}, {Smol{\v{c}}i{\'c}}, {Ao}, {Aravena}, {Bournaud}, {Collier}, {Randriamampandry}, {Randriamanakoto}, {Rodighiero}, {Schober}, {White}, \& {Zamorani}}]{Delvecchio+2021}
{Delvecchio}, I., {Daddi}, E., {Sargent}, M.~T., {et~al.} 2021, \aap, 647, A123, \dodoi{10.1051/0004-6361/202039647}

\bibitem[{{Dey} \& {NDWFS/MIPS Collaboration}(2009)}]{Dey+2009}
{Dey}, A., \& {NDWFS/MIPS Collaboration}. 2009, in Astronomical Society of the Pacific Conference Series, Vol. 408, The Starburst-AGN Connection, ed. W.~{Wang}, Z.~{Yang}, Z.~{Luo}, \& Z.~{Chen}, 411, \dodoi{10.48550/arXiv.0905.4531}

\bibitem[{{Dey} {et~al.}(2008){Dey}, {Soifer}, {Desai}, {Brand}, {Le Floc'h}, {Brown}, {Jannuzi}, {Armus}, {Bussmann}, {Brodwin}, {Bian}, {Eisenhardt}, {Higdon}, {Weedman}, \& {Willner}}]{Dey+2008}
{Dey}, A., {Soifer}, B.~T., {Desai}, V., {et~al.} 2008, \apj, 677, 943, \dodoi{10.1086/529516}

\bibitem[{{Di Matteo} {et~al.}(2005){Di Matteo}, {Springel}, \& {Hernquist}}]{DiMatteo+2005}
{Di Matteo}, T., {Springel}, V., \& {Hernquist}, L. 2005, \nat, 433, 604, \dodoi{10.1038/nature03335}

\bibitem[{{Doe} {et~al.}(2007){Doe}, {Nguyen}, {Stawarz}, {Refsdal}, {Siemiginowska}, {Burke}, {Evans}, {Evans}, {McDowell}, {Houck}, \& {Nowak}}]{Doe+2007}
{Doe}, S., {Nguyen}, D., {Stawarz}, C., {et~al.} 2007, in Astronomical Society of the Pacific Conference Series, Vol. 376, Astronomical Data Analysis Software and Systems XVI, ed. R.~A. {Shaw}, F.~{Hill}, \& D.~J. {Bell}, 543

\bibitem[{{Donnari} {et~al.}(2019){Donnari}, {Pillepich}, {Nelson}, {Vogelsberger}, {Genel}, {Weinberger}, {Marinacci}, {Springel}, \& {Hernquist}}]{Donnari+2019}
{Donnari}, M., {Pillepich}, A., {Nelson}, D., {et~al.} 2019, \mnras, 485, 4817, \dodoi{10.1093/mnras/stz712}

\bibitem[{{Draine} {et~al.}(2014){Draine}, {Aniano}, {Krause}, {Groves}, {Sandstrom}, {Braun}, {Leroy}, {Klaas}, {Linz}, {Rix}, {Schinnerer}, {Schmiedeke}, \& {Walter}}]{Draine+2014}
{Draine}, B.~T., {Aniano}, G., {Krause}, O., {et~al.} 2014, \apj, 780, 172, \dodoi{10.1088/0004-637X/780/2/172}

\bibitem[{{Driver} {et~al.}(2019){Driver}, {Liske}, {Davies}, {Robotham}, {Baldry}, {Brown}, {Cluver}, {Kuijken}, {Loveday}, {McMahon}, {Meyer}, {Norberg}, {Owers}, {Power}, {Taylor}, \& {WAVES Team}}]{Driver+2019}
{Driver}, S.~P., {Liske}, J., {Davies}, L.~J.~M., {et~al.} 2019, The Messenger, 175, 46, \dodoi{10.18727/0722-6691/5126}

\bibitem[{{Eisenhardt} {et~al.}(2012){Eisenhardt}, {Wu}, {Tsai}, {Assef}, {Benford}, {Blain}, {Bridge}, {Condon}, {Cushing}, {Cutri}, {Evans}, {Gelino}, {Griffith}, {Grillmair}, {Jarrett}, {Lonsdale}, {Masci}, {Mason}, {Petty}, {Sayers}, {Stanford}, {Stern}, {Wright}, \& {Yan}}]{Eisenhardt+2012}
{Eisenhardt}, P. R.~M., {Wu}, J., {Tsai}, C.-W., {et~al.} 2012, \apj, 755, 173, \dodoi{10.1088/0004-637X/755/2/173}

\bibitem[{{Euclid Collaboration} {et~al.}(2024){Euclid Collaboration}, {Mellier}, {Abdurro'uf}, {Acevedo Barroso}, {Ach{\'u}carro}, {Adamek}, {Adam}, {Addison}, {Aghanim}, {Aguena}, {Ajani}, {Akrami}, {Al-Bahlawan}, {Alavi}, {Albuquerque}, {Alestas}, {Alguero}, {Allaoui}, {Allen}, {Allevato}, {Alonso-Tetilla}, {Altieri}, {Alvarez-Candal}, {Alvi}, {Amara}, {Amendola}, {Amiaux}, {Andika}, {Andreon}, {Andrews}, {Angora}, {Angulo}, {Annibali}, {Anselmi}, {Anselmi}, {Arcari}, {Archidiacono}, {Aric{\`o}}, {Arnaud}, {Arnouts}, {Asgari}, {Asorey}, {Atayde}, {Atek}, {Atrio-Barandela}, {Aubert}, {Aubourg}, {Auphan}, {Auricchio}, {Aussel}, {Aussel}, {Avelino}, {Avgoustidis}, {Avila}, {Awan}, {Azzollini}, {Baccigalupi}, {Bachelet}, {Bacon}, {Baes}, {Bagley}, {Bahr-Kalus}, {Balaguera-Antolinez}, {Balbinot}, {Balcells}, {Baldi}, {Baldry}, {Balestra}, {Ballardini}, {Ballester}, {Balogh}, {Ba{\~n}ados}, {Barbier}, {Bardelli}, {Baron}, {Barreiro}, {Barrena}, {Barriere}, {Barros}, {Barthelemy}, {Bartolo}, {Basset},
  {Battaglia}, {Battisti}, {Baugh}, {Baumont}, {Bazzanini}, {Beaulieu}, {Beckmann}, {Belikov}, {Bel}, {Bellagamba}, {Bella}, {Bellini}, {Benabed}, {Bender}, {Benevento}, {Bennett}, {Benson}, {Bergamini}, {Bermejo-Climent}, {Bernardeau}, {Bertacca}, {Berthe}, {Berthier}, {Bethermin}, {Beutler}, {Bevillon}, {Bhargava}, {Bhatawdekar}, {Bianchi}, {Bisigello}, {Biviano}, {Blake}, {Blanchard}, {Blazek}, {Blot}, {Bosco}, {Bodendorf}, {Boenke}, {B{\"o}hringer}, {Boldrini}, {Bolzonella}, {Bonchi}, {Bonici}, {Bonino}, {Bonino}, {Bonvin}, {Bon}, {Booth}, {Borgani}, {Borlaff}, {Borsato}, {Bosco}, {Bose}, {Botticella}, {Boucaud}, {Bouche}, {Boucher}, {Boutigny}, {Bouvard}, {Bouwens}, {Bouy}, {Bowler}, {Bozza}, {Bozzo}, {Branchini}, {Brando}, {Brau-Nogue}, {Brekke}, {Bremer}, {Brescia}, {Breton}, {Brinchmann}, {Brinckmann}, {Brockley-Blatt}, {Brodwin}, {Brouard}, {Brown}, {Bruton}, {Bucko}, {Buddelmeijer}, {Buenadicha}, {Buitrago}, {Burger}, {Burigana}, {Busillo}, {Busonero}, {Cabanac}, {Cabayol-Garcia}, {Cagliari},
  {Caillat}, {Caillat}, {Calabrese}, {Calabro}, {Calderone}, {Calura}, {Camacho Quevedo}, {Camera}, {Campos}, {Canas-Herrera}, {Candini}, {Cantiello}, {Capobianco}, {Cappellaro}, {Cappelluti}, {Cappi}, {Caputi}, {Cara}, {Carbone}, {Cardone}, {Carella}, {Carlberg}, {Carle}, {Carminati}, {Caro}, {Carrasco}, {Carretero}, {Carrilho}, {Carron Duque}, {Carry}, {Carvalho}, {Carvalho}, {Casas}, {Casas}, {Casenove}, {Casey}, {Cassata}, {Castander}, {Castelao}, {Castellano}, {Castiblanco}, {Castignani}, {Castro}, {Cavet}, {Cavuoti}, {Chabaud}, {Chambers}, {Charles}, {Charlot}, {Chartab}, {Chary}, {Chaumeil}, {Cho}, {Chon}, {Ciancetta}, {Ciliegi}, {Cimatti}, {Cimino}, {Cioni}, {Claydon}, {Cleland}, {Cl{\'e}ment}, {Clements}, {Clerc}, {Clesse}, {Codis}, {Cogato}, {Colbert}, {Cole}, {Coles}, {Collett}, {Collins}, {Colodro-Conde}, {Colombo}, {Combes}, {Conforti}, {Congedo}, {Conseil}, {Conselice}, {Contarini}, {Contini}, {Conversi}, {Cooray}, {Copin}, {Corasaniti}, {Corcho-Caballero}, {Corcione}, {Cordes}, {Corpace},
  {Correnti}, {Costanzi}, {Costille}, {Courbin}, {Courcoult Mifsud}, {Courtois}, {Cousinou}, {Covone}, {Cowell}, {Cragg}, {Cresci}, {Cristiani}, {Crocce}, {Cropper}, {E Crouzet}, {Csizi}, {Cuby}, {Cucchetti}, {Cucciati}, {Cuillandre}, {Cunha}, {Cuozzo}, {Daddi}, {D'Addona}, {Dafonte}, {Dagoneau}, {Dalessandro}, {Dalton}, {D'Amico}, {Dannerbauer}, {Danto}, {Das}, {Da Silva}, {da Silva}, {d'Assignies Doumerg}, {Daste}, {Davies}, {Davini}, {Dayal}, {de Boer}, {Decarli}, {De Caro}, {Degaudenzi}, {Degni}, {de Jong}, {de la Bella}, {de la Torre}, {Delhaise}, {Delley}, {Delucchi}, {De Lucia}, {Denniston}, {De Paolis}, {De Petris}, {Derosa}, {Desai}, {Desjacques}, {Despali}, {Desprez}, {De Vicente-Albendea}, {Deville}, {Dias}, {D{\'\i}az-S{\'a}nchez}, {Diaz}, {Di Domizio}, {Diego}, {Di Ferdinando}, {Di Giorgio}, {Dimauro}, {Dinis}, {Dolag}, {Dolding}, {Dole}, {Dom{\'\i}nguez S{\'a}nchez}, {Dor{\'e}}, {Dournac}, {Douspis}, {Dreihahn}, {Droge}, {Dryer}, {Dubath}, {Duc}, {Ducret}, {Duffy}, {Dufresne}, {Duncan}, {Dupac},
  {Duret}, {Durrer}, {Durret}, {Dusini}, {Ealet}, {Eggemeier}, {Eisenhardt}, {Elbaz}, {Elkhashab}, {Ellien}, {Endicott}, {Enia}, {Erben}, {Escartin Vigo}, {Escoffier}, {Escudero Sanz}, {Essert}, {Ettori}, {Ezziati}, {Fabbian}, {Fabricius}, {Fang}, {Farina}, {Farina}, {Farinelli}, {Farrens}, {Faustini}, {Feltre}, {Ferguson}, {Ferrando}, {Ferrari}, {Ferr{\'e}-Mateu}, {Ferreira}, {Ferreras}, {Ferrero}, {Ferriol}, {Ferruit}, {Filleul}, {Finelli}, {Finkelstein}, {Finoguenov}, {Fiorini}, {Flentge}, {Focardi}, {Fonseca}, {Fontana}, {Fontanot}, {Fornari}, {Fosalba}, {Fossati}, {Fotopoulou}, {Fouchez}, {Fourmanoit}, {Frailis}, {Fraix-Burnet}, {Franceschi}, {Franco}, {Franzetti}, {Freihoefer}, {Frenk}, {Frittoli}, {Frugier}, {Frusciante}, {Fumagalli}, {Fumagalli}, {Fumana}, {Fu}, {Gabarra}, {Galeotta}, {Galluccio}, {Ganga}, {Gao}, {Garc{\'\i}a-Bellido}, {Garcia}, {Gardner}, {Garilli}, {Gaspar-Venancio}, {Gasparetto}, {Gautard}, {Gavazzi}, {Gaztanaga}, {Genolet}, {Genova Santos}, {Gentile}, {George}, {Gerbino},
  {Ghaffari}, {Giacomini}, {Gianotti}, {Gibb}, {Gillard}, {Gillis}, {Ginolfi}, {Giocoli}, {Girardi}, {Giri}, {Goh}, {G{\'o}mez-Alvarez}, {Gonzalez-Perez}, {Gonzalez}, {Gonzalez}, {Gonzalez}, {Gouyou Beauchamps}, {Gozaliasl}, {Gracia-Carpio}, {Grandis}, {Granett}, {Granvik}, {Grazian}, {Gregorio}, {Grenet}, {Grillo}, {Grupp}, {Gruppioni}, {Gruppuso}, {Guerbuez}, {Guerrini}, {Guidi}, {Guillard}, {Gutierrez}, {Guttridge}, {Guzzo}, {Gwyn}, {Haapala}, {Haase}, {Haddow}, {Hailey}, {Hall}, {Hall}, {Hamaus}, {Haridasu}, {Harnois-D{\'e}raps}, {Harper}, {Hartley}, {Hasinger}, {Hassani}, {Hatch}, {Haugan}, {H{\"a}u{\ss}ler}, {Heavens}, {Heisenberg}, {Helmi}, {Helou}, {Hemmati}, {Henares}, {Herent}, {Hern{\'a}ndez-Monteagudo}, {Heuberger}, {Hewett}, {Heydenreich}, {Hildebrandt}, {Hirschmann}, {Hjorth}, {Hoar}, {Hoekstra}, {Holland}, {Holliman}, {Holmes}, {Hook}, {Horeau}, {Hormuth}, {Hornstrup}, {Hosseini}, {Hu}, {Hudelot}, {Hudson}, {Huertas-Company}, {Huff}, {Hughes}, {Humphrey}, {Hunt}, {Huynh}, {Ibata}, {Ichikawa},
  {Iglesias-Groth}, {Ilbert}, {Ili{\'c}}, {Ingoglia}, {Iodice}, {Israel}, {Israelsson}, {Izzo}, {Jablonka}, {Jackson}, {Jacobson}, {Jafariyazani}, {Jahnke}, {Jain}, {Jansen}, {Jarvis}, {Jasche}, {Jauzac}, {Jeffrey}, {Jhabvala}, {Jimenez-Teja}, {Jimenez Mu{\~n}oz}, {Joachimi}, {Johansson}, {Joudaki}, {Jullo}, {Kajava}, {Kang}, {Kannawadi}, {Kansal}, {Karagiannis}, {K{\"a}rcher}, {Kashlinsky}, {Kazandjian}, {Keck}, {Keih{\"a}nen}, {Kerins}, {Kermiche}, {Khalil}, {Kiessling}, {Kiiveri}, {Kilbinger}, {Kim}, {King}, {Kirkpatrick}, {Kitching}, {Kluge}, {Knabenhans}, {Knapen}, {Knebe}, {Kneib}, {Kohley}, {Koopmans}, {Koskinen}, {Koulouridis}, {Kou}, {Kov{\'a}cs}, {Kova{\v{c}}i{\'c}}, {Kowalczyk}, {Koyama}, {Kraljic}, {Krause}, {Kruk}, {Kubik}, {Kuchner}, {Kuijken}, {K{\"u}mmel}, {Kunz}, {Kurki-Suonio}, {Lacasa}, {Lacey}, {La Franca}, {Lagarde}, {Lahav}, {Laigle}, {La Marca}, {La Marle}, {Lamine}, {Lam}, {Lan{\c{c}}on}, {Landt}, {Langer}, {Lapi}, {Larcheveque}, {Larsen}, {Lattanzi}, {Laudisio}, {Laugier}, {Laureijs},
  {Laurent}, {Lavaux}, {Lawrenson}, {Lazanu}, {Lazeyras}, {Le Boulc'h}, {Le Brun}, {Le Brun}, {Leclercq}, {Lee}, {Le Graet}, {Legrand}, {Leirvik}, {Le Jeune}, {Lembo}, {Le Mignant}, {Lepinzan}, {Lepori}, {Le Reun}, {Leroy}, {Lesci}, {Lesgourgues}, {Leuzzi}, {Levi}, {Liaudat}, {Libet}, {Liebing}, {Ligori}, {Lilje}, {Lin}, {Linde}, {Linder}, {Lindholm}, {Linke}, {Li}, {Liu}, {Lloro}, {Lobo}, {Lodieu}, {Lombardi}, {Lombriser}, {Lonare}, {Longo}, {L{\'o}pez-Caniego}, {Lopez Lopez}, {Alvarez}, {Loureiro}, {Loveday}, {Lusso}, {Macias-Perez}, {Maciaszek}, {Maggio}, {Magliocchetti}, {Magnard}, {Magnier}, {Magro}, {Mahler}, {Mainetti}, {Maino}, {Maiorano}, {Maiorano}, {Malavasi}, {Mamon}, {Mancini}, {Mandelbaum}, {Manera}, {Manj{\'o}n-Garc{\'\i}a}, {Mannucci}, {Mansutti}, {Manteiga Outeiro}, {Maoli}, {Maraston}, {Marcin}, {Marcos-Arenal}, {Margalef-Bentabol}, {Marggraf}, {Marinucci}, {Marinucci}, {Markovic}, {Marleau}, {Marpaud}, {Martignac}, {Mart{\'\i}n-Fleitas}, {Martin-Moruno}, {Martin}, {Martinelli}, {Martinet},
  {Martin}, {Martins}, {Marulli}, {Massari}, {Massey}, {Masters}, {Matarrese}, {Matsuoka}, {Matthew}, {Maughan}, {Mauri}, {Maurin}, {Maurogordato}, {McCarthy}, {McConnachie}, {McCracken}, {McDonald}, {McEwen}, {McPartland}, {Medinaceli}, {Mehta}, {Mei}, {Melchior}, {Melin}, {M{\'e}nard}, {Mendes}, {Mendez-Abreu}, {Meneghetti}, {Mercurio}, {Merlin}, {Metcalf}, {Meylan}, {Migliaccio}, {Mignoli}, {Miller}, {Miluzio}, {Milvang-Jensen}, {Mimoso}, {Miquel}, {Miyatake}, {Mobasher}, {Mohr}, {Monaco}, {Mongui{\'o}}, {Montoro}, {Mora}, {Moradinezhad Dizgah}, {Moresco}, {Moretti}, {Morgante}, {Morisset}, {Moriya}, {Morris}, {Mortlock}, {Moscardini}, {Mota}, {Mottet}, {Moustakas}, {Moutard}, {M{\"u}ller}, {Munari}, {Murphree}, {Murray}, {Murray}, {Musi}, {Nadathur}, {Nagam}, {Nagao}, {Naidoo}, {Nakajima}, {Nally}, {Natoli}, {Navarro-Alsina}, {Navarro Girones}, {Neissner}, {Nersesian}, {Nesseris}, {Nguyen-Kim}, {Nicastro}, {Nichol}, {Nielbock}, {Niemi}, {Nieto}, {Nilsson}, {Noller}, {Norberg}, {Nouri-Zonoz}, {Ntelis},
  {Nucita}, {Nugent}, {Nunes}, {Nutma}, {Ocampo}, {Odier}, {Oesch}, {Oguri}, {Magalhaes Oliveira}, {Onoue}, {Oosterbroek}, {Oppizzi}, {Ordenovic}, {Osato}, {Pacaud}, {Pace}, {Padilla}, {Paech}, {Pagano}, {Page}, {Palazzi}, {Paltani}, {Pamuk}, {Pandolfi}, {Paoletti}, {Paolillo}, {Papaderos}, {Pardede}, {Parimbelli}, {Parmar}, {Partmann}, {Pasian}, {Passalacqua}, {Paterson}, {Patrizii}, {Pattison}, {Paulino-Afonso}, {Paviot}, {Peacock}, {Pearce}, {Pedersen}, {Peel}, {Peletier}, {Pellejero Ibanez}, {Pello}, {Penny}, {Percival}, {Perez-Garrido}, {Perotto}, {Pettorino}, {Pezzotta}, {Pezzuto}, {Philippon}, {Pierre}, {Piersanti}, {Pietroni}, {Piga}, {Pilo}, {Pires}, {Pisani}, {Pizzella}, {Pizzuti}, {Plana}, {Polenta}, {Pollack}, {Poncet}, {P{\"o}ntinen}, {Pool}, {Popa}, {Popa}, {Popp}, {Porciani}, {Porth}, {Potter}, {Poulain}, {Pourtsidou}, {Pozzetti}, {Prandoni}, {Pratt}, {Prezelus}, {Prieto}, {Pugno}, {Quai}, {Quilley}, {Racca}, {Raccanelli}, {R{\'a}cz}, {Radinovi{\'c}}, {Radovich}, {Ragagnin}, {Ragnit}, {Raison},
  {Ramos-Chernenko}, {Ranc}, {Rasera}, {Raylet}, {Rebolo}, {Refregier}, {Reimberg}, {Reiprich}, {Renk}, {Renzi}, {Retre}, {Revaz}, {Reyl{\'e}}, {Reynolds}, {Rhodes}, {Ricci}, {Ricci}, {Riccio}, {Ricken}, {Rissanen}, {Risso}, {Rix}, {Robin}, {Rocca-Volmerange}, {Rocci}, {Rodenhuis}, {Rodighiero}, {Rodriguez Monroy}, {Rollins}, {Romanello}, {Roman}, {Romelli}, {Romero-Gomez}, {Roncarelli}, {Rosati}, {Rosset}, {Rossetti}, {Roster}, {Rottgering}, {Rozas-Fern{\'a}ndez}, {Ruane}, {Rubino-Martin}, {Rudolph}, {Ruppin}, {Rusholme}, {Sacquegna}, {S{\'a}ez-Casares}, {Saga}, {Saglia}, {Sahl{\'e}n}, {Saifollahi}, {Sakr}, {Salvalaggio}, {Salvaterra}, {Salvati}, {Salvato}, {Salvignol}, {S{\'a}nchez}, {Sanchez}, {Sanders}, {Sapone}, {Saponara}, {Sarpa}, {Sarron}, {Sartori}, {Sartoris}, {Sassolas}, {Sauniere}, {Sauvage}, {Sawicki}, {Scaramella}, {Scarlata}, {Scharr{\'e}}, {Schaye}, {Schewtschenko}, {Schindler}, {Schinnerer}, {Schirmer}, {Schmidt}, {Schmidt}, {Schmidt}, {Schneider}, {Schneider}, {Schneider}, {Sch{\"o}neberg},
  {Schrabback}, {Schultheis}, {Schulz}, {Schuster}, {Schwartz}, {Sciotti}, {Scodeggio}, {Scognamiglio}, {Scott}, {Scottez}, {Secroun}, {Sefusatti}, {Seidel}, {Seiffert}, {Sellentin}, {Selwood}, {Semboloni}, {Sereno}, {Serjeant}, {Serrano}, {Setnikar}, {Shankar}, {Sharples}, {Short}, {Shulevski}, {Shuntov}, {Sias}, {Sikkema}, {Silvestri}, {Simon}, {Sirignano}, {Sirri}, {Skottfelt}, {Slezak}, {Sluse}, {Smith}, {Smith}, {Smith}, {Smit}, {Soldano}, {Solheim}, {Sorce}, {Sorrenti}, {Soubrie}, {Spinoglio}, {Spurio Mancini}, {Stadel}, {Stagnaro}, {Stanco}, {Stanford}, {Starck}, {Stassi}, {Steinwagner}, {Stern}, {Stone}, {Strada}, {Strafella}, {Stramaccioni}, {Surace}, {Sureau}, {Suyu}, {Swindells}, {Szafraniec}, {Szapudi}, {Taamoli}, {Talia}, {Tallada-Cresp{\'\i}}, {Tanidis}, {Tao}, {Tarr{\'\i}o}, {Tavagnacco}, {Taylor}, {Taylor}, {Taylor}, {Teixeira}, {Tenti}, {Teodoro Idiago}, {Teplitz}, {Tereno}, {Tessore}, {Testa}, {Testera}, {Tewes}, {Teyssier}, {Theret}, {Thizy}, {Thomas}, {Toba}, {Toft}, {Toledo-Moreo},
  {Tolstoy}, {Tommasi}, {Torbaniuk}, {Torradeflot}, {Tortora}, {Tosi}, {Tosti}, {Trifoglio}, {Troja}, {Trombetti}, {Tronconi}, {Tsedrik}, {Tsyganov}, {Tucci}, {Tutusaus}, {Uhlemann}, {Ulivi}, {Urbano}, {Vacher}, {Vaillon}, {Valageas}, {Valdes}, {Valentijn}, {Valenziano}, {Valieri}, {Valiviita}, {Van den Broeck}, {Vassallo}, {Vavrek}, {Vega-Ferrero}, {Venemans}, {Venhola}, {Ventura}, {Verdoes Kleijn}, {Vergani}, {Verma}, {Vernizzi}, {Veropalumbo}, {Verza}, {Vescovi}, {Vibert}, {Viel}, {Vielzeuf}, {Viglione}, {Viitanen}, {Villaescusa-Navarro}, {Vinciguerra}, {Visticot}, {Voggel}, {von Wietersheim-Kramsta}, {Vriend}, {Wachter}, {Walmsley}, {Walth}, {Walton}, {Walton}, {Wander}, {Wang}, {Wang}, {Weaver}, {Weller}, {Wetzstein}, {Whalen}, {Whittam}, {Widmer}, {Wiesmann}, {Wilde}, {Williams}, {Winther}, {Wittje}, {Wong}, {Wright}, {Yankelevich}, {Yeung}, {Yoon}, {Youles}, {Yung}, {Zacchei}, {Zalesky}, {Zamorani}, {Zamorano Vitorelli}, {Zanoni Marc}, {Zennaro}, {Zerbi}, {Zinchenko}, {Zoubian}, {Zucca}, \&
  {Zumalacarregui}}]{EuclidCollaboration+2024}
{Euclid Collaboration}, {Mellier}, Y., {Abdurro'uf}, {et~al.} 2024, arXiv e-prints, arXiv:2405.13491, \dodoi{10.48550/arXiv.2405.13491}

\bibitem[{{Falocco} {et~al.}(2015){Falocco}, {Paolillo}, {Covone}, {De Cicco}, {Longo}, {Grado}, {Limatola}, {Vaccari}, {Botticella}, {Pignata}, {Cappellaro}, {Trevese}, {Vagnetti}, {Salvato}, {Radovich}, {Hsu}, {Capaccioli}, {Napolitano}, {Brandt}, {Baruffolo}, {Cascone}, \& {Schipani}}]{Falocco+2015}
{Falocco}, S., {Paolillo}, M., {Covone}, G., {et~al.} 2015, \aap, 579, A115, \dodoi{10.1051/0004-6361/201425111}

\bibitem[{{Fan} {et~al.}(2016){Fan}, {Han}, {Nikutta}, {Drouart}, \& {Knudsen}}]{Fan+2016}
{Fan}, L., {Han}, Y., {Nikutta}, R., {Drouart}, G., \& {Knudsen}, K.~K. 2016, \apj, 823, 107, \dodoi{10.3847/0004-637X/823/2/107}

\bibitem[{{Feigelson} \& {Babu}(2012)}]{Feigelson&Babu+2012}
{Feigelson}, E.~D., \& {Babu}, G.~J. 2012, {Modern Statistical Methods for Astronomy}, \dodoi{10.48550/arXiv.1205.2064}

\bibitem[{{Fiore} {et~al.}(2008){Fiore}, {Grazian}, {Santini}, {Puccetti}, {Brusa}, {Feruglio}, {Fontana}, {Giallongo}, {Comastri}, {Gruppioni}, {Pozzi}, {Zamorani}, \& {Vignali}}]{Fiore+2008}
{Fiore}, F., {Grazian}, A., {Santini}, P., {et~al.} 2008, \apj, 672, 94, \dodoi{10.1086/523348}

\bibitem[{{Fiore} {et~al.}(2009){Fiore}, {Puccetti}, {Brusa}, {Salvato}, {Zamorani}, {Aldcroft}, {Aussel}, {Brunner}, {Capak}, {Cappelluti}, {Civano}, {Comastri}, {Elvis}, {Feruglio}, {Finoguenov}, {Fruscione}, {Gilli}, {Hasinger}, {Koekemoer}, {Kartaltepe}, {Ilbert}, {Impey}, {Le Floc'h}, {Lilly}, {Mainieri}, {Martinez-Sansigre}, {McCracken}, {Menci}, {Merloni}, {Miyaji}, {Sanders}, {Sargent}, {Schinnerer}, {Scoville}, {Silverman}, {Smolcic}, {Steffen}, {Santini}, {Taniguchi}, {Thompson}, {Trump}, {Vignali}, {Urry}, \& {Yan}}]{Fiore+2009}
{Fiore}, F., {Puccetti}, S., {Brusa}, M., {et~al.} 2009, \apj, 693, 447, \dodoi{10.1088/0004-637X/693/1/447}

\bibitem[{{Franzen} {et~al.}(2015){Franzen}, {Banfield}, {Hales}, {Hopkins}, {Norris}, {Seymour}, {Chow}, {Herzog}, {Huynh}, {Lenc}, {Mao}, \& {Middelberg}}]{Franzen+2015}
{Franzen}, T.~M.~O., {Banfield}, J.~K., {Hales}, C.~A., {et~al.} 2015, \mnras, 453, 4020, \dodoi{10.1093/mnras/stv1866}

\bibitem[{{Freeman} {et~al.}(2001){Freeman}, {Doe}, \& {Siemiginowska}}]{Freeman+2001}
{Freeman}, P., {Doe}, S., \& {Siemiginowska}, A. 2001, in Society of Photo-Optical Instrumentation Engineers (SPIE) Conference Series, Vol. 4477, Astronomical Data Analysis, ed. J.-L. {Starck} \& F.~D. {Murtagh}, 76--87, \dodoi{10.1117/12.447161}

\bibitem[{{Fukuchi} {et~al.}(2023){Fukuchi}, {Ichikawa}, {Akiyama}, {Kimura}, {Toba}, {Inayoshi}, {Noboriguchi}, {Kawaguchi}, {Chen}, \& {Fitriana}}]{Fukuchi+2023}
{Fukuchi}, H., {Ichikawa}, K., {Akiyama}, M., {et~al.} 2023, arXiv e-prints, arXiv:2303.05605, \dodoi{10.48550/arXiv.2303.05605}

\bibitem[{{Georgakakis} {et~al.}(2017){Georgakakis}, {Salvato}, {Liu}, {Buchner}, {Brandt}, {Ananna}, {Schulze}, {Shen}, {LaMassa}, {Nandra}, {Merloni}, \& {McGreer}}]{Georgakakis+2017}
{Georgakakis}, A., {Salvato}, M., {Liu}, Z., {et~al.} 2017, \mnras, 469, 3232, \dodoi{10.1093/mnras/stx953}

\bibitem[{{Gilli} {et~al.}(2022){Gilli}, {Norman}, {Calura}, {Vito}, {Decarli}, {Marchesi}, {Iwasawa}, {Comastri}, {Lanzuisi}, {Pozzi}, {D'Amato}, {Vignali}, {Brusa}, {Mignoli}, \& {Cox}}]{Gilli+2022}
{Gilli}, R., {Norman}, C., {Calura}, F., {et~al.} 2022, \aap, 666, A17, \dodoi{10.1051/0004-6361/202243708}

\bibitem[{{Gillman} {et~al.}(2023){Gillman}, {Gullberg}, {Brammer}, {Vijayan}, {Lee}, {Bl{\'a}nquez}, {Brinch}, {Greve}, {Jermann}, {Jin}, {Kokorev}, {Liu}, {Magdis}, {Rizzo}, \& {Valentino}}]{Gillman+2023}
{Gillman}, S., {Gullberg}, B., {Brammer}, G., {et~al.} 2023, \aap, 676, A26, \dodoi{10.1051/0004-6361/202346531}

\bibitem[{{Goulding} {et~al.}(2018){Goulding}, {Zakamska}, {Alexandroff}, {Assef}, {Banerji}, {Hamann}, {Wylezalek}, {Brandt}, {Greene}, {Lansbury}, {P{\^a}ris}, {Richards}, {Stern}, \& {Strauss}}]{Goulding+2018}
{Goulding}, A.~D., {Zakamska}, N.~L., {Alexandroff}, R.~M., {et~al.} 2018, \apj, 856, 4, \dodoi{10.3847/1538-4357/aab040}

\bibitem[{{Guainazzi} \& {Bianchi}(2007)}]{Guainazzi&Bianchi+2007}
{Guainazzi}, M., \& {Bianchi}, S. 2007, \mnras, 374, 1290, \dodoi{10.1111/j.1365-2966.2006.11229.x}

\bibitem[{{Guo} {et~al.}(2021){Guo}, {Gu}, {Ding}, {Yu}, \& {Chen}}]{Guo+2021}
{Guo}, X., {Gu}, Q., {Ding}, N., {Yu}, X., \& {Chen}, Y. 2021, \apj, 908, 169, \dodoi{10.3847/1538-4357/abd0f5}

\bibitem[{{Hale} {et~al.}(2019){Hale}, {Williams}, {Jarvis}, {Hardcastle}, {Morabito}, {Shimwell}, {Tasse}, {Best}, {Harwood}, {Heywood}, {Prandoni}, {R{\"o}ttgering}, {Sabater}, {Smith}, \& {van Weeren}}]{Hale+2019}
{Hale}, C.~L., {Williams}, W., {Jarvis}, M.~J., {et~al.} 2019, \aap, 622, A4, \dodoi{10.1051/0004-6361/201833906}

\bibitem[{{Hales} {et~al.}(2014){Hales}, {Norris}, {Gaensler}, {Middelberg}, {Chow}, {Hopkins}, {Huynh}, {Lenc}, \& {Mao}}]{Hales+2014}
{Hales}, C.~A., {Norris}, R.~P., {Gaensler}, B.~M., {et~al.} 2014, \mnras, 441, 2555, \dodoi{10.1093/mnras/stu576}

\bibitem[{{Heywood} {et~al.}(2020){Heywood}, {Hale}, {Jarvis}, {Makhathini}, {Peters}, {Sebokolodi}, \& {Smirnov}}]{Heywood+2020}
{Heywood}, I., {Hale}, C.~L., {Jarvis}, M.~J., {et~al.} 2020, \mnras, 496, 3469, \dodoi{10.1093/mnras/staa1770}

\bibitem[{{Heywood} {et~al.}(2022){Heywood}, {Jarvis}, {Hale}, {Whittam}, {Bester}, {Hugo}, {Kenyon}, {Prescott}, {Smirnov}, {Tasse}, {Afonso}, {Best}, {Collier}, {Deane}, {Frank}, {Hardcastle}, {Knowles}, {Maddox}, {Murphy}, {Prandoni}, {Randriamampandry}, {Santos}, {Sekhar}, {Tabatabaei}, {Taylor}, \& {Thorat}}]{Heywood+2022}
{Heywood}, I., {Jarvis}, M.~J., {Hale}, C.~L., {et~al.} 2022, \mnras, 509, 2150, \dodoi{10.1093/mnras/stab3021}

\bibitem[{{Hopkins} {et~al.}(2006){Hopkins}, {Hernquist}, {Cox}, {Di Matteo}, {Robertson}, \& {Springel}}]{Hopkins+2006}
{Hopkins}, P.~F., {Hernquist}, L., {Cox}, T.~J., {et~al.} 2006, \apjs, 163, 1, \dodoi{10.1086/499298}

\bibitem[{{Hopkins} {et~al.}(2008){Hopkins}, {Hernquist}, {Cox}, \& {Kere{\v{s}}}}]{Hopkins+2008}
{Hopkins}, P.~F., {Hernquist}, L., {Cox}, T.~J., \& {Kere{\v{s}}}, D. 2008, \apjs, 175, 356, \dodoi{10.1086/524362}

\bibitem[{{Hou} {et~al.}(2009){Hou}, {Parker}, {Harris}, \& {Wilman}}]{Hou+2009}
{Hou}, A., {Parker}, L.~C., {Harris}, W.~E., \& {Wilman}, D.~J. 2009, \apj, 702, 1199, \dodoi{10.1088/0004-637X/702/2/1199}

\bibitem[{{Hudelot} {et~al.}(2012){Hudelot}, {Cuillandre}, {Withington}, {Goranova}, {McCracken}, {Magnard}, {Mellier}, {Regnault}, {Betoule}, {Aussel}, {Kavelaars}, {Fernique}, {Bonnarel}, {Ochsenbein}, \& {Ilbert}}]{Hudelot+2012}
{Hudelot}, P., {Cuillandre}, J.~C., {Withington}, K., {et~al.} 2012, {VizieR Online Data Catalog: The CFHTLS Survey (T0007 release) (Hudelot+ 2012)}, VizieR On-line Data Catalog: II/317. Originally published in: SPIE Conf. 2012

\bibitem[{{Inoue}(2011)}]{Inoue+2011}
{Inoue}, A.~K. 2011, \mnras, 415, 2920, \dodoi{10.1111/j.1365-2966.2011.18906.x}

\bibitem[{{Ivezi{\'c}} {et~al.}(2019){Ivezi{\'c}}, {Kahn}, {Tyson}, {Abel}, {Acosta}, {Allsman}, {Alonso}, {AlSayyad}, {Anderson}, {Andrew}, {Angel}, {Angeli}, {Ansari}, {Antilogus}, {Araujo}, {Armstrong}, {Arndt}, {Astier}, {Aubourg}, {Auza}, {Axelrod}, {Bard}, {Barr}, {Barrau}, {Bartlett}, {Bauer}, {Bauman}, {Baumont}, {Bechtol}, {Bechtol}, {Becker}, {Becla}, {Beldica}, {Bellavia}, {Bianco}, {Biswas}, {Blanc}, {Blazek}, {Blandford}, {Bloom}, {Bogart}, {Bond}, {Booth}, {Borgland}, {Borne}, {Bosch}, {Boutigny}, {Brackett}, {Bradshaw}, {Brandt}, {Brown}, {Bullock}, {Burchat}, {Burke}, {Cagnoli}, {Calabrese}, {Callahan}, {Callen}, {Carlin}, {Carlson}, {Chandrasekharan}, {Charles-Emerson}, {Chesley}, {Cheu}, {Chiang}, {Chiang}, {Chirino}, {Chow}, {Ciardi}, {Claver}, {Cohen-Tanugi}, {Cockrum}, {Coles}, {Connolly}, {Cook}, {Cooray}, {Covey}, {Cribbs}, {Cui}, {Cutri}, {Daly}, {Daniel}, {Daruich}, {Daubard}, {Daues}, {Dawson}, {Delgado}, {Dellapenna}, {de Peyster}, {de Val-Borro}, {Digel}, {Doherty}, {Dubois},
  {Dubois-Felsmann}, {Durech}, {Economou}, {Eifler}, {Eracleous}, {Emmons}, {Fausti Neto}, {Ferguson}, {Figueroa}, {Fisher-Levine}, {Focke}, {Foss}, {Frank}, {Freemon}, {Gangler}, {Gawiser}, {Geary}, {Gee}, {Geha}, {Gessner}, {Gibson}, {Gilmore}, {Glanzman}, {Glick}, {Goldina}, {Goldstein}, {Goodenow}, {Graham}, {Gressler}, {Gris}, {Guy}, {Guyonnet}, {Haller}, {Harris}, {Hascall}, {Haupt}, {Hernandez}, {Herrmann}, {Hileman}, {Hoblitt}, {Hodgson}, {Hogan}, {Howard}, {Huang}, {Huffer}, {Ingraham}, {Innes}, {Jacoby}, {Jain}, {Jammes}, {Jee}, {Jenness}, {Jernigan}, {Jevremovi{\'c}}, {Johns}, {Johnson}, {Johnson}, {Jones}, {Juramy-Gilles}, {Juri{\'c}}, {Kalirai}, {Kallivayalil}, {Kalmbach}, {Kantor}, {Karst}, {Kasliwal}, {Kelly}, {Kessler}, {Kinnison}, {Kirkby}, {Knox}, {Kotov}, {Krabbendam}, {Krughoff}, {Kub{\'a}nek}, {Kuczewski}, {Kulkarni}, {Ku}, {Kurita}, {Lage}, {Lambert}, {Lange}, {Langton}, {Le Guillou}, {Levine}, {Liang}, {Lim}, {Lintott}, {Long}, {Lopez}, {Lotz}, {Lupton}, {Lust}, {MacArthur}, {Mahabal},
  {Mandelbaum}, {Markiewicz}, {Marsh}, {Marshall}, {Marshall}, {May}, {McKercher}, {McQueen}, {Meyers}, {Migliore}, {Miller}, {Mills}, {Miraval}, {Moeyens}, {Moolekamp}, {Monet}, {Moniez}, {Monkewitz}, {Montgomery}, {Morrison}, {Mueller}, {Muller}, {Mu{\~n}oz Arancibia}, {Neill}, {Newbry}, {Nief}, {Nomerotski}, {Nordby}, {O'Connor}, {Oliver}, {Olivier}, {Olsen}, {O'Mullane}, {Ortiz}, {Osier}, {Owen}, {Pain}, {Palecek}, {Parejko}, {Parsons}, {Pease}, {Peterson}, {Peterson}, {Petravick}, {Libby Petrick}, {Petry}, {Pierfederici}, {Pietrowicz}, {Pike}, {Pinto}, {Plante}, {Plate}, {Plutchak}, {Price}, {Prouza}, {Radeka}, {Rajagopal}, {Rasmussen}, {Regnault}, {Reil}, {Reiss}, {Reuter}, {Ridgway}, {Riot}, {Ritz}, {Robinson}, {Roby}, {Roodman}, {Rosing}, {Roucelle}, {Rumore}, {Russo}, {Saha}, {Sassolas}, {Schalk}, {Schellart}, {Schindler}, {Schmidt}, {Schneider}, {Schneider}, {Schoening}, {Schumacher}, {Schwamb}, {Sebag}, {Selvy}, {Sembroski}, {Seppala}, {Serio}, {Serrano}, {Shaw}, {Shipsey}, {Sick}, {Silvestri},
  {Slater}, {Smith}, {Smith}, {Sobhani}, {Soldahl}, {Storrie-Lombardi}, {Stover}, {Strauss}, {Street}, {Stubbs}, {Sullivan}, {Sweeney}, {Swinbank}, {Szalay}, {Takacs}, {Tether}, {Thaler}, {Thayer}, {Thomas}, {Thornton}, {Thukral}, {Tice}, {Trilling}, {Turri}, {Van Berg}, {Vanden Berk}, {Vetter}, {Virieux}, {Vucina}, {Wahl}, {Walkowicz}, {Walsh}, {Walter}, {Wang}, {Wang}, {Warner}, {Wiecha}, {Willman}, {Winters}, {Wittman}, {Wolff}, {Wood-Vasey}, {Wu}, {Xin}, {Yoachim}, \& {Zhan}}]{Ivezic+2019}
{Ivezi{\'c}}, {\v{Z}}., {Kahn}, S.~M., {Tyson}, J.~A., {et~al.} 2019, \apj, 873, 111, \dodoi{10.3847/1538-4357/ab042c}

\bibitem[{{Jarvis} {et~al.}(2013){Jarvis}, {Bonfield}, {Bruce}, {Geach}, {McAlpine}, {McLure}, {Gonz{\'a}lez-Solares}, {Irwin}, {Lewis}, {Yoldas}, {Andreon}, {Cross}, {Emerson}, {Dalton}, {Dunlop}, {Hodgkin}, {Le}, {Karouzos}, {Meisenheimer}, {Oliver}, {Rawlings}, {Simpson}, {Smail}, {Smith}, {Sullivan}, {Sutherland}, {White}, \& {Zwart}}]{Jarvis+2013}
{Jarvis}, M.~J., {Bonfield}, D.~G., {Bruce}, V.~A., {et~al.} 2013, \mnras, 428, 1281, \dodoi{10.1093/mnras/sts118}

\bibitem[{{Johnson} {et~al.}(2021){Johnson}, {Leja}, {Conroy}, \& {Speagle}}]{Johnson+2021}
{Johnson}, B.~D., {Leja}, J., {Conroy}, C., \& {Speagle}, J.~S. 2021, \apjs, 254, 22, \dodoi{10.3847/1538-4365/abef67}

\bibitem[{{Kayal} \& {Singh}(2024)}]{Kayal+2024}
{Kayal}, A., \& {Singh}, V. 2024, \mnras, 531, 830, \dodoi{10.1093/mnras/stae1191}

\bibitem[{{Kellermann} {et~al.}(1994){Kellermann}, {Sramek}, {Schmidt}, {Green}, \& {Shaffer}}]{Kellermann+1994}
{Kellermann}, K.~I., {Sramek}, R.~A., {Schmidt}, M., {Green}, R.~F., \& {Shaffer}, D.~B. 1994, \aj, 108, 1163, \dodoi{10.1086/117145}

\bibitem[{{Kim} \& {Fabbiano}(2015)}]{Kim&Fabbiano+2015}
{Kim}, D.-W., \& {Fabbiano}, G. 2015, \apj, 812, 127, \dodoi{10.1088/0004-637X/812/2/127}

\bibitem[{{Lacy} {et~al.}(2021){Lacy}, {Surace}, {Farrah}, {Nyland}, {Afonso}, {Brandt}, {Clements}, {Lagos}, {Maraston}, {Pforr}, {Sajina}, {Sako}, {Vaccari}, {Wilson}, {Ballantyne}, {Barkhouse}, {Brunner}, {Cane}, {Clarke}, {Cooper}, {Cooray}, {Covone}, {D'Andrea}, {Evrard}, {Ferguson}, {Frieman}, {Gonzalez-Perez}, {Gupta}, {Hatziminaoglou}, {Huang}, {Jagannathan}, {Jarvis}, {Jones}, {Kimball}, {Lidman}, {Lubin}, {Marchetti}, {Martini}, {McMahon}, {Mei}, {Messias}, {Murphy}, {Newman}, {Nichol}, {Norris}, {Oliver}, {Perez-Fournon}, {Peters}, {Pierre}, {Polisensky}, {Richards}, {Ridgway}, {R{\"o}ttgering}, {Seymour}, {Shirley}, {Somerville}, {Strauss}, {Suntzeff}, {Thorman}, {van Kampen}, {Verma}, {Wechsler}, \& {Wood-Vasey}}]{Lacy+2021}
{Lacy}, M., {Surace}, J.~A., {Farrah}, D., {et~al.} 2021, \mnras, 501, 892, \dodoi{10.1093/mnras/staa3714}

\bibitem[{{Lansbury} {et~al.}(2020){Lansbury}, {Banerji}, {Fabian}, \& {Temple}}]{Lansbury+2020}
{Lansbury}, G.~B., {Banerji}, M., {Fabian}, A.~C., \& {Temple}, M.~J. 2020, \mnras, 495, 2652, \dodoi{10.1093/mnras/staa1220}

\bibitem[{{Lanzuisi} {et~al.}(2009){Lanzuisi}, {Piconcelli}, {Fiore}, {Feruglio}, {Vignali}, {Salvato}, \& {Gruppioni}}]{Lanzuisi+2009}
{Lanzuisi}, G., {Piconcelli}, E., {Fiore}, F., {et~al.} 2009, \aap, 498, 67, \dodoi{10.1051/0004-6361/200811282}

\bibitem[{{Lanzuisi} {et~al.}(2017){Lanzuisi}, {Delvecchio}, {Berta}, {Brusa}, {Comastri}, {Gilli}, {Gruppioni}, {Marchesi}, {Perna}, {Pozzi}, {Salvato}, {Symeonidis}, {Vignali}, {Vito}, {Volonteri}, \& {Zamorani}}]{Lanzuisi+2017}
{Lanzuisi}, G., {Delvecchio}, I., {Berta}, S., {et~al.} 2017, \aap, 602, A123, \dodoi{10.1051/0004-6361/201629955}

\bibitem[{{Lanzuisi} {et~al.}(2018){Lanzuisi}, {Civano}, {Marchesi}, {Comastri}, {Brusa}, {Gilli}, {Vignali}, {Zamorani}, {Brightman}, {Griffiths}, \& {Koekemoer}}]{Lanzuisi+2018}
{Lanzuisi}, G., {Civano}, F., {Marchesi}, S., {et~al.} 2018, \mnras, 480, 2578, \dodoi{10.1093/mnras/sty2025}

\bibitem[{{Le Bail} {et~al.}(2024){Le Bail}, {Daddi}, {Elbaz}, {Dickinson}, {Giavalisco}, {Magnelli}, {G{\'o}mez-Guijarro}, {Kalita}, {Koekemoer}, {Holwerda}, {Bournaud}, {de la Vega}, {Calabr{\`o}}, {Dekel}, {Cheng}, {Bisigello}, {Franco}, {Costantin}, {Lucas}, {P{\'e}rez-Gonz{\'a}lez}, {Lu}, {Wilkins}, {Arrabal Haro}, {Bagley}, {Finkelstein}, {Kartaltepe}, {Papovich}, {Pirzkal}, \& {Yung}}]{LeBail+2024}
{Le Bail}, A., {Daddi}, E., {Elbaz}, D., {et~al.} 2024, \aap, 688, A53, \dodoi{10.1051/0004-6361/202347465}

\bibitem[{{Lee} {et~al.}(2018){Lee}, {Giavalisco}, {Whitaker}, {Williams}, {Ferguson}, {Acquaviva}, {Koekemoer}, {Straughn}, {Guo}, {Kartaltepe}, {Lotz}, {Pacifici}, {Croton}, {Somerville}, \& {Lu}}]{Lee+2018}
{Lee}, B., {Giavalisco}, M., {Whitaker}, K., {et~al.} 2018, \apj, 853, 131, \dodoi{10.3847/1538-4357/aaa40f}

\bibitem[{{Lehmer} {et~al.}(2016){Lehmer}, {Basu-Zych}, {Mineo}, {Brandt}, {Eufrasio}, {Fragos}, {Hornschemeier}, {Luo}, {Xue}, {Bauer}, {Gilfanov}, {Ranalli}, {Schneider}, {Shemmer}, {Tozzi}, {Trump}, {Vignali}, {Wang}, {Yukita}, \& {Zezas}}]{Lehmer+2016}
{Lehmer}, B.~D., {Basu-Zych}, A.~R., {Mineo}, S., {et~al.} 2016, \apj, 825, 7, \dodoi{10.3847/0004-637X/825/1/7}

\bibitem[{{Leja} {et~al.}(2017){Leja}, {Johnson}, {Conroy}, {van Dokkum}, \& {Byler}}]{Leja+2017}
{Leja}, J., {Johnson}, B.~D., {Conroy}, C., {van Dokkum}, P.~G., \& {Byler}, N. 2017, \apj, 837, 170, \dodoi{10.3847/1538-4357/aa5ffe}

\bibitem[{{Leja} {et~al.}(2019){Leja}, {Johnson}, {Conroy}, {van Dokkum}, {Speagle}, {Brammer}, {Momcheva}, {Skelton}, {Whitaker}, {Franx}, \& {Nelson}}]{Leja+2019b}
{Leja}, J., {Johnson}, B.~D., {Conroy}, C., {et~al.} 2019, \apj, 877, 140, \dodoi{10.3847/1538-4357/ab1d5a}

\bibitem[{{Li} {et~al.}(2024){Li}, {Assef}, {Tsai}, {Wu}, {Eisenhardt}, {Stern}, {D{\'\i}az-Santos}, {Blain}, {Jun}, {Fern{\'a}ndez Arand{\'a}}, \& {Zewdie}}]{Li+2024}
{Li}, G., {Assef}, R.~J., {Tsai}, C.-W., {et~al.} 2024, arXiv e-prints, arXiv:2405.20479, \dodoi{10.48550/arXiv.2405.20479}

\bibitem[{{Li} {et~al.}(2020){Li}, {Xue}, {Sun}, {Brandt}, {Yang}, {Vito}, {Tozzi}, {Vignali}, {Comastri}, {Shu}, {Fang}, {Fan}, {Luo}, {Chen}, \& {Zheng}}]{Li+2020}
{Li}, J., {Xue}, Y., {Sun}, M., {et~al.} 2020, \apj, 903, 49, \dodoi{10.3847/1538-4357/abb6e7}

\bibitem[{{Liu} {et~al.}(2017){Liu}, {Tozzi}, {Wang}, {Brandt}, {Vignali}, {Xue}, {Schneider}, {Comastri}, {Yang}, {Bauer}, {Paolillo}, {Luo}, {Gilli}, {Wang}, {Giavalisco}, {Ji}, {Alexander}, {Mainieri}, {Shemmer}, {Koekemoer}, \& {Risaliti}}]{Liu+2017}
{Liu}, T., {Tozzi}, P., {Wang}, J.-X., {et~al.} 2017, \apjs, 232, 8, \dodoi{10.3847/1538-4365/aa7847}

\bibitem[{{Lonsdale} {et~al.}(2004){Lonsdale}, {Polletta}, {Surace}, {Shupe}, {Fang}, {Xu}, {Smith}, {Siana}, {Rowan-Robinson}, {Babbedge}, {Oliver}, {Pozzi}, {Davoodi}, {Owen}, {Padgett}, {Frayer}, {Jarrett}, {Masci}, {O'Linger}, {Conrow}, {Farrah}, {Morrison}, {Gautier}, {Franceschini}, {Berta}, {Perez-Fournon}, {Hatziminaoglou}, {Afonso-Luis}, {Dole}, {Stacey}, {Serjeant}, {Pierre}, {Griffin}, \& {Puetter}}]{Lonsdale+2004}
{Lonsdale}, C., {Polletta}, M. d.~C., {Surace}, J., {et~al.} 2004, \apjs, 154, 54, \dodoi{10.1086/423206}

\bibitem[{{Lonsdale} {et~al.}(2003){Lonsdale}, {Smith}, {Rowan-Robinson}, {Surace}, {Shupe}, {Xu}, {Oliver}, {Padgett}, {Fang}, {Conrow}, {Franceschini}, {Gautier}, {Griffin}, {Hacking}, {Masci}, {Morrison}, {O'Linger}, {Owen}, {P{\'e}rez-Fournon}, {Pierre}, {Puetter}, {Stacey}, {Castro}, {Polletta}, {Farrah}, {Jarrett}, {Frayer}, {Siana}, {Babbedge}, {Dye}, {Fox}, {Gonzalez-Solares}, {Salaman}, {Berta}, {Condon}, {Dole}, \& {Serjeant}}]{Lonsdale+2003}
{Lonsdale}, C.~J., {Smith}, H.~E., {Rowan-Robinson}, M., {et~al.} 2003, \pasp, 115, 897, \dodoi{10.1086/376850}

\bibitem[{{Lower} {et~al.}(2020){Lower}, {Narayanan}, {Leja}, {Johnson}, {Conroy}, \& {Dav{\'e}}}]{Lower+2020}
{Lower}, S., {Narayanan}, D., {Leja}, J., {et~al.} 2020, \apj, 904, 33, \dodoi{10.3847/1538-4357/abbfa7}

\bibitem[{{Luo} {et~al.}(2017){Luo}, {Brandt}, {Xue}, {Lehmer}, {Alexander}, {Bauer}, {Vito}, {Yang}, {Basu-Zych}, {Comastri}, {Gilli}, {Gu}, {Hornschemeier}, {Koekemoer}, {Liu}, {Mainieri}, {Paolillo}, {Ranalli}, {Rosati}, {Schneider}, {Shemmer}, {Smail}, {Sun}, {Tozzi}, {Vignali}, \& {Wang}}]{Luo+2017}
{Luo}, B., {Brandt}, W.~N., {Xue}, Y.~Q., {et~al.} 2017, \apjs, 228, 2, \dodoi{10.3847/1538-4365/228/1/2}

\bibitem[{{Martin} {et~al.}(2005){Martin}, {Fanson}, {Schiminovich}, {Morrissey}, {Friedman}, {Barlow}, {Conrow}, {Grange}, {Jelinsky}, {Milliard}, {Siegmund}, {Bianchi}, {Byun}, {Donas}, {Forster}, {Heckman}, {Lee}, {Madore}, {Malina}, {Neff}, {Rich}, {Small}, {Surber}, {Szalay}, {Welsh}, \& {Wyder}}]{Martin+2005}
{Martin}, D.~C., {Fanson}, J., {Schiminovich}, D., {et~al.} 2005, \apjl, 619, L1, \dodoi{10.1086/426387}

\bibitem[{{Martocchia} {et~al.}(2017){Martocchia}, {Piconcelli}, {Zappacosta}, {Duras}, {Vietri}, {Vignali}, {Bianchi}, {Bischetti}, {Bongiorno}, {Brusa}, {Lanzuisi}, {Marconi}, {Mathur}, {Miniutti}, {Nicastro}, {Bruni}, \& {Fiore}}]{Martocchia+2017}
{Martocchia}, S., {Piconcelli}, E., {Zappacosta}, L., {et~al.} 2017, \aap, 608, A51, \dodoi{10.1051/0004-6361/201731314}

\bibitem[{{McConnell} {et~al.}(2020){McConnell}, {Hale}, {Lenc}, {Banfield}, {Heald}, {Hotan}, {Leung}, {Moss}, {Murphy}, {O'Brien}, {Pritchard}, {Raja}, {Sadler}, {Stewart}, {Thomson}, {Whiting}, {Allison}, {Amy}, {Anderson}, {Ball}, {Bannister}, {Bell}, {Bock}, {Bolton}, {Bunton}, {Chippendale}, {Collier}, {Cooray}, {Cornwell}, {Diamond}, {Edwards}, {Gupta}, {Hayman}, {Heywood}, {Jackson}, {Koribalski}, {Lee-Waddell}, {McClure-Griffiths}, {Ng}, {Norris}, {Phillips}, {Reynolds}, {Roxby}, {Schinckel}, {Shields}, {Tremblay}, {Tzioumis}, {Voronkov}, \& {Westmeier}}]{McConnell+2020}
{McConnell}, D., {Hale}, C.~L., {Lenc}, E., {et~al.} 2020, \pasa, 37, e048, \dodoi{10.1017/pasa.2020.41}

\bibitem[{{Melbourne} {et~al.}(2012){Melbourne}, {Soifer}, {Desai}, {Pope}, {Armus}, {Dey}, {Bussmann}, {Jannuzi}, \& {Alberts}}]{Melbourne+2012}
{Melbourne}, J., {Soifer}, B.~T., {Desai}, V., {et~al.} 2012, \aj, 143, 125, \dodoi{10.1088/0004-6256/143/5/125}

\bibitem[{{Merloni} {et~al.}(2014){Merloni}, {Bongiorno}, {Brusa}, {Iwasawa}, {Mainieri}, {Magnelli}, {Salvato}, {Berta}, {Cappelluti}, {Comastri}, {Fiore}, {Gilli}, {Koekemoer}, {Le Floc'h}, {Lusso}, {Lutz}, {Miyaji}, {Pozzi}, {Riguccini}, {Rosario}, {Silverman}, {Symeonidis}, {Treister}, {Vignali}, \& {Zamorani}}]{Merloni+2014}
{Merloni}, A., {Bongiorno}, A., {Brusa}, M., {et~al.} 2014, \mnras, 437, 3550, \dodoi{10.1093/mnras/stt2149}

\bibitem[{{Mountrichas} {et~al.}(2021){Mountrichas}, {Buat}, {Yang}, {Boquien}, {Burgarella}, {Ciesla}, {Malek}, \& {Shirley}}]{Mountrichas+2021}
{Mountrichas}, G., {Buat}, V., {Yang}, G., {et~al.} 2021, \aap, 653, A74, \dodoi{10.1051/0004-6361/202140630}

\bibitem[{{Mountrichas} {et~al.}(2024){Mountrichas}, {Masoura}, {Corral}, \& {Carrera}}]{Mountrichas+2024}
{Mountrichas}, G., {Masoura}, V.~A., {Corral}, A., \& {Carrera}, F.~J. 2024, \aap, 683, A143, \dodoi{10.1051/0004-6361/202348952}

\bibitem[{{Mountrichas} {et~al.}(2017){Mountrichas}, {Georgantopoulos}, {Secrest}, {Ordov{\'a}s-Pascual}, {Corral}, {Akylas}, {Mateos}, {Carrera}, \& {Batziou}}]{Mountrichas+2017}
{Mountrichas}, G., {Georgantopoulos}, I., {Secrest}, N.~J., {et~al.} 2017, \mnras, 468, 3042, \dodoi{10.1093/mnras/stx572}

\bibitem[{{Murphy} \& {Yaqoob}(2009)}]{Yaqoob&Murphy+2011}
{Murphy}, K.~D., \& {Yaqoob}, T. 2009, \mnras, 397, 1549, \dodoi{10.1111/j.1365-2966.2009.15025.x}

\bibitem[{{Narayanan} {et~al.}(2010){Narayanan}, {Dey}, {Hayward}, {Cox}, {Bussmann}, {Brodwin}, {Jonsson}, {Hopkins}, {Groves}, {Younger}, \& {Hernquist}}]{Narayanan+2010}
{Narayanan}, D., {Dey}, A., {Hayward}, C.~C., {et~al.} 2010, \mnras, 407, 1701, \dodoi{10.1111/j.1365-2966.2010.16997.x}

\bibitem[{{Netzer}(2015)}]{Netzer+2015}
{Netzer}, H. 2015, \araa, 53, 365, \dodoi{10.1146/annurev-astro-082214-122302}

\bibitem[{{Netzer} {et~al.}(2016){Netzer}, {Lani}, {Nordon}, {Trakhtenbrot}, {Lira}, \& {Shemmer}}]{Netzer+2016}
{Netzer}, H., {Lani}, C., {Nordon}, R., {et~al.} 2016, \apj, 819, 123, \dodoi{10.3847/0004-637X/819/2/123}

\bibitem[{{Ni} {et~al.}(2019){Ni}, {Timlin}, {Brandt}, \& {Yang}}]{Ni+2019}
{Ni}, Q., {Timlin}, J., {Brandt}, W.~N., \& {Yang}, G. 2019, Research Notes of the American Astronomical Society, 3, 5, \dodoi{10.3847/2515-5172/aaf8af}

\bibitem[{Ni {et~al.}(2021)Ni, Brandt, Chen, Luo, Nyland, Yang, Zou, Aird, Alexander, Bauer, Lacy, Lehmer, Mallick, Salvato, Schneider, Tozzi, Traulsen, Vaccari, Vignali, Vito, Xue, Banerji, Chow, Comastri, Moro, Gilli, Mullaney, Paolillo, Schwope, Shemmer, Sun, III, \& Trump}]{Ni+2021}
Ni, Q., Brandt, W.~N., Chen, C.-T., {et~al.} 2021, \apj, 256, 21, \dodoi{10.3847/1538-4365/ac0dc6}

\bibitem[{{Noboriguchi} {et~al.}(2019){Noboriguchi}, {Nagao}, {Toba}, {Niida}, {Kajisawa}, {Onoue}, {Matsuoka}, {Yamashita}, {Chang}, {Kawaguchi}, {Komiyama}, {Nobuhara}, {Terashima}, \& {Ueda}}]{Noboriguchi+2019}
{Noboriguchi}, A., {Nagao}, T., {Toba}, Y., {et~al.} 2019, \apj, 876, 132, \dodoi{10.3847/1538-4357/ab1754}

\bibitem[{{Noboriguchi} {et~al.}(2022){Noboriguchi}, {Nagao}, {Toba}, {Ichikawa}, {Kajisawa}, {Kato}, {Kawaguchi}, {Matsuhara}, {Matsuoka}, {Onishi}, {Onoue}, {Tamada}, {Terao}, {Terashima}, {Ueda}, \& {Yamashita}}]{Noboriguchi+2022}
---. 2022, \apj, 941, 195, \dodoi{10.3847/1538-4357/aca403}

\bibitem[{{Norris} {et~al.}(2006){Norris}, {Afonso}, {Appleton}, {Boyle}, {Ciliegi}, {Croom}, {Huynh}, {Jackson}, {Koekemoer}, {Lonsdale}, {Middelberg}, {Mobasher}, {Oliver}, {Polletta}, {Siana}, {Smail}, \& {Voronkov}}]{Norris+2006}
{Norris}, R.~P., {Afonso}, J., {Appleton}, P.~N., {et~al.} 2006, \aj, 132, 2409, \dodoi{10.1086/508275}

\bibitem[{{Nyland} {et~al.}(2023){Nyland}, {Lacy}, {Brandt}, {Yang}, {Ni}, {Sajina}, {Zou}, \& {Vaccari}}]{Nyland+2023}
{Nyland}, K., {Lacy}, M., {Brandt}, W.~N., {et~al.} 2023, Research Notes of the American Astronomical Society, 7, 33, \dodoi{10.3847/2515-5172/acbc72}

\bibitem[{{Nyland} {et~al.}(2017){Nyland}, {Lacy}, {Sajina}, {Pforr}, {Farrah}, {Wilson}, {Surace}, {H{\"a}u{\ss}ler}, {Vaccari}, \& {Jarvis}}]{Nyland+2017}
{Nyland}, K., {Lacy}, M., {Sajina}, A., {et~al.} 2017, \apjs, 230, 9, \dodoi{10.3847/1538-4365/aa6fed}

\bibitem[{{Oliver} {et~al.}(2012){Oliver}, {Bock}, {Altieri}, {Amblard}, {Arumugam}, {Aussel}, {Babbedge}, {Beelen}, {B{\'e}thermin}, {Blain}, {Boselli}, {Bridge}, {Brisbin}, {Buat}, {Burgarella}, {Castro-Rodr{\'\i}guez}, {Cava}, {Chanial}, {Cirasuolo}, {Clements}, {Conley}, {Conversi}, {Cooray}, {Dowell}, {Dubois}, {Dwek}, {Dye}, {Eales}, {Elbaz}, {Farrah}, {Feltre}, {Ferrero}, {Fiolet}, {Fox}, {Franceschini}, {Gear}, {Giovannoli}, {Glenn}, {Gong}, {Gonz{\'a}lez Solares}, {Griffin}, {Halpern}, {Harwit}, {Hatziminaoglou}, {Heinis}, {Hurley}, {Hwang}, {Hyde}, {Ibar}, {Ilbert}, {Isaak}, {Ivison}, {Lagache}, {Le Floc'h}, {Levenson}, {Faro}, {Lu}, {Madden}, {Maffei}, {Magdis}, {Mainetti}, {Marchetti}, {Marsden}, {Marshall}, {Mortier}, {Nguyen}, {O'Halloran}, {Omont}, {Page}, {Panuzzo}, {Papageorgiou}, {Patel}, {Pearson}, {P{\'e}rez-Fournon}, {Pohlen}, {Rawlings}, {Raymond}, {Rigopoulou}, {Riguccini}, {Rizzo}, {Rodighiero}, {Roseboom}, {Rowan-Robinson}, {S{\'a}nchez Portal}, {Schulz}, {Scott}, {Seymour}, {Shupe},
  {Smith}, {Stevens}, {Symeonidis}, {Trichas}, {Tugwell}, {Vaccari}, {Valtchanov}, {Vieira}, {Viero}, {Vigroux}, {Wang}, {Ward}, {Wardlow}, {Wright}, {Xu}, \& {Zemcov}}]{Oliver+2012}
{Oliver}, S.~J., {Bock}, J., {Altieri}, B., {et~al.} 2012, \mnras, 424, 1614, \dodoi{10.1111/j.1365-2966.2012.20912.x}

\bibitem[{{P{\'e}rez-Gonz{\'a}lez} {et~al.}(2005){P{\'e}rez-Gonz{\'a}lez}, {Rieke}, {Egami}, {Alonso-Herrero}, {Dole}, {Papovich}, {Blaylock}, {Jones}, {Rieke}, {Rigby}, {Barmby}, {Fazio}, {Huang}, \& {Martin}}]{Perez-Gonzalez+2005}
{P{\'e}rez-Gonz{\'a}lez}, P.~G., {Rieke}, G.~H., {Egami}, E., {et~al.} 2005, \apj, 630, 82, \dodoi{10.1086/431894}

\bibitem[{{Polletta} {et~al.}(2006){Polletta}, {Wilkes}, {Siana}, {Lonsdale}, {Kilgard}, {Smith}, {Kim}, {Owen}, {Efstathiou}, {Jarrett}, {Stacey}, {Franceschini}, {Rowan-Robinson}, {Babbedge}, {Berta}, {Fang}, {Farrah}, {Gonz{\'a}lez-Solares}, {Morrison}, {Surace}, \& {Shupe}}]{Polletta+2006}
{Polletta}, M. d.~C., {Wilkes}, B.~J., {Siana}, B., {et~al.} 2006, \apj, 642, 673, \dodoi{10.1086/500821}

\bibitem[{{Poulain} {et~al.}(2020){Poulain}, {Paolillo}, {De Cicco}, {Brandt}, {Bauer}, {Falocco}, {Vagnetti}, {Grado}, {Ragosta}, {Botticella}, {Cappellaro}, {Pignata}, {Vaccari}, {Schipani}, {Covone}, {Longo}, \& {Napolitano}}]{Poulain+2020}
{Poulain}, M., {Paolillo}, M., {De Cicco}, D., {et~al.} 2020, \aap, 634, A50, \dodoi{10.1051/0004-6361/201937108}

\bibitem[{{Pyrzas} {et~al.}(2015){Pyrzas}, {Steenbrugge}, \& {Blundell}}]{Pyrzas+2015}
{Pyrzas}, S., {Steenbrugge}, K.~C., \& {Blundell}, K.~M. 2015, \aap, 574, A30, \dodoi{10.1051/0004-6361/201425061}

\bibitem[{{Ricci} {et~al.}(2017){Ricci}, {Assef}, {Stern}, {Nikutta}, {Alexander}, {Asmus}, {Ballantyne}, {Bauer}, {Blain}, {Boggs}, {Boorman}, {Brandt}, {Brightman}, {Chang}, {Chen}, {Christensen}, {Comastri}, {Craig}, {D{\'\i}az-Santos}, {Eisenhardt}, {Farrah}, {Gandhi}, {Hailey}, {Harrison}, {Jun}, {Koss}, {LaMassa}, {Lansbury}, {Markwardt}, {Stalevski}, {Stanley}, {Treister}, {Tsai}, {Walton}, {Wu}, {Zappacosta}, \& {Zhang}}]{Ricci+2017}
{Ricci}, C., {Assef}, R.~J., {Stern}, D., {et~al.} 2017, \apj, 835, 105, \dodoi{10.3847/1538-4357/835/1/105}

\bibitem[{{Riguccini} {et~al.}(2019){Riguccini}, {Treister}, {Men{\'e}ndez-Delmestre}, {Cardamone}, {Civano}, {Gon{\c{c}}alves}, {Hasinger}, {Koekemoer}, {Lanzuisi}, {Le Floc'h}, {Lusso}, {Lutz}, {Marchesi}, {Miyaji}, {Pozzi}, {Ricci}, {Rodighiero}, {Salvato}, {Sanders}, {Schawinski}, \& {Suh}}]{Riguccini+2019}
{Riguccini}, L.~A., {Treister}, E., {Men{\'e}ndez-Delmestre}, K., {et~al.} 2019, \aj, 157, 233, \dodoi{10.3847/1538-3881/ab16cd}

\bibitem[{{Rovilos} {et~al.}(2014){Rovilos}, {Georgantopoulos}, {Akylas}, {Aird}, {Alexander}, {Comastri}, {Del Moro}, {Gandhi}, {Georgakakis}, {Harrison}, \& {Mullaney}}]{Rovilos+2014}
{Rovilos}, E., {Georgantopoulos}, I., {Akylas}, A., {et~al.} 2014, \mnras, 438, 494, \dodoi{10.1093/mnras/stt2228}

\bibitem[{{Ruiz} {et~al.}(2022){Ruiz}, {Georgakakis}, {Gerakakis}, {Saxton}, {Kretschmar}, {Akylas}, \& {Georgantopoulos}}]{Ruiz+2022}
{Ruiz}, A., {Georgakakis}, A., {Gerakakis}, S., {et~al.} 2022, \mnras, 511, 4265, \dodoi{10.1093/mnras/stac272}

\bibitem[{{Sanders} {et~al.}(1988){Sanders}, {Soifer}, {Elias}, {Madore}, {Matthews}, {Neugebauer}, \& {Scoville}}]{Sanders+1988}
{Sanders}, D.~B., {Soifer}, B.~T., {Elias}, J.~H., {et~al.} 1988, \apj, 325, 74, \dodoi{10.1086/165983}

\bibitem[{{Sargent} {et~al.}(2010){Sargent}, {Schinnerer}, {Murphy}, {Carilli}, {Helou}, {Aussel}, {Le Floc'h}, {Frayer}, {Ilbert}, {Oesch}, {Salvato}, {Smol{\v{c}}i{\'c}}, {Kartaltepe}, \& {Sanders}}]{Sargent+2010}
{Sargent}, M.~T., {Schinnerer}, E., {Murphy}, E., {et~al.} 2010, \apjl, 714, L190, \dodoi{10.1088/2041-8205/714/2/L190}

\bibitem[{{Schartmann} {et~al.}(2005){Schartmann}, {Meisenheimer}, {Camenzind}, {Wolf}, \& {Henning}}]{Schartmann+2005}
{Schartmann}, M., {Meisenheimer}, K., {Camenzind}, M., {Wolf}, S., \& {Henning}, T. 2005, \aap, 437, 861, \dodoi{10.1051/0004-6361:20042363}

\bibitem[{{Scott} {et~al.}(2011){Scott}, {Stewart}, {Mateos}, {Alexander}, {Hutton}, \& {Ward}}]{Scott+2011}
{Scott}, A.~E., {Stewart}, G.~C., {Mateos}, S., {et~al.} 2011, \mnras, 417, 992, \dodoi{10.1111/j.1365-2966.2011.19325.x}

\bibitem[{{Stalevski} {et~al.}(2012){Stalevski}, {Fritz}, {Baes}, {Nakos}, \& {Popovi{\'c}}}]{Stalevski+2012}
{Stalevski}, M., {Fritz}, J., {Baes}, M., {Nakos}, T., \& {Popovi{\'c}}, L.~{\v{C}}. 2012, \mnras, 420, 2756, \dodoi{10.1111/j.1365-2966.2011.19775.x}

\bibitem[{{Stalevski} {et~al.}(2016){Stalevski}, {Ricci}, {Ueda}, {Lira}, {Fritz}, \& {Baes}}]{Stalevski+2016}
{Stalevski}, M., {Ricci}, C., {Ueda}, Y., {et~al.} 2016, \mnras, 458, 2288, \dodoi{10.1093/mnras/stw444}

\bibitem[{Stephens(1974)}]{Stephens+1974}
Stephens, M.~A. 1974, Journal of the American Statistical Association, 69, 730.
\newblock \url{http://www.jstor.org/stable/2286009}

\bibitem[{Stern(2015)}]{Stern+2015}
Stern, D. 2015, The Astrophysical Journal, 807, 129, \dodoi{10.1088/0004-637X/807/2/129}

\bibitem[{{Stern} {et~al.}(2014){Stern}, {Lansbury}, {Assef}, {Brandt}, {Alexander}, {Ballantyne}, {Balokovi{\'c}}, {Bauer}, {Benford}, {Blain}, {Boggs}, {Bridge}, {Brightman}, {Christensen}, {Comastri}, {Craig}, {Del Moro}, {Eisenhardt}, {Gandhi}, {Griffith}, {Hailey}, {Harrison}, {Hickox}, {Jarrett}, {Koss}, {Lake}, {LaMassa}, {Luo}, {Tsai}, {Urry}, {Walton}, {Wright}, {Wu}, {Yan}, \& {Zhang}}]{Stern+2014}
{Stern}, D., {Lansbury}, G.~B., {Assef}, R.~J., {et~al.} 2014, \apj, 794, 102, \dodoi{10.1088/0004-637X/794/2/102}

\bibitem[{{Suleiman} {et~al.}(2022){Suleiman}, {Noboriguchi}, {Toba}, {Bal{\'a}zs}, {Burgarella}, {Kov{\'a}cs}, {Marton}, {Talafha}, {Frey}, \& {T{\'o}th}}]{Suleiman+2022}
{Suleiman}, N., {Noboriguchi}, A., {Toba}, Y., {et~al.} 2022, \pasj, 74, 1157, \dodoi{10.1093/pasj/psac061}

\bibitem[{{Tabatabaei} {et~al.}(2017){Tabatabaei}, {Schinnerer}, {Krause}, {Dumas}, {Meidt}, {Damas-Segovia}, {Beck}, {Murphy}, {Mulcahy}, {Groves}, {Bolatto}, {Dale}, {Galametz}, {Sandstrom}, {Boquien}, {Calzetti}, {Kennicutt}, {Hunt}, {De Looze}, \& {Pellegrini}}]{Tabatabaei+2017}
{Tabatabaei}, F.~S., {Schinnerer}, E., {Krause}, M., {et~al.} 2017, \apj, 836, 185, \dodoi{10.3847/1538-4357/836/2/185}

\bibitem[{{Tacchella} {et~al.}(2022){Tacchella}, {Conroy}, {Faber}, {Johnson}, {Leja}, {Barro}, {Cunningham}, {Deason}, {Guhathakurta}, {Guo}, {Hernquist}, {Koo}, {McKinnon}, {Rockosi}, {Speagle}, {van Dokkum}, \& {Yesuf}}]{Tacchella+2022}
{Tacchella}, S., {Conroy}, C., {Faber}, S.~M., {et~al.} 2022, \apj, 926, 134, \dodoi{10.3847/1538-4357/ac449b}

\bibitem[{{Tadhunter}(2008)}]{Tadhunter+2008}
{Tadhunter}, C. 2008, \nar, 52, 227, \dodoi{10.1016/j.newar.2008.06.004}

\bibitem[{{Takada} {et~al.}(2014){Takada}, {Ellis}, {Chiba}, {Greene}, {Aihara}, {Arimoto}, {Bundy}, {Cohen}, {Dor{\'e}}, {Graves}, {Gunn}, {Heckman}, {Hirata}, {Ho}, {Kneib}, {Le F{\`e}vre}, {Lin}, {More}, {Murayama}, {Nagao}, {Ouchi}, {Seiffert}, {Silverman}, {Sodr{\'e}}, {Spergel}, {Strauss}, {Sugai}, {Suto}, {Takami}, \& {Wyse}}]{Takada+2014}
{Takada}, M., {Ellis}, R.~S., {Chiba}, M., {et~al.} 2014, \pasj, 66, R1, \dodoi{10.1093/pasj/pst019}

\bibitem[{{Toba} {et~al.}(2017){Toba}, {Bae}, {Nagao}, {Woo}, {Wang}, {Wagner}, {Sun}, \& {Chang}}]{Toba+2017}
{Toba}, Y., {Bae}, H.-J., {Nagao}, T., {et~al.} 2017, \apj, 850, 140, \dodoi{10.3847/1538-4357/aa918a}

\bibitem[{Toba \& Nagao(2016)}]{Toba+2016}
Toba, Y., \& Nagao, T. 2016, The Astrophysical Journal, 820, 46, \dodoi{10.3847/0004-637X/820/1/46}

\bibitem[{{Toba} {et~al.}(2015){Toba}, {Nagao}, {Strauss}, {Aoki}, {Goto}, {Imanishi}, {Kawaguchi}, {Terashima}, {Ueda}, {Bosch}, {Bundy}, {Doi}, {Inami}, {Komiyama}, {Lupton}, {Matsuhara}, {Matsuoka}, {Miyazaki}, {Morokuma}, {Nakata}, {Oi}, {Onoue}, {Oyabu}, {Price}, {Tait}, {Takata}, {Tanaka}, {Terai}, {Turner}, {Uchida}, {Usuda}, {Utsumi}, {Yamada}, \& {Wang}}]{Toba+2015}
{Toba}, Y., {Nagao}, T., {Strauss}, M.~A., {et~al.} 2015, \pasj, 67, 86, \dodoi{10.1093/pasj/psv057}

\bibitem[{{Toba} {et~al.}(2020{\natexlab{a}}){Toba}, {Wang}, {Nagao}, {Ueda}, {Ueda}, {Lim}, {Chang}, {Saito}, \& {Kawabe}}]{Toba+2020sofia}
{Toba}, Y., {Wang}, W.-H., {Nagao}, T., {et~al.} 2020{\natexlab{a}}, \apj, 889, 76, \dodoi{10.3847/1538-4357/ab616d}

\bibitem[{{Toba} {et~al.}(2020{\natexlab{b}}){Toba}, {Yamada}, {Ueda}, {Ricci}, {Terashima}, {Nagao}, {Wang}, {Tanimoto}, \& {Kawamuro}}]{Toba+2020nustar}
{Toba}, Y., {Yamada}, S., {Ueda}, Y., {et~al.} 2020{\natexlab{b}}, \apj, 888, 8, \dodoi{10.3847/1538-4357/ab5718}

\bibitem[{{Tsai} {et~al.}(2015){Tsai}, {Eisenhardt}, {Wu}, {Stern}, {Assef}, {Blain}, {Bridge}, {Benford}, {Cutri}, {Griffith}, {Jarrett}, {Lonsdale}, {Masci}, {Moustakas}, {Petty}, {Sayers}, {Stanford}, {Wright}, {Yan}, {Leisawitz}, {Liu}, {Mainzer}, {McLean}, {Padgett}, {Skrutskie}, {Gelino}, {Beichman}, \& {Juneau}}]{Tsai+2015}
{Tsai}, C.-W., {Eisenhardt}, P. R.~M., {Wu}, J., {et~al.} 2015, \apj, 805, 90, \dodoi{10.1088/0004-637X/805/2/90}

\bibitem[{{Urrutia} {et~al.}(2005){Urrutia}, {Lacy}, {Gregg}, \& {Becker}}]{Urrutia+2005}
{Urrutia}, T., {Lacy}, M., {Gregg}, M.~D., \& {Becker}, R.~H. 2005, \apj, 627, 75, \dodoi{10.1086/430165}

\bibitem[{{Vaccari} {et~al.}(2016){Vaccari}, {Covone}, {Radovich}, {Grado}, {Limatola}, {Botticella}, {Cappellaro}, {Paolillo}, {Pignata}, {De Cicco}, {Falocco}, {Marchetti}, {Brescia}, {Cavuoti}, {Longo}, {Capaccioli}, {Napolitano}, \& {Schipani}}]{Vaccari+2016}
{Vaccari}, M., {Covone}, G., {Radovich}, M., {et~al.} 2016, in The 4th Annual Conference on High Energy Astrophysics in Southern Africa (HEASA 2016), 26, \dodoi{10.22323/1.275.0026}

\bibitem[{{Vito} {et~al.}(2016){Vito}, {Gilli}, {Vignali}, {Brandt}, {Comastri}, {Yang}, {Lehmer}, {Luo}, {Basu-Zych}, {Bauer}, {Cappelluti}, {Koekemoer}, {Mainieri}, {Paolillo}, {Ranalli}, {Shemmer}, {Trump}, {Wang}, \& {Xue}}]{Vito+2016}
{Vito}, F., {Gilli}, R., {Vignali}, C., {et~al.} 2016, \mnras, 463, 348, \dodoi{10.1093/mnras/stw1998}

\bibitem[{{Vito} {et~al.}(2018){Vito}, {Brandt}, {Stern}, {Assef}, {Chen}, {Brightman}, {Comastri}, {Eisenhardt}, {Garmire}, {Hickox}, {Lansbury}, {Tsai}, {Walton}, \& {Wu}}]{Vito+2018}
{Vito}, F., {Brandt}, W.~N., {Stern}, D., {et~al.} 2018, \mnras, 474, 4528, \dodoi{10.1093/mnras/stx3120}

\bibitem[{{Volonteri} {et~al.}(2016){Volonteri}, {Dubois}, {Pichon}, \& {Devriendt}}]{Volonteri+2016}
{Volonteri}, M., {Dubois}, Y., {Pichon}, C., \& {Devriendt}, J. 2016, \mnras, 460, 2979, \dodoi{10.1093/mnras/stw1123}

\bibitem[{{Wang} {et~al.}(2013){Wang}, {Brandt}, {Luo}, {Smail}, {Alexander}, {Danielson}, {Hodge}, {Karim}, {Lehmer}, {Simpson}, {Swinbank}, {Walter}, {Wardlow}, {Xue}, {Chapman}, {Coppin}, {Dannerbauer}, {De Breuck}, {Menten}, \& {van der Werf}}]{Wang+2013}
{Wang}, S.~X., {Brandt}, W.~N., {Luo}, B., {et~al.} 2013, \apj, 778, 179, \dodoi{10.1088/0004-637X/778/2/179}

\bibitem[{{Whitaker} {et~al.}(2017){Whitaker}, {Pope}, {Cybulski}, {Casey}, {Popping}, \& {Yun}}]{Whitaker+2017}
{Whitaker}, K.~E., {Pope}, A., {Cybulski}, R., {et~al.} 2017, \apj, 850, 208, \dodoi{10.3847/1538-4357/aa94ce}

\bibitem[{{Whitaker} {et~al.}(2012){Whitaker}, {van Dokkum}, {Brammer}, \& {Franx}}]{Whitaker+2012}
{Whitaker}, K.~E., {van Dokkum}, P.~G., {Brammer}, G., \& {Franx}, M. 2012, \apjl, 754, L29, \dodoi{10.1088/2041-8205/754/2/L29}

\bibitem[{{Whitaker} {et~al.}(2015){Whitaker}, {Franx}, {Bezanson}, {Brammer}, {van Dokkum}, {Kriek}, {Labb{\'e}}, {Leja}, {Momcheva}, {Nelson}, {Rigby}, {Rix}, {Skelton}, {van der Wel}, \& {Wuyts}}]{Whitaker+2015}
{Whitaker}, K.~E., {Franx}, M., {Bezanson}, R., {et~al.} 2015, \apjl, 811, L12, \dodoi{10.1088/2041-8205/811/1/L12}

\bibitem[{{Williams} {et~al.}(2009){Williams}, {Quadri}, {Franx}, {van Dokkum}, \& {Labb{\'e}}}]{Williams+2009}
{Williams}, R.~J., {Quadri}, R.~F., {Franx}, M., {van Dokkum}, P., \& {Labb{\'e}}, I. 2009, \apj, 691, 1879, \dodoi{10.1088/0004-637X/691/2/1879}

\bibitem[{{Wilson} {et~al.}(2020){Wilson}, {Abi-Saad}, {Ade}, {Aretxaga}, {Austermann}, {Ban}, {Bardin}, {Beall}, {Berthoud}, {Bryan}, {Bussan}, {Castillo}, {Chavez}, {Contente}, {DeNigris}, {Dober}, {Eiben}, {Ferrusca}, {Fissel}, {Gao}, {Golec}, {Golina}, {Gomez}, {Gordon}, {Gutermuth}, {Hilton}, {Hosseini}, {Hubmayr}, {Hughes}, {Kuczarski}, {Lee}, {Lunde}, {Ma}, {Mani}, {Mauskopf}, {McCrackan}, {McKenney}, {McMahon}, {Novak}, {Pisano}, {Pope}, {Ralston}, {Rodriguez}, {S{\'a}nchez-Arg{\"u}elles}, {Schloerb}, {Simon}, {Sinclair}, {Souccar}, {Torres Campos}, {Tucker}, {Ullom}, {Van Camp}, {Van Lanen}, {Velazquez}, {Vissers}, {Weeks}, \& {Yun}}]{Wilson+2020}
{Wilson}, G.~W., {Abi-Saad}, S., {Ade}, P., {et~al.} 2020, in Society of Photo-Optical Instrumentation Engineers (SPIE) Conference Series, Vol. 11453, Millimeter, Submillimeter, and Far-Infrared Detectors and Instrumentation for Astronomy X, ed. J.~{Zmuidzinas} \& J.-R. {Gao}, 1145302, \dodoi{10.1117/12.2562331}

\bibitem[{{Wright} {et~al.}(2010){Wright}, {Eisenhardt}, {Mainzer}, {Ressler}, {Cutri}, {Jarrett}, {Kirkpatrick}, {Padgett}, {McMillan}, {Skrutskie}, {Stanford}, {Cohen}, {Walker}, {Mather}, {Leisawitz}, {Gautier}, {McLean}, {Benford}, {Lonsdale}, {Blain}, {Mendez}, {Irace}, {Duval}, {Liu}, {Royer}, {Heinrichsen}, {Howard}, {Shannon}, {Kendall}, {Walsh}, {Larsen}, {Cardon}, {Schick}, {Schwalm}, {Abid}, {Fabinsky}, {Naes}, \& {Tsai}}]{Wright+2010}
{Wright}, E.~L., {Eisenhardt}, P. R.~M., {Mainzer}, A.~K., {et~al.} 2010, \aj, 140, 1868, \dodoi{10.1088/0004-6256/140/6/1868}

\bibitem[{{Wu} {et~al.}(2012){Wu}, {Tsai}, {Sayers}, {Benford}, {Bridge}, {Blain}, {Eisenhardt}, {Stern}, {Petty}, {Assef}, {Bussmann}, {Comerford}, {Cutri}, {Evans}, {Griffith}, {Jarrett}, {Lake}, {Lonsdale}, {Rho}, {Stanford}, {Weiner}, {Wright}, \& {Yan}}]{Wu+2012}
{Wu}, J., {Tsai}, C.-W., {Sayers}, J., {et~al.} 2012, \apj, 756, 96, \dodoi{10.1088/0004-637X/756/1/96}

\bibitem[{{Xue} {et~al.}(2010){Xue}, {Brandt}, {Luo}, {Rafferty}, {Alexander}, {Bauer}, {Lehmer}, {Schneider}, \& {Silverman}}]{Xue+2010}
{Xue}, Y.~Q., {Brandt}, W.~N., {Luo}, B., {et~al.} 2010, \apj, 720, 368, \dodoi{10.1088/0004-637X/720/1/368}

\bibitem[{{Yan} {et~al.}(2023){Yan}, {Brandt}, {Zou}, {Zhu}, {Chen}, {Hickox}, {Luo}, {Ni}, {Alexander}, {Bauer}, {Vignali}, \& {Vito}}]{Yan+2023}
{Yan}, W., {Brandt}, W.~N., {Zou}, F., {et~al.} 2023, \apj, 951, 27, \dodoi{10.3847/1538-4357/accea6}

\bibitem[{{Yang} {et~al.}(2018){Yang}, {Brandt}, {Vito}, {Chen}, {Trump}, {Luo}, {Sun}, {Xue}, {Koekemoer}, {Schneider}, {Vignali}, \& {Wang}}]{Yang+2018}
{Yang}, G., {Brandt}, W.~N., {Vito}, F., {et~al.} 2018, \mnras, 475, 1887, \dodoi{10.1093/mnras/stx2805}

\bibitem[{{Yang} {et~al.}(2020){Yang}, {Boquien}, {Buat}, {Burgarella}, {Ciesla}, {Duras}, {Stalevski}, {Brandt}, \& {Papovich}}]{Yang+2020}
{Yang}, G., {Boquien}, M., {Buat}, V., {et~al.} 2020, \mnras, 491, 740, \dodoi{10.1093/mnras/stz3001}

\bibitem[{{Yang} {et~al.}(2022){Yang}, {Boquien}, {Brandt}, {Buat}, {Burgarella}, {Ciesla}, {Lehmer}, {Ma{\l}ek}, {Mountrichas}, {Papovich}, {Pons}, {Stalevski}, {Theul{\'e}}, \& {Zhu}}]{Yang+2022}
{Yang}, G., {Boquien}, M., {Brandt}, W.~N., {et~al.} 2022, \apj, 927, 192, \dodoi{10.3847/1538-4357/ac4971}

\bibitem[{{Yang} {et~al.}(2023){Yang}, {Caputi}, {Papovich}, {Arrabal Haro}, {Bagley}, {Behroozi}, {Bell}, {Bisigello}, {Buat}, {Burgarella}, {Cheng}, {Cleri}, {Dav{\'e}}, {Dickinson}, {Elbaz}, {Ferguson}, {Finkelstein}, {Grogin}, {Hathi}, {Hirschmann}, {Holwerda}, {Huertas-Company}, {Hutchison}, {Iani}, {Kartaltepe}, {Kirkpatrick}, {Kocevski}, {Koekemoer}, {Kokorev}, {Larson}, {Lucas}, {P{\'e}rez-Gonz{\'a}lez}, {Rinaldi}, {Shen}, {Trump}, {de la Vega}, {Yung}, \& {Zavala}}]{Yang+2023}
{Yang}, G., {Caputi}, K.~I., {Papovich}, C., {et~al.} 2023, \apjl, 950, L5, \dodoi{10.3847/2041-8213/acd639}

\bibitem[{{Yutani} {et~al.}(2022){Yutani}, {Toba}, {Baba}, \& {Wada}}]{Yutani+2022}
{Yutani}, N., {Toba}, Y., {Baba}, S., \& {Wada}, K. 2022, \apj, 936, 118, \dodoi{10.3847/1538-4357/ac87a2}

\bibitem[{{Zhu} {et~al.}(2023){Zhu}, {Brandt}, {Zou}, {Luo}, {Ni}, {Xue}, \& {Yan}}]{Zhu+2023}
{Zhu}, S., {Brandt}, W.~N., {Zou}, F., {et~al.} 2023, \mnras, 522, 3506, \dodoi{10.1093/mnras/stad1178}

\bibitem[{{Zinn} {et~al.}(2012){Zinn}, {Middelberg}, {Norris}, {Hales}, {Mao}, \& {Randall}}]{Zinn+2012}
{Zinn}, P.~C., {Middelberg}, E., {Norris}, R.~P., {et~al.} 2012, \aap, 544, A38, \dodoi{10.1051/0004-6361/201219349}

\bibitem[{{Zou} {et~al.}(2020){Zou}, {Brandt}, {Vito}, {Chen (陳建廷)}, {Garmire}, {Stern}, \& {Ayubinia}}]{Zou+2020}
{Zou}, F., {Brandt}, W.~N., {Vito}, F., {et~al.} 2020, \mnras, 499, 1823, \dodoi{10.1093/mnras/staa2930}

\bibitem[{{Zou} {et~al.}(2019){Zou}, {Yang}, {Brandt}, \& {Xue}}]{Zou+2019}
{Zou}, F., {Yang}, G., {Brandt}, W.~N., \& {Xue}, Y. 2019, \apj, 878, 11, \dodoi{10.3847/1538-4357/ab1eb1}

\bibitem[{{Zou} {et~al.}(2024){Zou}, {Yu}, {Brandt}, {Tak}, {Yang}, \& {Ni}}]{Zou+2024}
{Zou}, F., {Yu}, Z., {Brandt}, W.~N., {et~al.} 2024, arXiv e-prints, arXiv:2404.00097, \dodoi{10.48550/arXiv.2404.00097}

\bibitem[{{Zou} {et~al.}(2021{\natexlab{a}}){Zou}, {Brandt}, {Lacy}, {Ni}, {Nyland}, {Yang}, {Bauer}, {Covone}, {Grado}, {Napolitano}, {Paolillo}, {Radovich}, {Spavone}, \& {Vaccari}}]{Zou+2021a}
{Zou}, F., {Brandt}, W.~N., {Lacy}, M., {et~al.} 2021{\natexlab{a}}, Research Notes of the American Astronomical Society, 5, 31, \dodoi{10.3847/2515-5172/abe769}

\bibitem[{{Zou} {et~al.}(2021{\natexlab{b}}){Zou}, {Yang}, {Brandt}, {Ni}, {Bauer}, {Covone}, {Lacy}, {Napolitano}, {Nyland}, {Paolillo}, {Radovich}, {Spavone}, \& {Vaccari}}]{Zou+2021b}
{Zou}, F., {Yang}, G., {Brandt}, W.~N., {et~al.} 2021{\natexlab{b}}, Research Notes of the American Astronomical Society, 5, 56, \dodoi{10.3847/2515-5172/abf050}

\bibitem[{{Zou} {et~al.}(2022){Zou}, {Brandt}, {Chen}, {Leja}, {Ni}, {Yan}, {Yang}, {Zhu}, {Luo}, {Nyland}, {Vito}, \& {Xue}}]{Zou+2022}
{Zou}, F., {Brandt}, W.~N., {Chen}, C.-T., {et~al.} 2022, \apjs, 262, 15, \dodoi{10.3847/1538-4365/ac7bdf}

\bibitem[{{Zou} {et~al.}(2023){Zou}, {Brandt}, {Ni}, {Zhu}, {Alexander}, {Bauer}, {Chen}, {Luo}, {Sun}, {Vignali}, {Vito}, {Xue}, \& {Yan}}]{Zou+2023}
{Zou}, F., {Brandt}, W.~N., {Ni}, Q., {et~al.} 2023, \apj, 950, 136, \dodoi{10.3847/1538-4357/acce39}

\end{thebibliography}
\bibliographystyle{aasjournal}
\end{CJK*}
\end{document}